\documentclass[12pt]{article}
\begin{document}
\begin{titlepage}
\begin{flushright}
hep-th/0401012\\
TIT/HEP-516\\
January, 2004\\
\end{flushright}
\vspace{0.5cm}
\begin{center}
{\Large \bf ${\cal N}=2$ Supersymmetric $U(1)$ Gauge Theory\\
in Noncommutative Harmonic Superspace
}
\lineskip .75em
\vskip2.5cm
{\large Takeo Araki},\ \
{\large Katsushi Ito} \
and {\large Akihisa Ohtsuka}
\vskip 2.5em
 {\large\it Department of Physics\\
Tokyo Institute of Technology\\
Tokyo, 152-8551, Japan}  \vskip 4.5em
\end{center}
%\vskip1cm
\begin{abstract}
We study ${\cal N}=2$ supersymmetric $U(1)$ gauge theory in the 
noncommutative harmonic superspace 
with nonanticommutative fermionic coordinates.
We examine the gauge transformation which preserves the Wess-Zumino
gauge by harmonic expansions of component fields. 
The gauge transformation is shown to depend on the deformation
parameters and the anti-holomorphic scalar field.
We compute the action explicitly 
up to the third order in component fields
and discuss the field redefinitions so that the component fields
transform canonically. 
\end{abstract}
\end{titlepage}
\baselineskip=0.7cm
\section{Introduction}
Field theories on noncommutative superspace\cite{ncsuper} have
been taken much attentions recently.
In superstring theory, they correspond to the 
field theories on the D-branes in the graviphoton background\cite{OoVa}.
The Ramond-Ramond fields induces the non(anti)commutativity for the
worldsheet fermionic fields in the hybrid formalism of superstrings
\cite{OoVa,BeSe,DeGrNi}.
When we consider the superstring theory compactified on a Calabi-Yau
threefold, 
the field theory on the D-branes is described by ${\cal N}=1$ super Yang-Mills
theory on the noncommutative superspace.
This theory has been also constructed in terms of chiral
superfields\cite{Se},  
which has been shown to have ${\cal N}={1\over2}$ supersymmetry.

{}From the field theoretical viewpoint, noncommutative superspace
provides various interesting problems.
Compare to the field theories on noncommutative spacetime,
only a finite number of the deformation terms are necessary in the case of 
${\cal N}=1$ noncommutative superspace.
This property makes the deformed Lagrangian simple, so that the
perturbative\cite{Pert} and nonperturbative structure\cite{Inst} 
of field theories can be studied explicitly.

It is an interesting problem to 
study the supersymmetric gauge theory on the 
noncommutative extended superspace\cite{KlPeTa,IvLeZu,FeSo}.
In a previous paper \cite{ArItOh}, 
we investigated the ${\cal N}=2$ super Yang-Mills theory
on the ${\cal N}=1$ noncommutative superspace by adding adjoint chiral
superfields to ${\cal N}=1$ super Yang-Mills theory.
Since the deformation of ${\cal N}=1$ superspace breaks the ${\cal R}$-symmetry
explicitly,
the theory has only ${\cal N}={1\over2}$ supersymmetry.
Noncommutative ${\cal N}=2$ superspace provides another approach to
study the
deformation of ${\cal N}=2$ supersymmetric gauge theories.
${\cal N}=2$ rigid superspace  is 
convenient to construct the on-shell Lagrangian of 
${\cal N}=2$ supersymmetric Yang-Mills theory\cite{GrSoWe}.
The Lagrangian is defined by chiral superfields, which obeys 
certain constrains.
These constraints connects the chiral part and anti-chiral 
parts.
One may consider the noncommutative deformation of rigid superspace.
It turns out that the constraint equations are highly nontrivial 
to solve.

Harmonic superspace \cite{GaIvOgSo} is known to provide a good 
off-shell formulation of  quantum field theory with extended supersymmetry.
This superspace is obtained by adding the harmonic coordinates of
$S^2$ to the rigid ${\cal N}=2$ superspace.
In this formalism, analytic superfields play an important role
instead of ${\cal N}=2$ chiral superfields.
These superfields are unconstrained but includes
infinitely many fields in their components arising from the 
harmonic expansions.

In refs. \cite{FeSo,IvLeZu}, the noncommutative harmonic superspace
has been considered.
Ferrara and Sokatchev \cite{FeSo}
studied ${\cal N}=2$ $U(1)$ gauge theories based on the singlet deformation 
whose Poisson structure includes only the
supercovariant derivatives (singlet $D$ deformation).
This Poisson structure preserves ${\cal N}=2$ supersymmetry and 
analyticity
but
does not preserve chiral structure of the theory.
The Lagrangian depends on certain function of anti-holomorphic scalar field.
Ivanov {\it et al.} have studied the
deformation of ${\cal N}=2$ harmonic superspace\cite{IvLeZu},
whose Poisson structure is characterized by the bi-differential
operator with supercharges called the $Q$ deformation in \cite{FeSo}.
The analytic superfields are also well-defined in this deformation.
But the only the chiral part of ${\cal N}=2$ supersymmetry are realized.

In this paper, we study the $Q$ deformation of ${\cal N}=2$ harmonic superspace
and ${\cal N}=2$ supersymmetric $U(1)$ gauge theory in this noncommutative 
superspace.
We will consider  the most general deformation in the harmonic
superspace.
We examine the gauge transformation of the theory in detail
in terms of component fields.
The action of ${\cal N}=2$ $U(1)$ gauge theory, as observed 
in the case of the singlet $D$ deformation \cite{FeSo},
is shown to contain arbitrary power of anti-holomorphic scalar fields.
This means also the action contains infinite number of
deformation parameters.
In this paper, we will examine the action up to the third order contributions
in component fields.

This paper is organized as follows:
In section 2, we review ${\cal N}=2$ harmonic superspace and 
study its deformation and the Poisson structure.
In section 3, we discuss the Lagrangian of the ${\cal N}=2$ supersymmetric
$U(1)$ gauge theory on the ${\cal N}=2$ harmonic superspace.
We will calculate the Lagrangian up to the
third order in component fields explicitly.
In section 4, 
we will study the gauge transformation in the deformed harmonic
superspace
and discuss the field redefinitions.
In the appendix A, we summarize the Poisson structure in the analytic
basis.
In the appendix B, we will calculate the action of the third order
in component fields explicitly.

\section{Noncommutative Harmonic Superspace}
In this section, we introduce noncommutative harmonic superspace and
study the Poisson structure.
We begin with reviewing the ${\cal N}=2$ harmonic superspace which has been
explained in detail in \cite{GaIvOgSo}.
Let $(x^{\mu}, \theta^{\alpha}_{i},\bar{\theta}^{\dot{\alpha} i})$ be the 
coordinates of the ${\cal N}=2$ rigid superspace, where $\mu=0,1,2,3$ are
indices of spacetime coordinates, $\alpha,\dot{\alpha}=1,2$ spinor
indices and $i=1,2$ labels the $SU(2)$ doublet.
The signature of spacetime is Euclidean. 
Concerning spinor indices, we will follow the notation in ref. \cite{WeBa}.
Namely spinors with upper and lower indices are related through the
$\varepsilon$-tensor with $\varepsilon^{12}=-\varepsilon_{12}=1$.
On the other hand, raising and lowering $SU(2)$ indices are done with
the help of
antisymmetric tensor $\epsilon_{ij}$ with
$\epsilon^{12}=-\epsilon_{12}=-1$,
 $\theta^{i}=\epsilon^{ij}\theta_{j}$
and  $\theta_{i}=\epsilon_{ij}\theta^{j}$ .

The supersymmetry generators $Q^{i}_{\alpha}$,
$\bar{Q}_{\dot{\alpha}i}$ and the supercovariant
derivatives $D_{\alpha}^{i}$, $\bar{D}_{\dot{\alpha}i}$ are
defined by
\begin{eqnarray}
Q_{\alpha}^{i}&=&{\partial\over\partial\theta^{\alpha}_{i}}
-i(\sigma^{\mu})_{\alpha\dot{\alpha}}\bar{\theta}^{\dot{\alpha} i}
{\partial\over\partial x^{\mu}}, \quad
\bar{Q}_{\dot{\alpha} i}= -{\partial\over\partial\bar{\theta}^{\dot{\alpha} i}}
+i\theta^{\alpha}_{i}(\sigma^{\mu})_{\alpha\dot{\alpha}}
{\partial\over \partial x^{\mu}}
\nonumber\\
D_{\alpha}^{i}&=&{\partial\over\partial\theta^{\alpha}_{i}}
+i(\sigma^{\mu})_{\alpha\dot{\alpha}}\bar{\theta}^{\dot{\alpha} i}
{\partial\over\partial x^{\mu}}, \quad
\bar{D}_{\dot{\alpha} i}= -{\partial\over\partial\bar{\theta}^{\dot{\alpha} i}}
-i\theta^{\alpha}_{i}(\sigma^{\mu})_{\alpha\dot{\alpha}}
{\partial\over \partial x^{\mu}}.
\end{eqnarray}
They satisfy the anticommutation relations
\begin{eqnarray}
 \{ D^{i}_{\alpha}, D^{j}_{\beta}\}&=& 
\{\bar{D}_{\dot{\alpha} i},\bar{D}_{\dot{\beta} j}\}=0,\quad
\{ D^{i}_{\alpha},\bar{D}_{\dot{\beta} j}\}=
-2i \delta^{i}_{j}(\sigma^{\mu})_{\alpha\dot{\beta}}
{\partial\over \partial x^{\mu}},\nonumber\\
 \{ Q^{i}_{\alpha}, Q^{j}_{\beta}\}&=& 
\{\bar{Q}_{\dot{\alpha} i},\bar{Q}_{\dot{\beta} j}\}=0,\quad
\{ Q^{i}_{\alpha},\bar{Q}_{\dot{\beta} j}\}=
2i \delta^{i}_{j}(\sigma^{\mu})_{\alpha\dot{\beta}}
{\partial\over \partial x^{\mu}},\nonumber\\
 \{D^{i}_{\alpha},Q^{j}_{\beta} \}
&=&\{ D^{i}_{\alpha}, \bar{Q}_{\dot{\beta} j}\}
=\{ \bar{D}_{\dot{\alpha} i}, Q^{j}_{\beta} \}
=\{ \bar{D}_{\dot{\alpha} i},\bar{Q}_{\dot{\beta} j}\}=0.
\end{eqnarray}
We introduce the left-handed chiral basis 
$(x^{\mu}_{L},\theta^{\alpha}_{i},\bar{\theta}^{\dot{\alpha} i})$
in superspace, where
\begin{equation}
 x^\mu_{L}= x^\mu+i\theta_i \sigma^\mu \bar{\theta}^i.
\end{equation}
In this basis, supercharges and supercovariant derivatives
take the form
\begin{eqnarray}
 Q^{i}_{\alpha}&=& {\partial \over \partial\theta^{\alpha}_{i}},\quad
\bar{Q}_{i\dot{\alpha}}= -{\partial\over\partial \bar{\theta}^{i\dot{\alpha} }}
+2i \theta^{\alpha}_{i}\sigma^{\mu}_{\alpha\dot{\alpha}}
{\partial\over \partial x^{\mu}_{L}}\nonumber\\
 D^{i}_{\alpha}&=& {\partial\over\partial\theta^{\alpha}_{i}}
+2i \sigma^{\mu}_{\alpha\dot{\alpha}}\bar{\theta}^{\dot{\alpha} i}
{\partial\over \partial x^{\mu}_{L}}\quad
\bar{D}_{\dot{\alpha} i}= -{\partial \over \partial 
\bar{\theta}^{\dot{\alpha} i}}.
\label{eq:chiral1}
\end{eqnarray}
In the ${\cal N}=2$ rigid superspace, we introduce nonanticommutative 
deformation:
\begin{equation}
 \{ \theta^{\alpha}_{i}, \theta^{\beta}_{j}\}_{*}=C^{\alpha\beta}_{ij},
\label{eq:nac1}
\end{equation} 
Here the anticommutation relation is calculated by using 
the $*$-product, which is defined  by
\begin{equation}
 f*g(\theta)=f(\theta)\exp\left\{-{1\over2}
\overleftarrow{\partial\over\partial \theta^{\alpha}_{i}}
C^{\alpha\beta}_{ij}\overrightarrow{\partial\over\partial\theta^{\beta}_{j}}
\right\}g(\theta)
\label{eq:moyal1}
\end{equation}
for functions $f$ and $g$ of $\theta$.
We assume that $x_{L}^{\mu}$  and $\bar{\theta}_{\dot{\alpha}i}$
(anti-)commute with other coordinates
\begin{eqnarray}
 [x^{\mu}_{L}, x^{\nu}_{L}]_{*}&=&[x^{\mu}_{L}, \theta^{\alpha}_{i}]_{*}=
[x^{\mu}_{L},\bar{\theta}^{\dot{\alpha} i}]_{*}=0, \nonumber\\
\{ \bar{\theta}^{\dot{\alpha} i}, \bar{\theta}^{\dot{\beta} j}\}_{*}&=&
\{ \bar{\theta}^{\dot{\alpha} i}, \theta^{\alpha}_{j}\}_{*}=0,
\label{eq:nac2}
\end{eqnarray}
which is a generalization of  ${\cal N}=1$ noncommutative 
superspace\cite{OoVa,Se}.
$C^{\alpha\beta}_{ij}$ is symmetric
under the exchange of pairs of indices $(\alpha i)$,$(\beta j)$:
 $C^{\alpha\beta}_{ij}=C^{\beta\alpha}_{ji}$.
We decompose the nonanticommutative parameter $C^{\alpha\beta}_{ij}$
into symmetric and antisymmetric parts with respect to $SU(2)$ indices,
such as
\begin{equation}
C^{\alpha\beta}_{ij}=C^{\alpha\beta}_{(ij)}
+{1\over4}\epsilon_{ij}\varepsilon^{\alpha\beta}C_{s}.
\end{equation}
Here $C_{s}$ corresponds to the singlet deformation parameter\cite{FeSo}.
$C^{\alpha\beta}_{(ij)}$ is symmetric with respect to $i$ and $j$, 
and also $\alpha$ and $\beta$.
Since $x^{\mu}_{L}$ are commuting variables, spacetime coordinates 
$x^{\mu}$ have nontrivial (anti-)commutations:
\begin{eqnarray}
 [ x^{\mu},x^{\nu}]_{*}&=&
C^{\mu\nu}_{(ij)}\bar{\theta}^{i}\bar{\theta}^{j}
+{1\over2}C_s \epsilon_{ij}\bar{\theta}^i\bar{\sigma}^{\mu\nu}\bar{\theta}^j
\nonumber\\
 \mbox{[} x^{\mu},\theta^{\alpha}_{i}\mbox{]}_{*}&=& i C^{\alpha\beta}_{ij} 
(\sigma^{\mu})_{\beta\dot{\beta}}\bar{\theta}^{\dot{\beta} j}
\label{eq:nac3}
\end{eqnarray}
where
\begin{equation}
C^{\mu\nu}_{(ij)}=C^{\alpha\beta}_{(ij)}
(\sigma^{\mu\nu})_{\alpha} {}^{\gamma} \varepsilon_{\beta\gamma}
\end{equation}
and $\sigma^{\mu\nu}={1\over4}
(\sigma^{\mu}\bar{\sigma}^{\nu}-\sigma^{\nu}\bar{\sigma}^{\mu})$,
$\bar{\sigma}^{\mu\nu}={1\over4}
(\bar{\sigma}^{\mu}\sigma^{\nu}-\bar{\sigma}^{\nu}\sigma^{\mu})$.

We next discuss ${\cal N}=2$ harmonic superspace. It is formulated by
introducing harmonic variables $u^{\pm}_{i}$
which form a $2\times 2$ $SU(2)$ matrix.
The harmonic variables satisfy the condition
\begin{equation}
 u^{+i}u^{-}_{i}=1, \quad
\overline{u^{+i}}=u^{-}_{i}.
\end{equation}
The coordinates of the harmonic superspace are 
$(x^{\mu},\theta^{\alpha}_{i},\bar{\theta}_{\dot{\alpha}i},
u^{\pm}_{i})$.
Using $u^{\pm}_{i}$, $SU(2)$ indices can be projected into two parts
which have $\pm 1$ $U(1)$ charges,
\begin{equation}
 D^{\pm}_{\alpha}=u^{\pm}_{i}D^{i}_{\alpha}, \quad
\bar{D}^{\pm}_{\alpha}=u^{\pm}_{i}\bar{D}^{i}_{\alpha}.
\end{equation}
$D^{i}_{\alpha}$ is solved by $D^{\pm}_{\alpha}$ such as
$D_{\alpha}{}_i=u^{+}_{i}D^{-}_{\alpha}-u^{-}_{i}D^{+}_{\alpha}$
with the help of the completeness condition
$u^{+}_{i}u^{-}_{j}-u^{+}_{j}u^{-}_{i}=\epsilon_{ij}$.
In the harmonic superspace, an important notion is an analytic
superfield rather than chiral superfield satisfying 
$\bar{D}_{\dot{\alpha}}^{i}\Phi=0$.
An analytic superfield $\Phi(x,\theta,\bar{\theta},u)$ is defined by
\begin{equation}
 D^{+}_{\alpha}\Phi=\bar{D}^{+}_{\dot{\alpha}}\Phi=0,
\end{equation}
which can be conveniently written in terms of analytic coordinates:
\begin{eqnarray}
 x_{A}^{\mu}&=& 
x^{\mu}-i (\theta^i \sigma^\mu \bar{\theta}^j +\theta^j
\sigma^\mu\bar{\theta}^i)
u^{+}_{i}u^{-}_{j}
=x^{\mu}-i (\theta^{+}\sigma^{\mu}\bar{\theta}^{-}+\theta^{-}
\sigma^{\mu}\bar{\theta}^{+}),
\\
\theta^{\pm}_{\alpha}&=& u^{\pm }_{i}\theta^{i}_{\alpha},\quad
\bar{\theta}^{\pm}_{\dot{\alpha}}= u^{\pm}_{i}\bar{\theta}^{i}_{\dot{\alpha}}.
\end{eqnarray}
The supercharges and supercovariant
derivatives take the form
\begin{eqnarray}
 Q^{+}_{\alpha}&=& {\partial\over\partial\theta^{-\alpha}}
-2i \sigma^{\mu}_{\alpha\dot{\alpha}}\bar{\theta}^{+\dot{\alpha}}
{\partial\over\partial x^{\mu}_{A}},\quad
Q^{-}_{\alpha}=-{\partial\over\partial \theta^{+\alpha}}\nonumber\\
\bar{Q}^{+}_{\dot{\alpha}}&=&{\partial\over\partial
\bar{\theta}^{-\dot{\alpha}}}
+2i \theta^{+\alpha}\sigma^{\mu}_{\alpha\dot{\alpha}}
{\partial\over \partial x^{\mu}_{A}},
\quad
\bar{Q}^{-}_{\dot{\alpha}}=
-{\partial\over\partial\bar{\theta}^{+\dot{\alpha}}}\nonumber\\
 D^{+}_{\alpha}&=& {\partial\over\partial\theta^{-\alpha}},\quad
D^{-}_{\alpha}= -{\partial\over\partial\theta^{+\alpha}}
+2i(\sigma^\mu)_{\alpha\dot{\alpha}}\bar{\theta}^{-\dot{\alpha}}
{\partial\over\partial x_{A}^{\mu}},\nonumber\\
\bar{D}^{+}_{\dot{\alpha}}&=& {\partial\over\partial 
\bar{\theta}^{-\dot{\alpha}}},\quad
\bar{D}^{-}_{\dot{\alpha}}=
-{\partial\over\partial\bar{\theta}^{+\dot{\alpha}}}
-2i \theta^{-\alpha}\sigma^{\mu}_{\alpha\dot{\alpha}}
{\partial\over\partial x_{A}^{\mu}}.
\end{eqnarray}
Their nontrivial anticommutation relations are
\begin{eqnarray}
\{ Q^{+}_{\alpha},\bar{Q}^{-}_{\dot{\alpha}}\}&=& 2i 
\sigma^{\mu}_{\alpha\dot{\alpha}}
{\partial\over \partial x^{\mu}_{A}}, \quad
\{ Q^{-}_{\alpha}, \bar{Q}^{+}_{\dot{\alpha}}\}=
-2i  \sigma^{\mu}_{\alpha\dot{\alpha}}
{\partial\over \partial x^{\mu}_{A}}, \nonumber\\
\{ D^{+}_{\alpha}, \bar{D}^{-}_{\dot{\alpha}}\}
&=& -2i \sigma^{\mu}_{\alpha\dot{\alpha}}{\partial\over
\partial x_{A}^{\mu}},\quad
\{ \bar{D}^{+}_{\dot{\alpha}}, D^{-}_{\alpha} \}=
2i \sigma^{\mu}_{\alpha\dot{\alpha}}{\partial\over\partial x_{A}^{\mu}}.
\end{eqnarray}
An analytic superfield $\Phi$ is functions
of $(x^{\mu}_{A},\theta^{+},\bar{\theta}^{+},u)$,
\begin{equation}
 \Phi=\Phi(x^{\mu}_{A},\theta^{+},\bar{\theta}^{+},u).
\end{equation}
Expanding in $\theta$, $\Phi$ takes the form
\begin{eqnarray}
\Phi&=& \phi(x_A,u)+\theta^{+}\psi(x_A,u)+
\bar{\theta}^{+}\bar{\chi}(x_A,u)
+(\theta^{+})^2 M(x_A,u)
+(\bar{\theta}^{+})^2 N(x_A,u)\nonumber\\
&& +\theta^{+}\sigma^{\mu}\bar{\theta}^{+}A_{\mu}(x_A,u)
+(\bar{\theta}^{+})^2\theta^{+}\lambda(x_A,u)
+(\theta^{+})^2\bar{\theta}^{+}\bar{\kappa}(x_A,u)
+(\theta^{+})^2(\bar{\theta}^{+})^2 D(x_A,u).\nonumber\\
\label{eq:analytic1}
\end{eqnarray}
and each $\theta$-components are expanded in the harmonic variables,
e.g.,
\begin{equation}
 \phi(x_A,u)=\sum_{n=0}^{\infty}
\phi^{(i_1\cdots i_{q+n}j_{1}\cdots j_{n})}(x_{A})
u^{+}_{(i_1}\cdots u^{+}_{i_{q+n}}u^{-}_{j_1}\cdots u^{-}_{j_n)}
\end{equation}
for the field with $U(1)$ charge $q\geq 0$.
Similar formula holds for $q<0$ .
Here we define $A_{(i_1\cdots i_{n})}$ as the symmetrized sum over the
permutation of indices $i_1, \cdots i_n$:
\begin{equation}
 A_{(i_{1}\cdots i_{n})}={1\over n!}
\sum_{\sigma\in S_{n}} A_{i_{\sigma(1)}\cdots i_{\sigma(n)}}
\end{equation}
and $S_n$ denotes the permutation group of degree $n$.
We define the harmonic derivatives $D^{\pm\pm}$ and $D^{0}$ by
\begin{equation}
 D^{\pm\pm}=u^{+i}{\partial\over \partial u^{-i}},\quad
D^{0}=u^{+i}{\partial \over \partial u^{+i}}
-u^{-i}{\partial \over \partial u^{-i}}.
\end{equation}
They form an $SU(2)$ algebra.
In the analytic basis, they are expressed as
\begin{eqnarray}
  D^{\pm\pm}&=&\partial^{++}
-2i\theta^{\pm}\sigma^{\mu}\bar{\theta}^{\pm}
{\partial\over \partial x^{\mu}_{A}}
+\theta^{\pm \alpha}{\partial\over \partial \theta^{\mp\alpha}}
+\bar{\theta}^{\pm\dot{\alpha}}{\partial\over \partial
\bar{\theta}^{\mp\dot{\alpha}}},\nonumber\\
D^{0}&=& \partial^{0}
+\theta^{+\alpha}{\partial\over \partial \theta^{+\alpha}}
-\theta^{-\alpha}{\partial\over \partial \theta^{-\alpha}}
+\bar{\theta}^{+\dot{\alpha}}{\partial\over \partial
\bar{\theta}^{+\dot{\alpha}}}
-\bar{\theta}^{-\dot{\alpha}}{\partial\over \partial
\bar{\theta}^{-\dot{\alpha}}},
\end{eqnarray}
where $\partial^{\pm\pm}$ and $\partial^{0}$ denote the harmonic
derivatives with fixed $x_{A}$, $\theta^{\pm}$ and $\bar{\theta}^{\pm}$.

When we introduce the deformation (\ref{eq:nac1}) 
into the harmonic superspace, we note that the derivative in 
the $*$-product (\ref{eq:moyal1}) 
is taken with the fixed chiral coordinates $x_{L}$.
Therefore it is convenient to write the $*$-product 
of the form
\begin{equation}
 f*g(\theta)=f(\theta)\exp(P) g(\theta),\quad
P=-{1\over2}
\overleftarrow{Q^{i}_{\alpha}}
C^{\alpha\beta}_{ij}\overrightarrow{Q^{j}_{\beta}}.
\label{eq:moyal2}
\end{equation}
In the chiral basis, this formula leads to (\ref{eq:moyal1}) due to
(\ref{eq:chiral1}).
We call $P$ the Poisson structure.
In the analytic basis, one can compute the $*$-product by using
$Q^{i}_{\alpha}=u^{+i}Q^{-}_{\alpha}-u^{-i}Q^{+}_{\alpha}$.
For example we have
\begin{eqnarray}
\{ \theta^{\eta},\theta^{\eta'} \}_{*}&=&C^{\eta \eta' \alpha\beta},
\quad (\eta,\eta'=\pm)
\nonumber\\
 \mbox{[}x^{\mu}_{A}, x^{\nu}_{A}\mbox{]}_{*}&=& 4 C^{--\mu\nu}
(\bar{\theta}^{+})^2
\nonumber\\
\mbox{[}x^{\mu}_{A},\theta^{\pm}_{\alpha} \mbox{]}_{*}&=&
-2i C^{-\eta \beta\alpha}(\sigma^{\mu}\bar{\theta}^{+})_{\beta}
\end{eqnarray}
where
$C^{\eta\eta'\mu\nu}=u^{\eta i}u^{\eta' j}C^{\mu\nu}_{ij}$.
For superfields $A$ and $B$, the $*$-product takes the form
\begin{equation}
 A*B=AB+APB+{1\over2}AP^2B+{1\over6}AP^3B+{1\over24}AP^4B,\quad
P^5=0.
\label{eq:star2}
\end{equation}
We note that the Poisson structure $P$ commutes with 
the supercovariant derivatives. Therefore the analytic structure is 
preserved when we introduce noncommutativity in harmonic superspace.
Concerning supersymmetry, only $Q^{i}_{\alpha}$ are 
conserved. 
Field theory on this noncommutative superspace has chiral ${\cal N}=1$
 (or ${\cal N}=(1,0)$
in the sense of \cite{IvLeZu}) supersymmetry.

We now investigate the $*$-product for analytic superfields in detail 
for later use.
There are two ways of calculating this $*$-product.
One way is to use the formula (\ref{eq:moyal2}) directly for
analytic superfields.
This approach is helpful to keep the Grassmann analytic structure 
of the theory manifestly.
Since $D^{+}_{\alpha}$ and $\bar{D}^{+}_{\dot{\alpha}}$ commute with
the Poisson structure $P$, the product $A*B$ is also analytic
if $A$ and $B$ are analytic.
Another way is to expand the analytic superfields in terms of chiral
basis and to apply the formula (\ref{eq:moyal1}).
The latter approach is relatively 
easier to perform and be useful for the calculation of the action, 
which will be explained in detail in the next section.

Firstly we study the $*$-product in the analytic basis.
Using $P$ in (\ref{eq:moyal2}), $\exp P$ becomes
\begin{eqnarray}
 \exp P
&=&
1-{1\over2}
\overleftarrow{Q^{\alpha}_{i}}C^{\alpha\beta}_{ij}
\overrightarrow{Q^{j}_{\beta}}
+
{1\over8} \overleftarrow{Q^{\alpha_1}_{i_1}}
\overleftarrow{Q^{\alpha_2}_{i_2}}C^{\alpha_1\beta_1}_{i_1j_1}
C^{\alpha_2\beta_2}_{i_2j_2}
\overrightarrow{Q^{j_2}_{\beta_2}}
\overrightarrow{Q^{j_1}_{\beta_1}}
\nonumber\\
&&-{1\over48}
\overleftarrow{Q^{\alpha_1}_{i_1}}
\overleftarrow{Q^{\alpha_2}_{i_2}}
\overleftarrow{Q^{\alpha_3}_{i_3}}
C^{\alpha_1\beta_1}_{i_1j_1}
C^{\alpha_2\beta_2}_{i_2 j_2}
C^{\alpha_3\beta_3}_{i_3 j_3}
\overrightarrow{Q^{j_3}_{\beta_3}}
\overrightarrow{Q^{j_2}_{\beta_2}}
\overrightarrow{Q^{j_1}_{\beta_1}}\nonumber\\
&&
+{1\over384}
\overleftarrow{Q^{\alpha_1}_{i_1}}
\overleftarrow{Q^{\alpha_2}_{i_2}}
\overleftarrow{Q^{\alpha_3}_{i_3}}
\overleftarrow{Q^{\alpha_4}_{i_4}}
C^{\alpha_1\beta_1}_{i_1j_1}
C^{\alpha_2\beta_2}_{i_2 j_2}
C^{\alpha_3\beta_3}_{i_3 j_3}
C^{\alpha_4\beta_4}_{i_4 j_4}
\overrightarrow{Q^{j_4}_{\beta_4}}
\overrightarrow{Q^{j_3}_{\beta_3}}
\overrightarrow{Q^{j_2}_{\beta_2}}
\overrightarrow{Q^{j_1}_{\beta_1}}.
\end{eqnarray}
By computing the action of supercharges on the analytic superfield,
one can compute the $*$-product.
For the analytic superfield $\Phi$ in (\ref{eq:analytic1}), we obtain
\begin{eqnarray}
 Q^{i}_{\alpha}\Phi
&=&
-u^{+i}\psi_{\alpha}
-2u^{+i}\theta^{+}_{\alpha}M
+\bar{\theta}^{+\dot{\alpha}}\left\{-u^{+i}
\sigma^{\mu}_{\alpha\dot{\alpha}}A_{\mu}
+2i u^{-i}\sigma^{\mu}_{\alpha\dot{\alpha}}\partial_{\mu}\phi\right\}
\nonumber\\
&&
\!\!\!\!
-(\bar{\theta}^{+})^2\left\{u^{+i}\lambda_{\alpha}
+iu^{-i}\sigma^{\mu}_{\alpha\dot{\alpha}}
\partial_{\mu}\bar{\chi}^{\dot{\alpha}}
\right\}
+(\theta^{+}\sigma^{\mu}\bar{\theta}^{+})
\left\{
u^{+i} (\sigma^{\mu}\bar{\kappa})_{\alpha}
-iu^{-i}
(\partial_{\nu}\psi\sigma^{\mu}\bar{\sigma}^{\nu})^{\gamma}
\varepsilon_{\alpha\gamma}
\right\}
\nonumber\\
&&
+(\bar{\theta}^{+})^2\theta^{+\beta}
\left\{-u^{+i}2\varepsilon_{\alpha\beta}D
-i u^{-i}(\sigma^{\nu}\bar{\sigma}^{\mu})_{\beta}{}^{\gamma}
\varepsilon_{\alpha\gamma}\partial_{\mu}A_{\nu}
\right\}
+2i u^{-i} (\theta^+)^2 \sigma^{\mu}_{\alpha\dot{\alpha}}
\bar{\theta}^{+\dot{\alpha}}\partial_{\mu}M
\nonumber\\
&&-u^{-i}i
(\theta^+)^2(\bar{\theta}^{+})^2 \sigma^{\mu}_{\alpha\dot{\alpha}}
\partial_{\mu}\bar{\kappa}^{\dot{\alpha}}.
\label{eq:susy1}
\end{eqnarray}
Since $Q^{i}_{\alpha}\Phi$ is also of the form (\ref{eq:analytic1}), 
we can apply the above formula to compute 
$Q^{i_2}_{\alpha_2}Q^{i_1}_{\alpha_1}\Phi$ etc., which is given in the
appendix A.
The $*$-product $A*B$ can be expressed as
\begin{eqnarray}
A*B&=& AB-(-1)^{|A|}
{1\over2}Q^{i}_{\alpha}A C^{\alpha\beta}_{ij} Q^{j}_{\beta}B
-{1\over8} (Q^{i_2}_{\alpha_2} Q^{i_1}_{\alpha_1} A)
C^{\alpha_1\beta_1}_{i_1j_1} C^{\alpha_2\beta_2}_{i_2j_2}
(Q^{j_2}_{\beta_2} Q^{j_1}_{\beta_1}) B\nonumber\\
&& 
+{1\over48}(-1)^{|A|}
(Q^{i_3}_{\alpha_3}  Q^{i_2}_{\alpha_2} Q^{i_1}_{\alpha_1}  A)
C^{\alpha_1\beta_1}_{i_1j_1}
C^{\alpha_2\beta_2}_{i_2 j_2}
C^{\alpha_3\beta_3}_{i_3 j_3}
(Q^{j_3}_{\beta_3} Q^{j_2}_{\beta_2} Q^{j_1}_{\beta_1}B)  \nonumber\\
&&
+{1\over384}
(Q^{i_4}_{\alpha_4} Q^{i_3}_{\alpha_3} Q^{i_2}_{\alpha_2} Q^{i_1}_{\alpha_1} A)
C^{\alpha_1\beta_1}_{i_1j_1}
C^{\alpha_2\beta_2}_{i_2 j_2}
C^{\alpha_3\beta_3}_{i_3 j_3}
C^{\alpha_4\beta_4}_{i_4 j_4}
(Q^{j_4}_{\beta_4}Q^{j_3}_{\beta_3}Q^{j_2}_{\beta_2} Q^{j_1}_{\beta_1}B) .
\end{eqnarray}
where $|A|=0 (1)$  if $A$ is Grassmann even (odd).

We next compute the $*$-product in the chiral basis.
The analytic coordinates $x^{\mu}_{A}$ is related to the chiral
coordinates $x^{\mu}_{L}$ by 
\begin{equation}
 x^{\mu}_{A}=x^{\mu}_{L}-2i \theta^{-}\sigma^{\mu}\bar{\theta}^{+}.
\end{equation}
A function $f(x_{A})$ of $x_{A}$ is expanded in the chiral basis
\begin{equation}
  f(x_{A})
= f(x_{L})-2i\theta^{-}\sigma^{\mu}\bar{\theta}^{+}\partial_{\mu}f(x_{L})
+(\theta^{-})^2(\bar{\theta}^{+})^2\partial^{\mu}\partial_{\mu}f(x_{L}).
\end{equation}
Using this formula, we can express an analytic superfield $\Phi$ 
in (\ref{eq:analytic1})
in the chiral basis:
\begin{eqnarray}
\Phi&=& \phi(x_{L})
+\theta^{+}\psi(x_{L})+\bar{\theta}^{+}\bar{\chi}(x_{L})\nonumber\\
&& +(\theta^{+})^2 M(x_{L})+(\bar{\theta}^{+})^2N(x_{L})
+\theta^{+}\sigma^{\mu}\bar{\theta}^{+}A_{\mu}(x_{L})
-2i \theta^{-}\sigma^{\mu}\bar{\theta}^{+}\partial_{\mu}\phi(x_{L})\nonumber\\
&&
-2i (\theta^{-}\sigma^{\mu}\bar{\theta}^{+})\theta^{+}\partial_{\mu}\psi(x_{L})
+i(\bar{\theta}^{+})^2\theta^{-}\sigma^{\mu}\partial_{\mu}\bar{\chi}(x_{L})
+(\bar{\theta}^{+})^2\theta^{+}\lambda+(\theta^{+})^2
\bar{\theta}^{+}\bar{\kappa}(x_{L})\nonumber\\
&&
+(\theta^{-})^2(\bar{\theta}^{+})^2\partial^{\mu}\partial_{\mu}\phi(x_{L})
-2i (\theta^{+})^2 (\theta^{-}\sigma^{\mu}\bar{\theta}^{+})
\partial_{\mu}M(x_{L})
-i (\bar{\theta}^{+})^2\theta^{+}\sigma^{\mu}\bar{\sigma}^{\nu}\theta^{-}
\partial_{\nu}A_{\mu}(x_{L})\nonumber\\
&&+(\theta^{+})^2(\bar{\theta}^{+})^2 D(x_{L})
+(\theta^{-})^2 (\bar{\theta}^{+})^2 \theta^{+}\partial^{\mu}
\partial_{\mu}\psi(x_{L})
+i (\theta^{+})^2 (\bar{\theta}^{+})^2 \theta^{-}\sigma^{\mu}
\partial_{\mu}\bar{\kappa}(x_{L})
\nonumber\\
&&
+(\theta^{+})^2 (\theta^{-})^2 (\bar{\theta}^{+})^2\partial^{\mu}
\partial_{\mu}M(x_{L}).
\end{eqnarray}
In the chiral basis, 
${1\over n!}P^n$ for $n=1,2,3,4$ in (\ref{eq:star2}) become
\begin{eqnarray}
P
&=&
-\frac12 \overleftarrow{{\partial\over \partial \theta_i^\alpha}}
 C_{ij}^{\alpha\beta}
\overrightarrow{{\partial \over \partial \theta_j^\beta}},
\end{eqnarray}
\begin{eqnarray}
\frac12 P^2
&=&
\overleftarrow{{\partial \over \partial \theta_1 \theta_1}}
\left( -\det  C_{11}{} \overrightarrow{
{\partial \over \partial \theta_1 \theta_1}}
-\det  C_{12}{} \overrightarrow{{\partial \over \partial \theta_2 \theta_2}}
		- \frac12 (C_{11} \  \varepsilon C_{12} )^{\beta \delta}
%\overrightarrow{\partial^2}_{\delta}
\overrightarrow{{\partial\over \partial\theta_2^\delta}}
%\overrightarrow{\partial^1}_{\beta}
\overrightarrow{{\partial\over \partial\theta_1^\beta}}
	\right)
	\nonumber\\
&&{}
+ \overleftarrow{{\partial \over \partial \theta_2 \theta_2}}
\left( -\det  C_{21}{} \overrightarrow{{\partial \over 
\partial \theta_1 \theta_1}}
-\det  C_{22}{} \overrightarrow{{\partial \over \partial \theta_2 \theta_2}}
		+ \frac12 (C_{22} \  \varepsilon C_{21} )^{\delta \beta}
\overrightarrow{{\partial\over \partial\theta_2^\delta}}
\overrightarrow{{\partial\over \partial\theta_1^\beta}}
	\right)
	\nonumber\\
&&{}
+\overleftarrow{{\partial\over \partial\theta_1^\alpha}}
\overleftarrow{{\partial\over \partial\theta_2^\gamma}}
		\left(
		- \frac12  (C_{21} \  \varepsilon C_{11} )^{\gamma \alpha } 
		\overrightarrow{{\partial \over \partial \theta_1 \theta_1}}
		+ \frac12  (C_{12} \  \varepsilon C_{22} )^{\alpha \gamma } 
		\overrightarrow{{\partial \over \partial \theta_2 \theta_2}}
		\right. \nonumber\\
&&\left. {}\qquad 
+ \frac14  C_{11}^{\alpha\beta}  C_{22}^{\gamma\delta} 
\overrightarrow{{\partial\over \partial\theta_2^\delta}}
\overrightarrow{{\partial\over \partial\theta_1^\beta}}
		+ \frac14  C_{12}^{\alpha\beta}  C_{21}^{\gamma\delta} 
\overrightarrow{{\partial\over \partial\theta_1^\delta}}
\overrightarrow{{\partial\over \partial\theta_2^\beta}}
		\right)
	,
\end{eqnarray}
\begin{eqnarray}
{1\over 3!} P^3
&=& 
	\frac12 \left[
	\left( \det  C_{11}{} \   C_{22}^{\alpha\beta}
+ ( C_{21} \  \varepsilon C_{11} \  \varepsilon C_{12} )^{\alpha\beta} 
		\right) 
\overleftarrow{{\partial \over \partial \theta_1 \theta_1}} 
\overleftarrow{{\partial\over \partial\theta_2^\alpha}}
\overrightarrow{{\partial\over \partial\theta_2^\beta}}
		\overrightarrow{{\partial \over \partial \theta_1 \theta_1}}
		\right.\nonumber\\
&&{}
	+ \left( \det  C_{12}{} \   C_{21}^{\alpha\beta}
+ ( C_{22} \  \varepsilon C_{21} \  \varepsilon C_{11} )^{\alpha\beta} 
		\right) 
		\overleftarrow{{\partial \over \partial \theta_1 \theta_1}} 
\overleftarrow{{\partial\over \partial\theta_2^\alpha}}
\overrightarrow{{\partial\over \partial\theta_1^\beta}} 
		\overrightarrow{{\partial \over \partial \theta_2 \theta_2}}
		\nonumber\\
&&{}
	+ \left( \det  C_{21}{} \   C_{12}^{\alpha\beta}
+ ( C_{11} \  \varepsilon C_{12} \  \varepsilon C_{22} )^{\alpha\beta} 
		\right) 
		\overleftarrow{{\partial \over \partial \theta_2 \theta_2}} 
\overleftarrow{{\partial\over \partial\theta_1^\alpha}}
\overrightarrow{{\partial\over \partial\theta_2^\beta}}
		\overrightarrow{{\partial \over \partial \theta_1 \theta_1}}
		\nonumber\\
&&\left. {}
	+ \left( \det  C_{22}{} \   C_{11}^{\alpha\beta}
+ ( C_{12} \  \varepsilon C_{22} \  \varepsilon C_{21} )^{\alpha\beta} 
		\right) 
		\overleftarrow{{\partial \over \partial \theta_2 \theta_2}} 
\overleftarrow{{\partial\over \partial\theta_1^\alpha}}
\overrightarrow{{\partial\over \partial\theta_1^\beta}} 
		\overrightarrow{{\partial \over \partial \theta_2 \theta_2}}
	\right],
\end{eqnarray}
\begin{eqnarray}
{1\over 4!} P^4
&=&
	\left[
		\det  C_{11}{} \  \det  C_{22}{}
		+ (\det  C_{12})^2
	\right.\nonumber\\
&&\left. {}
		- (\varepsilon C_{11} \  \varepsilon C_{12} 
\  \varepsilon C_{22} \  \varepsilon C_{21} )_\alpha{}^\alpha
	\right]
	\overleftarrow{{\partial \over \partial \theta_1 \theta_1}} 
	\overleftarrow{{\partial \over \partial \theta_2 \theta_2}} 
	\  \overrightarrow{{\partial \over \partial \theta_2 \theta_2}} 
	\overrightarrow{{\partial \over \partial \theta_1 \theta_1}}
	. 
\end{eqnarray}
Here
$
(\varepsilon C)_\alpha{}^\beta 
\equiv 
\varepsilon_{\alpha\gamma} C^{\gamma \beta}
$
and
\begin{equation}
{{\partial} \over \partial \theta_{i} \theta_{i}}
\equiv 
	\frac14 \varepsilon^{\alpha\beta} 
	{\partial \over \partial \theta_{i}^{\alpha}}
	{\partial \over \partial \theta_{i}^{\beta}},
\end{equation}
for $i=1,2$.
Using the Poisson bracket in the chiral basis, we obtain for example, 
\begin{eqnarray}
\theta_1^+{}^{\alpha} \ast (\theta_2^+)^2 \theta_2^-{}^{\beta} 
&=& \frac{1}{2} \varepsilon^{\alpha\beta} (u_2^+u_1^+) \theta^4 
   +(C_{12}^{++} \theta_2^+)^{\alpha} \theta_2^-{}^{\beta} 
   +\frac{1}{2} C_{12}^{+-}{}^{\alpha\beta} (\theta_2^+)^2, 
                                                         \nonumber\\
 \theta^+_1{}^{\alpha}\theta^-_1{}^{\beta} \ast 
 \theta^+_2{}^{\gamma} \theta_2^-{}^{\delta} 
&=& \theta^+_1{}^{\alpha}\theta^-_1{}^{\beta}
 \theta^+_2{}^{\gamma} \theta_2^-{}^{\delta}
                                           \nonumber\\ 
&&{}+\frac{1}{2} 
    \left( 
     -C_{12}^{++}{}^{\alpha\gamma} 
      \theta^-_1{}^{\beta} \theta^-_2{}^{\delta} 
     +C_{12}^{+-}{}^{\alpha\delta} 
      \theta^-_1{}^{\beta} \theta^+_2{}^{\gamma} 
    \right.                                              \nonumber\\ 
&&{}\left. \qquad 
     +C_{12}^{-+}{}^{\beta\gamma} 
      \theta^+_1{}^{\alpha} \theta^-_2{}^{\delta} 
     -C_{12}^{--}{}^{\beta\delta} 
      \theta^+_1{}^{\alpha} \theta^+_2{}^{\gamma} 
    \right)                                              \nonumber\\ 
& &+\frac{1}{4} 
    \left( 
     -C_{12}^{++}{}^{\alpha\gamma} 
      C_{12}^{--}{}^{\beta\delta} 
     +C_{12}^{+-}{}^{\alpha\delta} 
      C_{12}^{-+}{}^{\beta\gamma} 
    \right) 
\end{eqnarray}
etc. 

\section{${\cal N}=2$ Supersymmetric $U(1)$ Gauge Theory}
In this section, we study the ${\cal N}=2$ supersymmetric $U(1)$ gauge theory
on the noncommutative harmonic superspace.

The action of ${\cal N}=2$ super Yang-Mill theory on the harmonic superspace
has been constructed in \cite{Zu,GaIvOgSo}.
The gauge superfield $V^{++}(x_{A},\theta^{+},\bar{\theta}^{+},u)$ 
is introduced by gauging the ${\cal N}=2$ matter action.
The covariant derivative $\nabla^{++}=D^{++}+i V^{++}$ satisfies
\begin{equation}
 \nabla^{++}\rightarrow e^{i\Lambda}\nabla^{++}e^{-i\Lambda}
\end{equation}
for the analytic superfield $\Lambda$.
$V^{++}$ and $\Lambda$ take values in the Lie 
algebra of the gauge group.
The gauge transformation for $V^{++}$ becomes
\begin{equation}
 \delta V^{++}=-D^{++}\Lambda+i [ \Lambda, V^{++}]
\label{eq:gauge1}
\end{equation}
in the infinitesimal form
\footnote{
We note that $V^{++}$ obeys 
a different gauge transformation 
from the vector superfield $V$ in ${\cal N}=1$ gauge theory 
in which $e^{V}$ transforms canonically. 
}.
Using the gauge superfield $V^{++}$ the action of ${\cal N}=2$  super
Yang-Mills theory is given by \cite{Zu,Oh}
\begin{equation}
 S={1\over2}\sum_{n=2}^{\infty}{(-i)^n\over n}
\int d^4 x d^{8}\theta du_{1}\cdots du_{n}
{{\rm tr} V^{++}(\zeta_{1},u_1)\cdots V^{++}(\zeta_{n},u_n)
\over (u^{+}_{1}u^{+}_{2})\cdots (u^{+}_{n}u^{+}_{1})}
\end{equation}
where $\zeta_{i}=(x_{A},\theta^{+}_{i},\bar{\theta}^{+}_{i})$ and 
$d^8\theta=d^4 \theta^+ d^4\theta^- $ with
$d^4\theta^{\pm}=d^2\theta^{\pm}d^2\bar{\theta}^{\pm}$.
The harmonic integral is defined by the rules:\\
(i)
\begin{equation}
 \int du f(u)=0
\end{equation}
for $f(u)$ with non-zero $U(1)$ charge.\\
 (ii)
\begin{equation}
 \int du 1=1.
\end{equation}
(iii)
\begin{equation}
 \int du u^{+}_{(i_1}\cdots u^{+}_{i_n} u^{-}_{j_1}\cdots u^{-}_{j_n)}=0,
\quad (n\geq 1).
\end{equation}

Since the gauge superfield contain infinite number of (auxiliary)
fields, it is not convenient to discuss the off-shell theory 
in the component formalism.
Fortunately the gauge parameters also contain infinite number of 
fields. 
We can eliminate unnecessary fields and take 
the Wess-Zumino(WZ) gauge:
\begin{eqnarray}
 V^{++}_{WZ}(x_{A},\theta^{+},\bar{\theta}^{+},u)
&=& 
-i\sqrt{2}(\theta^{+})^2 \bar{\phi}(x_{A})
+i\sqrt{2}(\bar{\theta}^{+})^2 \phi(x_{A})
-2i \theta^{+}\sigma^{\mu}\bar{\theta}^{+}A_{\mu}(x_{A})\nonumber\\
&&+4(\bar{\theta}^{+})^2\theta^{+}\psi^{i}(x_{A}) u^{-}_{i}
-4(\theta^{+})^2\bar{\theta}^{+}\bar{\psi}^{i}(x_{A})u^{-}_{i}\nonumber\\
&&
+3(\theta^{+})^2(\bar{\theta}^{+})^2 D^{ij}(x_{A})u^{-}_{i} u^{-}_{j}.
\label{eq:wz1}
\end{eqnarray}
The gauge symmetry with the superfield $\Lambda(x_{A})$ remains, which
corresponds to the canonical gauge transformation for component fields.
In the WZ gauge the sum in the action ends up to the fourth order.
The action in the component formalism leads to the well-known ${\cal
N}=2$
 action.

In the present paper, we will consider the simplest case {\it i.e.} 
$U(1)$ gauge group.
The gauge transformation (\ref{eq:gauge1}) is 
\begin{equation}
 \delta V^{++}=-D^{++}\Lambda
\label{eq:u1gaugetr}
\end{equation}
and the action is quadratic in $V^{++}_{WZ}$
\begin{equation}
S={1\over4}\int d^4 x d^{8}\theta du_1 du_2
{V^{++}_{WZ}(x_{A},\theta_{1}^{+},\bar{\theta}_{1}^{+},u_{1})
V^{++}_{WZ}(x_{A},\theta_{2}^{+},\bar{\theta}_{2}^{+},u_{2})
\over (u^{+}_{1}u^{+}_{2})^2}.
\label{eq:abelian}
\end{equation}
In terms of component fields, the action becomes
\begin{equation}
S=\int d^4 x\left\{
{1\over4}D^{ij}D_{ij}-i\psi^{i}\sigma^{\mu}\partial_{\mu}\bar{\psi}_{i}
-\partial^{\mu}\phi \partial_{\mu}\bar{\phi}
-{1\over 4}F_{\mu\nu}F^{\mu\nu}-{1\over4}\tilde{F}^{\mu\nu}F_{\mu\nu}
\right\}
\label{eq:abelianaction}
\end{equation}
where
$F_{\mu\nu}=\partial_{\mu}A_{\nu}-\partial_{\nu}A_{\mu}$,
$\tilde{F}^{\mu\nu}={i\over2}\varepsilon^{\mu\nu\lambda\rho}F_{\lambda\rho}$.

We now study the $U(1)$ gauge theory on the noncommutative harmonic
superspace by replacing the product by the $*$-product.
One finds that that the quadratic action (\ref{eq:abelian}) is not 
invariant under the ordinary gauge transformation (\ref{eq:u1gaugetr})
due to the deformation parameter.
In fact, as discussed in \cite{IvLeZu,FeSo},
the gauge invariance of ${\cal N}=2$ matter Lagrangian leads to
the gauge transformation (\ref{eq:u1gaugetr})
\begin{equation}
 \delta^{*} V^{++}=-D^{++}\Lambda+i [ \Lambda, V^{++}]_{*}.
\label{eq:gauge2}
\end{equation}
We find that the action which is invariant under
the gauge transformation (\ref{eq:gauge2}) is given by
\footnote{
${\cal N}=2$ supersymmetric Yang-Mills theory in 
${\cal N}=2$ harmonic superspace with spacetime noncommutativity 
has been studied in \cite{Bu}. 
}
\begin{equation}
S_{*}
= 
\frac12 \sum_{n=2}^{\infty} \int d^4xd^8\theta du_1\dots du_n {(-i)^n \over n}
{ V^{++}(1) * \cdots * V^{++}(n) 
\over (u_1^+ u_2^+)\cdots (u_{n}^+ u_1^+) }
. 
\label{eq:StarDeformedAction:Gen}
\end{equation} 
where $V^{++}(i)=V^{++}(\zeta_i, u_i)$.
The gauge invariance of the action is shown 
in the same way as in \cite{GaIvOgSo}.
In order to prove this,
we notice that 
\begin{equation}
\int d^4x d^4\theta \ A * B = \int d^4x d^4\theta \ AB
\label{eq:StarToOrdinaryProd}
\end{equation}
holds in the chiral-superspace integral,
where $d^4\theta=d^2\theta^+ d^2\theta^-$.
%upto total derivatives of the integrand. 
Then in the Abelian case,  we are allowed to use the formula
\begin{equation}
\int d^4x d^4\theta \ A * B * C = \int d^4x d^4\theta \ B * C * A
. 
\end{equation}
Together with the cyclic property of the denominator 
$(u^{+}_1 u^{+}_2) (u^{+}_2 u^{+}_3)\cdots (u^{+}_{n-1} u^{+}_n) 
(u^{+}_n u^{+}_1)$
in eq.(\ref{eq:StarDeformedAction:Gen}), 
the gauge variation of the action can be written as 
\begin{eqnarray}
\delta^{*} S_{*}
&=& 
	\frac12 \sum_{n=2}^{\infty} \int d^4xd^8\theta du_1\dots du_n {(-i)^n}
	\Big( \delta^{*} V^{++} (1) \Big)  
%	\cdot 
	{ V^{++} (2) * \cdots * V^{++}(n) 
		\over (u_1^+ u_2^+)\cdots (u_{n}^+ u_1^+) }
		\nonumber\\
&=& 
	\frac12 \int d^4xd^8\theta du_1 
	\Big( \delta^{*} V^{++} (X, u_1) \Big)  
%	\cdot
 V^{--}_{*} (X, u_1)
\end{eqnarray}
%(without trace, in spite of the star products), 
where 
\begin{equation}
V^{--}_{*} (X, u)
\equiv  
\sum_{n=1}^{\infty} \int du_1\dots du_n {(-i)^{n+1}}
{ V^{++}(1) * \cdots * V^{++}(n) 
\over (u^+ u_1^+)(u_1^+ u_2^+)\cdots (u_{n}^+ u_1^+)(u_1^+ u^+) }
. 
\end{equation} 
%It is as same as the ordinary case that
$V^{--}_{*}$ satisfies the relation 
\begin{equation}
D^{++} V_{*}^{--} - D^{--} V^{++} + i [ V_{*}^{--} , V^{++} ]_{*} = 0
. 
\end{equation} 
Substituting $\delta^{*} V^{++}$ given in eq.(\ref{eq:gauge2}) into $\delta^{*} S_{*}$, 
integrating by parts, 
with the use of the above relation, we find 
\begin{equation}
\delta^{*} S_{*} = \frac12 \int d^4xd^8\theta \int du 
\ \Lambda D^{--} V^{++}_{*}
. 
\end{equation}
As in the non-Abelian case \cite{GaIvOgSo}, 
%As usual, 
the analyticity of the gauge parameter $\Lambda$ 
ensures that this integral vanishes. 

%It is important to write down the action in WZ-gauge, which is given by 
In the WZ gauge, the action becomes
\begin{equation}
S_{*}
= 
\frac12 \sum_{n=2}^{\infty} \int d^4xd^8\theta du_1\dots du_n {(-i)^n \over n}
{ V_{WZ}^{++}(1) * \cdots * V_{WZ}^{++}(n) 
\over (u_1^+ u_2^+)\cdots (u_{n}^+ u_1^+) }
. 
\label{eq:StarDeformedAction}
\end{equation} 
In the following we examine the action in terms of component fields.
In the present work, however, we will calculate the action
explicitly up to the third order in $V_{WZ}^{++}$. 
Although it is possible to calculate higher order terms, 
explicit calculation is very complicated due to the harmonic
integrals, which is left for future works.

The terms coming from the second order of $V_{WZ}^{++}$ 
are the same as the commutative gauge theory (\ref{eq:abelianaction})
due to eq.(\ref{eq:StarToOrdinaryProd}). 
The second order action $S_{*,2} $ is given by
\begin{eqnarray}
S_{*,2}
&\equiv& 
{1 \over 4} 
\int d^4xd^8\theta du_1 du_2 {
	V^{++}_{WZ}(1) * V^{++}_{WZ}(2) 
	\over (u^{+}_1 u^{+}_2)^2 }
	\nonumber\\
&=&
	\int d^4x \left[
	- {1\over 4} F_{\mu\nu} ( F^{\mu\nu} + \tilde{F}^{\mu\nu} )
	-i \psi^i \sigma^\mu \partial_\mu \bar{\psi}_i
	+ \phi \partial^{\mu}\partial_{\mu} \bar{\phi}
	+ \frac14 D_{ij} D^{ij}
	\right]
	. 
	\label{eq:S_2}
\end{eqnarray}
The second order terms do not include the deformation parameter.
On the other hand,  in the Abelian case, 
all the terms coming from the third order of $V_{WZ}^{++}$, 
which we will denote $S_{*,3}$,  
contain the deformation parameter $C$. 
They are all the newly emerged terms due to the deformation. 
This is true for the terms coming from the higher order contributions
in terms of $V_{WZ}^{++}$.  

Note, however, that the ${\cal O}(C^2)$ and ${\cal O}(C^4)$-terms in 
$S_{*,3}$ vanish: 
for Grassmann even %bosonic (Abelian) 
functions $A_1$ and $A_2$, the equation
\begin{equation}
A_1 P^n A_2 = (-1)^n A_2 P^n A_1
\label{eq:OrderReversion} 
\end{equation}
holds, 
while the harmonic distribution 
$(u^{+}_1 u^{+}_2)^{-1} \* (u^{+}_2 u^{+}_3)^{-1} \*  (u^{+}_3 u^{+}_1)^{-1}$ 
is completely antisymmetric under the permutations of $(u_1,u_2,u_3)$. 

In addition to this, when we calculate the terms in $S_{*,3}$ 
it is useful to notice a relation between distinct contributions, which 
is given below.  
In the chiral-superspace integral, 
because of eq.(\ref{eq:StarToOrdinaryProd}), 
it holds that 
\begin{equation}
\int d^4x d^4\theta \ A_1 * A_2 * A_3 
= \int d^4x d^4\theta \ (A_1 * A_2) \ A_3
= \int d^4x d^4\theta \ A_1 \ (A_2 * A_3)
\end{equation}
up to total derivatives, 
so that 
\begin{equation}
\int d^4x d^4\theta \ (A_1 P A_2) \ A_3
= \int d^4x d^4\theta \ A_1 \ (A_2 P A_3)
\end{equation} 
also holds. 
This shows that these potentially distinct contributions are the same.
%either in the Abelian or the non-Abelian case. 
Moreover, in the Abelian case, 
together with eq.(\ref{eq:OrderReversion}), 
we have  
\begin{eqnarray}
&&
\int d^4xd^8\theta 
{du_1 du_2 du_3 \over (u^{+}_1 u^{+}_2) (u^{+}_2 u^{+}_3) (u^{+}_3 u^{+}_1) }
	\ {1\over 6!} \sum_{\sigma\in S_3}
	\left( A_{\sigma(1)}(u_1) P A_{\sigma(2)}(u_2) \right)) 
	\ A_{\sigma(3)}(u_3)
	\nonumber\\ 
&&= 
	\int d^4xd^8\theta  
{du_1 du_2 du_3 \over (u^{+}_1 u^{+}_2) (u^{+}_2 u^{+}_3) (u^{+}_3 u^{+}_1) }
	\left( A_1(u_1) P A_2(u_2) \right) \ A_3(u_3)
, 
\end{eqnarray}
where $S_3$ is the permutation group of degree $3$.  
Thus, among the terms in the left hand side 
we are allowed to choose the most simple combination to calculate. 

The terms coming from the third order of $V_{WZ}^{++}$ 
are followings :  
\begin{eqnarray}
S_{*,3}
&\equiv& 
{i \over 6} 
\int d^4xd^8\theta du_1 du_2 du_3 {
	V^{++}_{WZ}(1) * V^{++}_{WZ}(2) * V^{++}_{WZ}(3)
	\over (u^{+}_1 u^{+}_2)(u^{+}_2 u^{+}_3)(u^{+}_3 u^{+}_1)}
\nonumber\\
&=&{}
	\int d^4x \left[
%(10)(\partial \bar{\phi}, \partial A_\mu, A_\mu ;P^1)&:& 
+ {1\over \sqrt{2}} C_s A_\nu \partial_\mu \bar{\phi}
( F^{\mu\nu} + \tilde{F}^{\mu\nu} ) 
\right.\nonumber\\ 
&&{}
%(7)(\partial \bar{\psi}, \bar{\phi}, \psi ;P^1)&:& 
+ {i\over \sqrt{2}} C_s \bar{\phi} ( \psi^k \sigma^\nu \partial_\nu \bar{\psi}_k )
- { 2 \sqrt{2} \over 3} i C_{(ij)}^{\alpha\beta}
	\psi^i_\alpha ( \sigma^\nu \partial_\nu \bar{\psi}^j)_\beta 
	\bar{\phi}
\nonumber\\ 
&&{}
%(4)(\partial \bar{\phi}, \bar{\psi}, \psi ;P^1)&:& 
+ { i \over \sqrt{2}} C_s (\psi^k \sigma^\nu \bar{\psi}_k)
	\partial_\nu \bar{\phi}
- 2\sqrt{2} i C_{(ij)}^{\alpha\beta}
	\psi^i_\alpha (\sigma^\nu \bar{\psi}^j)_\beta 
	\partial_\nu \bar{\phi}
	\nonumber\\
&&{}
%(8)(\partial \bar{\psi}, A_\mu, \bar{\psi} ;P^1)&:& 
- {i \over 2} C_s \varepsilon^{\alpha\beta} A_\mu 
	(\sigma^\mu \bar{\psi}^k)_\alpha (\sigma^\nu \partial_\nu \bar{\psi}_k)_\beta
+{2\over 3} i C_{(ij)}^{\alpha\beta} A_\mu
	(\sigma^\mu \bar{\psi}^i)_\alpha (\sigma^\nu \partial_\nu \bar{\psi}^j)_\beta
\nonumber\\ 
&&{}
+ {i \over 2} C_s \bar{\psi}^i \bar{\psi}^j  D_{ij}
- i C_{(ij)}^{\mu\nu} \bar{\psi}^i \bar{\psi}^j  F_{\mu\nu}
\nonumber\\ 
&&{}
+ {\sqrt{2} \over 4} C_s A_\mu A^\mu \partial^2 \bar{\phi} 
\nonumber\\ 
&&\left. {} 
- { \sqrt{2} \over 4} C_s \bar{\phi} D^{ij} D_{ij}  
+ \sqrt{2} C_{(ij)}^{\mu\nu} D^{ij} A_\mu \partial_\nu \bar{\phi}
+ {1\over \sqrt{2}} C_{(ij)}^{\mu\nu} D^{ij} F_{\mu\nu} \bar{\phi} 
\right] 
. 
\label{eq:thirdorderLagrangian}
\end{eqnarray}
Detailed calculations of these terms are explained in Appendix B.

At first sight there are terms which are not gauge invariant 
in the sense of the ordinary gauge transformation. 
Of course the deformed gauge variation $\delta^{*} V^{++}$ of the gauge 
superfield 
contains the deformation parameter $C$, 
so that $\delta^{*} S_{*}$ also depends on $C$. 
As the gauge invariance of $S_{*}$ has been shown in the superfield formalism, 
the ${\cal O}(C^1)$ part of $\delta^{*} S_{*,2}$, for example, should
cancel with 
the ${\cal O}(C^1)$ part of $\delta^{*} S_{*,3}$. 
In the following, we would like to see 
how such a cancellation is actually realized in the component 
action in the WZ gauge.

\section{The deformed gauge transformation} 
In this section we study the gauge transformation 
of ${\cal N}=2$ supersymmetric $U(1)$ gauge theory on the noncommutative 
harmonic superspace.

The deformed gauge transformation of the gauge superfield in the WZ gauge, 
$V^{++}_{WZ}$, 
with an analytic gauge parameter 
$\lambda(\zeta,u)$ 
($\zeta=(x_A, \theta^{+}, \bar{\theta}^{+})$) is given by 
\begin{equation}
\delta^{*}_{\lambda} V^{++}_{WZ} (\zeta, u)
= 
-D^{++}_A \lambda (\zeta,u) + i [ \lambda, V^{++}_{WZ} ] _{*}(\zeta, u)
	. 
\end{equation}
If we choose the gauge parameter 
$\lambda(\zeta,u)=\lambda(x_A)$ as in the commutative case, 
the corresponding gauge variation becomes
\begin{eqnarray}
\delta^{*}_{\lambda} V^{++}_{WZ} (\zeta, u)
&=& 
	-2i\ \theta^{+}\! \sigma^\mu \bar{\theta}^{+} 
		\left(
		- \partial_\mu \lambda 
		- {1\over \sqrt{2}} C_s \partial_\mu \lambda \ \bar{\phi}
		+ 2 \sqrt{2} \eta_{\mu\rho} C^{(+-)}{}^{\rho\sigma} 
			\partial_\sigma \lambda \ \bar{\phi}
		\right) (x_A)
		\nonumber\\
&&\quad {}
	+ \sqrt{2} i \ (\bar{\theta}^{+})^2
		\left( 
		{1\over\sqrt{2}} C_s \partial_\mu \lambda \ A^\mu 
		+ 2\sqrt{2} C^{(+-)}{}^{\mu\nu} 
			\partial_\mu \lambda \ A_\nu 
		\right) (x_A)
		\nonumber\\
&&\quad {}
	+ 4\ (\bar{\theta}^{+})^2 \theta^{+}{}^{\alpha}
		\left(
		-\frac12 C_s \partial_\mu \lambda \ 
			(\sigma^\mu \bar{\psi}^{-} )_\alpha
		- 2 \partial_\mu \lambda 
		( \varepsilon C^{(+-)} \sigma^\mu \bar{\psi}^{-} )_\alpha 
		\right) (x_A) 
		\nonumber\\
&&\quad {}
	+ 3\ (\theta^{+})^2 (\bar{\theta}^{+})^2 
		\left(
		{4 \sqrt{2} \over 3} C^{--}{}^{\mu\nu} 
		\partial_\mu \lambda \ \partial_\nu \bar{\phi} 
		\right) (x_A) 
, 
\label{eq:DeformedTransf1}
\end{eqnarray}
where $C^{(+-)}{}^{\alpha\beta} \equiv C_{(ij)}^{\alpha\beta} u^{+i} u^{-j}$. 
This transformation obviously does not preserve the WZ gauge because 
each $\theta$-components in (\ref{eq:DeformedTransf1}) contain
extra $u$-dependent terms.
Therefore we need to regard the gauge parameter as more general 
analytic superfield.
If we choose an analytic gauge parameter of the form 
\begin{eqnarray}
\lambda_C (\zeta, u; C)
&=&
	\lambda(x_A)
	+ \theta^{+}\!\sigma^\mu \bar{\theta}^{+}
		\lambda_\mu^{(-2)} (x_A, u; C)
	+ (\bar{\theta}^{+})^2 
		\lambda^{(-2)} (x_A, u; C)
		\nonumber\\
&&{} 
	+ (\bar{\theta}^{+})^2 \theta^{+}{}^{\alpha} 
		\lambda_{\alpha}^{(-3)} (x_A, u; C)
	+ (\theta^{+})^2 (\bar{\theta}^{+})^2 
		\lambda^{(-4)} (x_A, u; C)
		, 
\end{eqnarray}
then the corresponding gauge variation is   
\begin{eqnarray}
\delta^{*}_{\lambda_C} V^{++}_{WZ} (\zeta, u)
&=& 
	\theta^{+}\! \sigma^\mu \bar{\theta}^{+} 
		\left(
		+ 2 i \partial_\mu \lambda 
		+ \sqrt{2} i C_s \partial_\mu \lambda \ \bar{\phi}
		+ 2 \sqrt{2} i C^{(+-)}{}^{\alpha\beta} 
		( \sigma_{\mu} \bar{\sigma}^{\nu} \varepsilon )_{\alpha\beta}  
			\partial_\nu \lambda \ \bar{\phi}
			\right.\nonumber\\
&&\left. {}\qquad 
	- \partial^{++} \lambda_\mu^{(-2)} 
	- \sqrt{2} C^{++}{}^{\alpha\beta} 
		( \sigma_\mu \bar{\sigma}^\nu \varepsilon)_{\alpha\beta}
		\lambda_\nu^{(-2)} \bar{\phi}
		\right) (x_A)
		\nonumber\\
&& {}
	+ (\bar{\theta}^{+})^2
		\left( 
		i  C_s \partial_\mu \lambda \ A^\mu 
		- 2 i C^{(+-)}{}^{\alpha\beta} 
		(\sigma^\nu \bar{\sigma}^\mu \varepsilon )_{\alpha\beta}
			\partial_\mu \lambda \ A_\nu 
		\right.\nonumber\\
&&\left.{}\qquad 
		+ C^{++}{}^{\alpha\beta} 
		(\sigma^\nu \bar{\sigma}^\mu \varepsilon )_{\alpha\beta} 
		\lambda_\mu^{(-2)} A_\nu 
		- \partial^{++} \lambda^{(-2)}
		\right) (x_A)
		\nonumber\\
&& {}
	+ (\bar{\theta}^{+})^2 \theta^{+}{}^{\alpha}
		\left(
		- 2 C_s \partial_\mu \lambda \ 
			(\sigma^\mu \bar{\psi}^{-} )_\alpha
		- 4 (\sigma^\mu \bar{\psi}^{-} )_\alpha 
			C^{(+-)}{}^{\gamma\delta} 
		(\sigma_\mu \bar{\sigma}^\nu \varepsilon )_{\gamma\delta} 
			\partial_\nu \lambda \ 
		\right.\nonumber\\
&&\left.{}\qquad 
		- 2 i (\sigma^\mu \bar{\psi}^{-} )_\alpha 
			C^{++}{}^{\gamma\delta} 
		(\sigma_\mu \bar{\sigma}^\nu \varepsilon )_{\gamma\delta}
			\lambda_\nu^{(-2)} 
			\right.\nonumber\\
&&\left.{}\qquad 
		- \partial^{++} \lambda_\alpha^{(-3)} 
	- 2 \sqrt{2} ( \varepsilon C^{++} \lambda^{(-3)} )_\alpha \bar{\phi} 
		\right) (x_A) 
		\nonumber\\
&& {}
	+ (\theta^{+})^2 (\bar{\theta}^{+})^2 
		\left(
		{4 \sqrt{2}} C^{--}{}^{\mu\nu} 
		\partial_\mu \lambda \ \partial_\nu \bar{\phi} 
		\right.\nonumber\\
&&\left.{}\qquad 
		- i \partial^\mu \lambda_\mu^{(-2)} 
	- \sqrt{2} i C^{+-}{}^{\alpha\beta} 
		( \sigma^\mu \bar{\sigma}^\nu \varepsilon )_{\alpha\beta} 
		\partial_\nu ( \lambda_\mu^{(-2)} \bar{\phi} )
		- \partial^{++} \lambda^{(-4)}
		\right) (x_A) 
		, \nonumber\\
\label{eq:DeformedTransf2}
\end{eqnarray}
where $(\sigma^{\mu}\bar{\sigma}^{\nu}\varepsilon)_{\alpha\beta}=
(\sigma^{\mu}\bar{\sigma}^{\nu})_{\alpha}{}^{\gamma}
\varepsilon_{\gamma\beta}$.
Note that other $\theta$ 
components will never emerge as long as we use the gauge 
parameter 
$\lambda_C$ of the form given above. 
Moreover, the degrees of freedom contained in 
$\lambda_{\mu}^{(-2)}$, $\lambda^{(-2)}$, 
$\lambda_{\alpha}^{(-3)}$ and $\lambda^{(-4)}$ 
are not only necessary but also sufficient 
to consistently gauge away all of the terms with extra $u$-dependence. 
Therefore, with an appropriate tuning of the analytic gauge parameter $\lambda_C$, 
this variation can be brought into the form 
\begin{eqnarray}
\delta^{*}_{\lambda_C} V^{++}_{WZ} (\zeta, u)
&=& 
	- 2 i \ \theta^{+}\! \sigma^\mu \bar{\theta}^{+} 
		\left(
		\delta^{*}_{\lambda_C} A_\mu (x_A)
		\right) 
		\nonumber\\
&&{} 
	+ \sqrt{2} i \ (\bar{\theta}^{+})^2
		\left( 
		\delta^{*}_{\lambda_C} \phi (x_A)
		\right) 
		\nonumber\\
&&{} 
	+ 4 \ (\bar{\theta}^{+})^2 \theta^{+}{}^{\alpha}
		\left(
		\delta^{*}_{\lambda_C} \psi^{i}_\alpha (x_A) u^{-}_{i}
		\right)  
		\nonumber\\
&&{} 
	+ 3 \ (\theta^{+})^2 (\bar{\theta}^{+})^2 
		\left(
		\delta^{*}_{\lambda_C} D^{ij} (x_A) u^{-}_{i} u^{-}_{j} 
		\right) 
		, 
\label{eq:DeformedTransf3}
\end{eqnarray}
where $\delta^{*}_{\lambda_C} A_\mu$, $\delta^{*}_{\lambda_C} \phi$, 
$\delta^{*}_{\lambda_C} \psi^{i}_\alpha$ and $\delta^{*}_{\lambda_C} D^{ij}$ 
are $u$-independent quantities, so that each gives the proper 
gauge transformation law 
for the respective component fields in this deformed theory. 

In the following, we will determine $\delta^{*}_{\lambda_C} A_\mu$, 
$\delta^{*}_{\lambda_C} \phi$ and so on.

\subsection{The deformed gauge transformation of the gauge field}
First of all, we will concentrate on the 
$\theta^{+} \bar{\theta}^{+}$-component of the gauge variation 
$\delta^{*}_{\lambda_C} V_{WZ}^{++}$ 
to determine a properly deformed gauge transformation of the gauge field, 
$\delta^{*}_{\lambda_C} A_\mu$, 
and the correspondingly deformed component of the gauge parameter, 
$\lambda_{\mu}^{(-2)}$. 
This is because the other components depend on $\lambda_{\mu}^{(-2)}$. 
The rest of the components will be determined in much the same way. 

Comparing the $\theta^{+} \bar{\theta}^{+}$-components of 
eqs.(\ref{eq:DeformedTransf2}) and (\ref{eq:DeformedTransf3}), 
we find the equation which determines 
$\delta^{*}_{\lambda_C} A_\mu$ and $\lambda_{\mu}^{(-2)}$: 
\begin{eqnarray}
	- 2 i \delta^{*}_{\lambda_C} A_\mu &=& 
2i \partial_\mu \lambda
	+ \sqrt{2} i C_s \partial_\mu \lambda \ \bar{\phi} 
+ 2 \sqrt{2} i C^{(+-)}{}^{\alpha\beta} 
	(\sigma_\mu \bar{\sigma}^\nu \varepsilon )_{\alpha\beta} 
		\partial_\nu \lambda \ \bar{\phi}
	\nonumber\\
&&{} 
	- \partial^{++} \lambda_\mu^{(-2)} 
	- \sqrt{2} C^{++}{}^{\alpha\beta} 
	(\sigma_\mu \bar{\sigma}^\nu \varepsilon )_{\alpha\beta} 
	\lambda_\nu^{(-2)} \bar{\phi},
\label{eq:(1,1)Comp}
\end{eqnarray}
where $\delta^{*}_{\lambda_C} A_\mu$ is identified 
with the lowest order term in the harmonic expansion of the left hand side 
(i.e. the $U(1)$ charge 0, isospin 0 part).  
Note that the gauge parameters $\lambda_{\mu}^{(-2)}$ 
admit not only the harmonic expansion but also the $C$-expansion at the 
same time, 
while $\delta^{*}_{\lambda_C} A_\mu$ can be expanded only in terms of $C$ 
by its definition. 
Here we first formally expand both of them in terms of $C$, 
\begin{eqnarray}
\delta^{*}_{\lambda_C} A_\mu
&\equiv& 
	\sum_{N=0}^{\infty} (\delta^{*}_{\lambda_C} A_\mu )_{(N)} 
	, 
	\\
\lambda_\mu^{(-2)} 
&\equiv& 
	\sum_{N=1}^{\infty} \lambda_\mu^{(-2)}{}_{(N)}
, 
\end{eqnarray}
where $(\delta^{*}_{\lambda_C} A_\mu )_{(N)}$ and 
$\lambda_\mu^{(-2)}{}_{(N)}$ are ${\cal O}(C^N)$-quantities. 
Note also that the $C$-independent part of $\lambda_{\mu}^{(-2)}$ is absent, 
because in the $C \rightarrow 0$ limit this should vanish. 
Substituting these into eq.(\ref{eq:(1,1)Comp}), 
we find the following equations: 
\begin{eqnarray}
(\delta^{*}_{\lambda_C} A_\mu )_{(0)}
&=& 
	- \partial_\mu \lambda
	\label{eq:DeltaAmu0}
	, \\
(\delta^{*}_{\lambda_C} A_\mu )_{(1)} 
&=& 
	- {1 \over \sqrt{2}} C_s \partial_\mu \lambda \ \bar{\phi} 
	\label{eq:DeltaAmu1}
	,\\
\partial^{++} \lambda_\mu^{(-2)}{}_{(1)}
&=&
	2 \sqrt{2} i C^{(+-)}{}^{\alpha\beta} 
		(\sigma_\mu\bar{\sigma}^\nu \varepsilon )_{\alpha\beta}
		\partial_\nu \lambda \ \bar{\phi} 
\label{eq:Recursion(1,1):Initial}
		, \\
\partial^{++} \lambda_\mu^{(-2)}{}_{(N)}
&=& 
	- \sqrt{2} C^{++}{}^{\alpha\beta} 
		(\sigma^\mu \bar{\sigma}^\nu \varepsilon )_{\alpha\beta} 
		\lambda_\nu^{(-2)}{}_{(N-1)} \bar{\phi} 
	+ 2 i (\delta^{*}_{\lambda_C} A_\mu )_{(N)}
	\quad ( N\ge 2 )
. 
\label{eq:Recursion(1,1)}
\end{eqnarray} 
The first two equations are immediately read from 
the left hand side of eq.(\ref{eq:(1,1)Comp}). 
They are giving 
the lowest and the next lowest order terms in the $C$-expansion of 
$\delta^{*}_{\lambda_C} A_\mu$. 
The rest of two equations are obtained 
by equating the terms of the same order in $C$ 
in both sides of eq.(\ref{eq:(1,1)Comp}). 

The latter two equations can be regarded as a recursion relation 
to determine $\lambda_\mu^{(-2)}{}_{(N)}$ and 
$(\delta^{*}_{\lambda_C} A_\mu)_{(N)}$ 
from $\lambda_\mu^{(-2)}{}_{(N-1)}$ 
with an initial condition given by $\lambda_\mu^{(-2)}{}_{(1)}$.  
It is useful to treat explicitly 
the coefficient functions in the harmonic expansion of 
$\lambda_\mu^{(-2)}{}_{(N)}$: 
\begin{eqnarray}
\lambda_{\mu}^{(-2)}{}_{(N)}
&=& 
	\sum_{n=0}^{\infty} 
	( \lambda_{\mu}^{(-2)}{}_{(N)} )^{(i_1 \dots i_{n} k_1 \dots k_{n+2})}
	u^{+}_{i_1} \dots u^{+}_{i_n} u^{-}_{k_1} \dots u^{-}_{k_{n+2}}
	\nonumber\\
&=& 
	\sum_{n=0}^{\infty} 
	( \lambda_{\mu}^{(-2)}{}_{(N)} )_{(i_1 \dots i_{n} k_1 \dots k_{n+2})}
	u^{+}{}^{i_1} \dots u^{+}{}^{i_n} u^{-}{}^{k_1} \dots u^{-}{}^{k_{n+2}}
. 
\end{eqnarray}
We first note that, 
from the general reduction identity of harmonic variables \cite{GaIvOgSo}, 
the following reduction formula holds for any term in this harmonic expansion: 
\begin{eqnarray}
&&
	C^{++}{}^{\alpha\beta} 
	(\lambda_\mu^{(-2)}{}_{(N)})_{( j_{n+m} j_{n+m-1} \dots j_{1} )}
	\overbrace{
	u^{+}{}^{(j_{n+m}} \dots u^{+}{}^{j_{m+1}} 
	}^{n}
	\overbrace{
	u^{-}{}^{j_{m}} \dots u^{-}{}^{j_{1})} 
	}^{m}
	\nonumber\\
&&= 
	C_{(kl)}^{\alpha\beta} 
	(\lambda_\mu^{(-2)}{}_{(N)})_{( j_{n+m} j_{n+m-1} \dots j_{1} )}
	u^{+}{}^{(k} u^{+}{}^{l)} 
	\overbrace{
	u^{+}{}^{(j_{n+m}} \dots u^{+}{}^{j_{m+1}} 
	}^{n}
	\overbrace{
	u^{-}{}^{j_{m}} \dots u^{-}{}^{j_{1})} 
	}^{m}
	\nonumber\\
&&= 
	C_{(kl)}^{\alpha\beta} 
	(\lambda_\mu^{(-2)}{}_{(N)})_{( j_{n+m} j_{n+m-1} \dots j_{1} )} 
	\left(
	\overbrace{
	u^{+}{}^{(k} u^{+}{}^{l} u^{+}{}^{j_{n+m}} \dots u^{+}{}^{j_{m+1}} 
	}^{n+2}
	\overbrace{
	u^{-}{}^{j_{m}} \dots u^{-}{}^{j_{1})}
	}^{m}
	\right.\nonumber\\
&&\left. {}
	- { 2m \over n+m+2 }
		\epsilon^{l j_{1}} 
		\overbrace{ 
		u^{+}{}^{(k} u^{+}{}^{j_{n+m}} \dots u^{+}{}^{j_{m+1}} 
		}^{n+1}
		\overbrace{
		u^{-}{}^{j_{m}} \dots u^{-}{}^{j_{2})}
		}^{m-1}
	\right.\nonumber\\
&&\left. {}
	+ { m(m-1) \over (n+m+1) (n+m) } 
		\epsilon^{k j_{2}} \epsilon^{l j_{1}} 
		\overbrace{
		u^{+}{}^{(j_{n+m}} \dots u^{+}{}^{j_{m+1}}
		}^{n} 
		\overbrace{
		u^{-}{}^{j_{m}} \dots u^{-}{}^{j_{3})}
		}^{m-2}
	\right) 
	. 
\label{eq:Reduction}
\end{eqnarray} 
Then 
eqs.(\ref{eq:Recursion(1,1):Initial}) and (\ref{eq:Recursion(1,1)}) 
can be rewritten as the following equations
\footnote{
Strictly speaking, the expression (\ref{eq:Recursion(1,1)2}) is only valid 
for $m \ge 2$, 
because the first term in the square bracket is absent if $m=1$. 
Thus for $m=1$, instead of eq.(\ref{eq:Recursion(1,1)2}), we should use  
\begin{eqnarray}
&&
(\lambda_\mu^{(-2)}{}_{(N)})_{(ij)} 
		u^{-}{}^{i} u^{-}{}^{j}
	\nonumber\\
&&=    
	\left[
	- \frac12 \epsilon^{kl}
		C_{(ik)}^{\alpha\beta}
		(\lambda_\nu^{(-2)}{}_{(N-1)})_{(jl)}
	+ {3\over 20} 
		\epsilon^{km} \epsilon^{ln} 
		C_{(kl)}^{\alpha\beta}
		(\lambda_\nu^{(-2)}{}_{(N-1)})_{(ijmn )}
		\right] 
		(\sigma_\mu \bar{\sigma}^\nu \varepsilon )_{\alpha\beta} 
		( - \sqrt{2} \bar{\phi} ) 
		u^{-}{}^{i} u^{-}{}^{j}
.  
\label{eq:Recursion(1,1)4}
\end{eqnarray}
} 
: 
\begin{eqnarray} 
&& 
\lambda_\mu^{(-2)}{}_{(1)}
= (\lambda_\mu^{(-2)}{}_{(1)} )_{(ij)} u^{-i} u^{-j}
= \sqrt{2} i C_{(ij)}^{\alpha\beta} 
	( \sigma_\mu \bar{\sigma}^\nu \varepsilon )_{\alpha\beta} 
	\partial_\nu \lambda \ \bar{\phi} 
	\ u^{-i} u^{-j}
, 
\label{eq:Lambda(1,1):1} 
	\\
&&
(\lambda_\mu^{(-2)}{}_{(N)})_{(j_{2m} j_{2m-1} \dots j_{1})} 
	\nonumber\\
&& \quad =    
	{1\over (2m)!} \sum_{\sigma \in S_{2m} } 
	\left[
	{1\over m+1}
		C_{(j_{\sigma(2m)} j_{\sigma(2m-1)})}^{\alpha\beta}
		(\lambda_\nu^{(-2)}{}_{(N-1)}
		)_{(j_{\sigma(2m-2)} j_{\sigma(2m-3)} \dots j_{\sigma(1)})}
		\right. \nonumber\\
&&\left. {} \quad 
	- {1\over m+1}
		\epsilon^{kl}
		C_{(k\, j_{\sigma(2m)})}^{\alpha\beta}
		(\lambda_\nu^{(-2)}{}_{(N-1)}
		)_{(j_{\sigma(2m-1)} j_{\sigma(2m-2)} \dots j_{\sigma(1)}\, l)}
		\right. \nonumber\\
&&\left. {} \quad 
	+ {(m+2) \over (2m+2) (2m+3)} 
		\epsilon^{km} \epsilon^{ln} 
		C_{(kl)}^{\alpha\beta}
		(\lambda_\nu^{(-2)}{}_{(N-1)}
	)_{(j_{\sigma(2m)} j_{\sigma(2m-1)} \dots j_{\sigma(1)} m \, n )}
		\right] 
		\nonumber\\
&&{} \quad 
	\times 
	(\sigma_\mu \bar{\sigma}^\nu \varepsilon )_{\alpha\beta} 
	( -\sqrt{2} \bar{\phi}) 
	\quad (m \le N) 
\label{eq:Recursion(1,1)2}
	, \\
&&
(\lambda_\mu^{(-2)}{}_{(N)})_{(j_{2m} j_{2m-1} \dots j_{1})} 
= 
	0 
	\quad (m > N)
\label{eq:Recursion(1,1)3}
	, \\
&&
( \delta^{*}_{\lambda_C} A_\mu )_{(N)}
= 
	- {i \over 3 \sqrt{2}} 
	\epsilon^{ik} \epsilon^{jl} 
	C_{(ij)}^{\alpha\beta} (\lambda_{\nu}^{(-2)}{}_{(N-1)} )_{(kl)} 
	(\sigma_\mu \bar{\sigma}^\nu \varepsilon )_{\alpha\beta} 
	\bar{\phi} 
	\quad (N\ge 2)
, 
\label{eq:DeltaAmu2}
\end{eqnarray} 
Here we have used the relation 
\begin{equation}
(\partial^{++})^{-1} \Big( 
	\overbrace{ 
		u^{+}_{(j_{2m}} u^{+}_{j_{2m-1}} \dots u^{+}_{j_{m+1}} 
		}^{m}
	\overbrace{
		u^{-}_{j_{m}} \dots u^{-}_{j_{1})}
		}^{m}
	\Big) 
= 
	{1\over m+1} 
	\overbrace{ 
		u^{+}_{(j_{2m}} u^{+}_{j_{2m-1}} \dots u^{+}_{j_{m+2}} 
		}^{m-1}
	\overbrace{
		u^{-}_{j_{m+1}} \dots u^{-}_{j_{1})}
		}^{m+1}
. 
\end{equation}
$S_{2m}$ is the permutation group of degree $2m$. 

{}From a direct calculation up to the second order in $C$,  
$\lambda_\mu^{(-2)}$ is determined as  
\begin{eqnarray}
\lambda_\mu^{(-2)}
&=& 
	-2 \sqrt{2} i C^{--}{}_\mu{}^\nu \partial_\nu \lambda \ \bar{\phi} 
- 4 i \left( C_{(ij)} \varepsilon C_{(kl)} \right)^{\alpha\beta} 
	\nonumber\\
&&{} \times
		( \sigma_\mu \bar{\sigma}^\nu \varepsilon )_{\alpha\beta} 
		\partial_\nu \lambda \ \bar{\phi}^2 
		\left( 
			\frac13 u^{+(i} u^{-j} u^{-k} u^{-l)} 
			- \frac12 \epsilon^{jl} u^{-(i} u^{-k)}
		\right) 
	+ {\cal O}(C^3)
, 
\label{eq:Lambda(1,1):DirectCal}
\end{eqnarray}
where we have used the identity
\begin{equation}
(\sigma^\mu \bar{\sigma}^\rho \varepsilon )_{\alpha\beta} 
	(\sigma_\rho \bar{\sigma}^\nu \varepsilon )_{\gamma\delta} 
= 
2 \varepsilon_{\beta\gamma} (\sigma^\mu \bar{\sigma}^\nu 
\varepsilon )_{\alpha\delta}
	. 
	\label{eq:GenRel}
\end{equation} 

Based on the observation of the result from direct calculation, 
(\ref{eq:Lambda(1,1):DirectCal}), 
we will assume 
$(\lambda_\mu^{(-2)}{}_{(N)})_{(j_1 \dots j_{2m})}$ ($m\le N$) has the form 
\begin{equation}
(\lambda_\mu^{(-2)}{}_{(N)})_{(j_1 \dots j_{2m})} 
= 
	\left[ 
	\sum \left(
\mbox{const.} \times {\cal C}_{(N)}^{\alpha\beta} {}_{(j_1 \dots j_{2m})} 
	\right)
	\right] 
	(\sigma_\mu \bar{\sigma}^\nu \varepsilon )_{\alpha\beta}
	\partial_\nu \lambda\  \bar{\phi}^{N}
	, 
\label{eq:Lambda(1,1):GenStr}
\end{equation}
where ${\cal C}_{(N)}^{\alpha\beta}{}_{(j_1 \dots j_{2m})}$ represents  
a term of the form
\begin{equation}
\left( 
C_{(j_1 j_2)} \cdot C_{(j_3 j_4)} \cdots C_{(j_{2N-1} j_{2N})} 
\right){}^{\alpha\beta} 
\equiv 
	C_{(j_1 j_2)}^{\alpha \alpha_2} 
	\varepsilon_{\alpha_2 \alpha_3} C_{(j_3 j_4)}^{\alpha_3 \alpha_4}
	\varepsilon_{\alpha_4 \alpha_5} 
	\cdots 
	\varepsilon_{\alpha_{2N-2} \alpha_{2N-1}} 
		C_{(j_{2N-1} j_{2N})}^{\alpha_{2N-1} \beta}
\label{eq:CCCC}
\end{equation}
with $2(N-m)$ $SU(2)$ indices contracted by $\epsilon$ tensors and the rest of $SU(2)$ 
indices completely symmetrized, which are renamed $j_1,\dots, j_{2m}$. 
The summation in (\ref{eq:Lambda(1,1):GenStr}) 
is taken over the appropriate contractions of the $SU(2)$ indices. 
{}From eq.(\ref{eq:Recursion(1,1)3}) and (\ref{eq:GenRel}), 
we can easily see that $(\lambda_\mu^{(-2)}{}_{(N+1)})_{(j_1 \dots j_{2m})}$ 
also has the same form as the assumption (\ref{eq:Lambda(1,1):GenStr}). 
$(\lambda_\mu^{(-2)}{}_{(1)})_{(ij)}$ also takes the same form as 
(\ref{eq:Lambda(1,1):1}). 
Therefore, the assumption (\ref{eq:Lambda(1,1):GenStr}) is true 
for any $N$. 

It is important to know $(\lambda_\mu^{(-2)}{}_{(N)})_{(ij)}$, 
because $\delta^{*}_{\lambda_C} A_\mu$ can be determined by them  
as was seen in eq.(\ref{eq:DeltaAmu2}). 
Based on the general form (\ref{eq:Lambda(1,1):GenStr}) and the fact 
stated below, 
the full form of $(\lambda_\mu^{(-2)}{}_{(N)})_{(ij)}$ can be written 
down explicitly,  
though we will actually give a formula for 
$(\lambda_\mu^{(-2)}{}_{(N)})_{(ij)} u^{-i} u^{-j}$. 
$(\lambda_\mu^{(-2)}{}_{(N-1)})_{(ij)}$ is determined 
by the recursion relations (\ref{eq:Recursion(1,1)2}), 
(\ref{eq:Recursion(1,1)3}) and (\ref{eq:Recursion(1,1)4}). 
Eq.(\ref{eq:Recursion(1,1)2}) tells us that any term 
in $(\lambda_\mu^{(-2)}{}_{(N-1)})_{(j_1 j_2 \dots j_{2m})}$ ($m\ge 2$) 
which has $2m$ uncontracted $SU(2)$ indices 
will split into three terms in $\lambda_\mu^{(-2)}{}_{(N)}$: 
terms with $2m+2$, $2m$ and $2m-2$ uncontracted $SU(2)$ indices. 
In this way, any specific term in
$(\lambda_\mu^{(-2)}{}_{(N-1)})_{(ij)}$ 
($N\ge 2$) 
can be traced back to $(\lambda_\mu^{(-2)}{}_{(1)})_{(ij)}$ 
given in (\ref{eq:Lambda(1,1):1}) 
through a unique track of increasing and decreasing of the number of 
uncontracted $SU(2)$ indices. 

Based on these facts, one of the full form of 
$(\lambda_\mu^{(-2)}{}_{(N)})_{(ij)}$ 
can be formulated as follows: 
Define 
a series of numbers
$\{ \eta_2, \eta_3, \dots , \eta_{N} \}$, 
where 
\begin{equation}
\eta_i=\{-1,0,+1\} 
\quad ( i= 2,\dots, N ) 
, 
\label{eq:Series}
\end{equation}
satisfying 
\begin{equation} 
m_n \ge 1  
	\quad 
	\mbox{for } {}^{\forall} n \le N-1 
, \qquad 
m_N = 1  
. 
\label{eq:Series:Condition:Lambda(1,1)}
\end{equation}
Here we have defined $m_n$ as follows: 
\begin{equation}
m_1 \equiv 1
, \qquad 
m_n 
\equiv
	1 + \sum_{i=2}^n \eta_i 
	\quad  (n\ge 2) 
	. 
\label{eq:m_n}
\end{equation}
Each series corresponds to ``a unique track'' stated above: 
$\eta_n$ determines which term of three terms in eq.(\ref{eq:Recursion(1,1)2}) 
is chosen in the given path, from the $(n-1)$-th step to the next. 
If $\eta_n$ is $+1$, the first term in eq.(\ref{eq:Recursion(1,1)2}) is chosen.   
Therefore the number of uncontracted $SU(2)$ indices will increase  
by two compared to the former step. 
Similarly, $\eta_n=0$ corresponds to the choice of the second term 
thus the number of uncontracted indices will be unchanged. 
$\eta_n=-1$ corresponds to the third term and the number of indices 
will decrease by two. 
$m_n$ denotes a half of the number of uncontracted $SU(2)$ indices at the $n$-th step. 
The latter condition in eq.(\ref{eq:Series:Condition:Lambda(1,1)}) 
restrict a path to the one which ends up with $(\lambda_\mu^{(-2)}{}_{(N)})_{(ij)}$: 
the number of uncontracted $SU(2)$ indices should be two at the $N$-th step. 
With the use of the set of such series, 
we can write 
$(\lambda_\mu^{(-2)})_{(ij)} \equiv \sum_{N} 
(\lambda_\mu^{(-2)}{}_{(N)})_{(ij)}$ 
as 
\begin{equation}
(\lambda_\mu^{(-2)})_{(ij)} \ u^{-i} u^{-j}  
= 
	- h^{--}{}_\mu{}^\nu (\bar{\phi}) \ \partial_\nu \lambda 
. 
\label{eq:Lambda(1,1):N:ij}
\end{equation}
Here we have defined 
\begin{eqnarray} 
&h^{--}{}_\mu{}^\nu (\bar{\phi}) &
= 
	h_{(ij)}{}_\mu{}^\nu (\bar{\phi}) \ u^{-i} u^{-j}
	\nonumber\\
&&\equiv 
	2 \sqrt{2} i C^{--}{}_\mu{}^\nu \bar{\phi} 
	\nonumber\\
&&{} \ 
	+ {i \over 2}  \sum_{N=2}^\infty 
%	\underbrace{
	\left(
		C_{(j_{2N} j_{2N-1})} \cdot C_{(j_{2N-2} j_{2N-3})} 
		\cdots  C_{(j_2  j_1)}
	\right)
%	}_{\sim {\cal O}(C^N)}
	{}^{\alpha\beta} 
	(\sigma_\mu \bar{\sigma}^\nu \varepsilon )_{\alpha\beta} 
	( - 2 \sqrt{2} \bar{\phi} )^N  
	\nonumber\\
&&{} \ 
\times \!\! 
	\sum_{ \{ \eta_2,\dots,\eta_{N} \} } \! 
	{1\over (2m_1)!} \cdots {1\over (2m_{N})!} 
	\sum_{\sigma_2} \dots \sum_{\sigma_{N+1}} 
	\nonumber\\
&&{} \ 
\times 
	\Big[
		\prod_{\{ n | \eta_n=+1 \} } 
		\alpha_{n}
	\Big] 
	\Big[
		\prod_{ \{ n | \eta_n= 0 \} } 
		\beta_{n}
		\epsilon^{j_{2n-1} j_{\sigma_n(k_n)}}
	\Big] 
	\Big[
		\prod_{ \{ n | \eta_n= -1 \} } 
		\gamma_{n}
		\epsilon^{j_{2n} j_{\sigma_n(l_n)}}
		\epsilon^{j_{2n-1} j_{\sigma_n(k_n)}}
	\Big] 
	\nonumber\\
&&{} \ 
\times 
	\  u^{-\, j_{\sigma_{N+1}(l_{N+1})}} u^{-\, j_{\sigma_{N+1}(k_{N+1})}} 
, 
\label{eq:h--}
\end{eqnarray}
where 
\begin{eqnarray}
\alpha_{n} 
&\equiv& 
	{1\over m_{n-1}+2}
\label{eq:alpha_n}
	, \\
\beta_{n}
&\equiv& 
	- {1\over m_{n-1}+1}
\label{eq:beta_n}
	, \\
\gamma_{n}
&\equiv& 
	{m_{n-1}+1 \over 2 m_{n-1} (2m_{n-1}+1)}
\label{eq:gamma_n}
. 
\end{eqnarray}
$\sigma_n$ means a permutation of the elements of $I_n \setminus K_n$, where 
\begin{eqnarray}
&& 
I 
\equiv 
\Bigl\{ 1,2,\dots, 2N \Bigr\} \setminus  \Bigl\{ 2i-1, 2j-1, 2j 
	\ \Big| \ {}^\forall i, \eta_i=0 
	;  
	\ {}^\forall j, \eta_j=-1 \Bigr\}
	, \\
&& 
I_1 
\equiv 
	\emptyset 
	, \quad 
	I_n \equiv 
		\Bigl\{ i \ \Big| \ {}^\forall i \le 2n-2 , i\in I \Bigr\} 
	\ ( 2 \le n \le N+1 )
	, \\
&& 
K_1, K_2 
\equiv 
	\emptyset 
	, \nonumber\\
&&{} \quad 
K_n 
\equiv 
	\Bigl\{ \sigma_p (k_p), \sigma_q (k_q), \sigma_q (l_q) 
	\ \Big| \ {}^\forall p<n , \eta_p=0, \  {}^\forall q<n , \eta_q=-1 \Bigr\}
		\nonumber\\
&&{} \qquad  
		\ ( 3 \le n \le N+1 ) 
		, \\
&&
k_n 
\equiv 
	\min (I_n \setminus K_n )
	\ (n \ge 2) 
	, \\
&& 
l_n 
\equiv 
	\min (I_n \setminus K_n \setminus \{ k_n \} ) 
	\ (n \ge 2) 
	. 
\end{eqnarray}
As was seen from eq.(\ref{eq:Recursion(1,1)2}) 
together with the definition of $\eta_n$, 
from the $(n-1)$-th step to the $n$-th step,  
$(C\cdots C)^{\alpha\beta}$ is multiplied by $C_{(ij)}$ from the left 
and they will become $(C_{(ij)}\cdot C\cdots C)^{\alpha\beta}$,  
while the number of uncontracted $SU(2)$ indices are not necessarily increased by two: 
indices of $C_{(ij)}$ may be contracted with 
uncontracted indices of $(C\cdots C)^{\alpha\beta}$ 
according to $\eta_n$ 
(zero for $\eta_n=+1$, one for $\eta_n=0$ and two for $\eta_n=-1$). 
Thus which of the $2N$ indices are contracted with an index in the right side 
is determined uniquely by the series $\{ \eta_2,\dots, \eta_N\}$ 
(up to the permutation of two indices of each $C$).   
$I$ is a set of the labels of the remaining indices. 
$I_n$ is a subset of $I$, where the indices are restricted 
to the ones placed at $n-1$ $C$'s 
from the right, that is, $(C\cdots C)^{\alpha\beta}$ above. 
In general, some of the indices specified by $I_n$ have already contracted 
at the $n$-th step  
and only $m_n$ indices of $I_n$ are uncontracted. 
$K_n$ specifies the indices of $I_n$ which have already contracted by $\epsilon$ tensors. 
Thus $I_n \setminus K_n$ specifies  
the $m_n$ uncontracted $SU(2)$ indices at the $(n-1)$-th step, 
which should be completely symmetrized. 
$k_n$ (and also $l_n$) has a meaning of the rightmost one of those indices  
(of course instead we can choose the leftmost one or another specific index 
because they are symmetrized). 
$h^{--}{}_\mu{}^\nu (\bar{\phi})$ is expanded as 
\begin{equation}
h^{--}{}_\mu{}^\nu (\bar{\phi})
= 
	2 \sqrt{2} i C^{--}{}_\mu{}^\nu \bar{\phi} 
	- 4 i \epsilon^{mn} \left( C_{(im)} \cdot C_{(jn)} \right){}\!^{\alpha\beta} 
		( \sigma_\mu{}^\nu \varepsilon )_{\alpha\beta} 
		\bar{\phi}^2 \ u^{-i} u^{-j}
	+ {\cal O}(C^3) 
	. 
\end{equation}

If we use the formula (\ref{eq:Lambda(1,1):N:ij}), 
eq.(\ref{eq:DeltaAmu2}) tells us that 
one of the full form of $\delta^{*}_{\lambda_C} A_\mu$ 
can be formulated as 
\begin{equation}
\delta^{*}_{\lambda_C} A_\mu 
= 
	- \partial_\mu \lambda 
	- f(\bar{\phi}) \bar{\phi} \ \partial_\mu \lambda
. 
\label{eq:DeltaAmu:Gen}
\end{equation} 
Here we have defined 
\begin{eqnarray}
f(\bar{\phi})
&\equiv& 
	{1\over \sqrt{2}} C_s 
	\nonumber\\
&&{} 
	- \frac14  \sum_{N=2}^\infty 
%	\underbrace{
	{\rm Tr} \left[
		C_{(j_{2N} j_{2N-1})} \cdot C_{(j_{2N-2} j_{2N-3})} 
		\cdots  C_{(j_2 j_1)}
	\right]
%	}_{\sim {\cal O}(C^N)}
	( - 2 \sqrt{2})^{N} \bar{\phi}^{N-1} 
	\nonumber\\
&&{} \ 
\times \!\! 
	\sum_{ \{ \eta_2,\dots,\eta_{N} \} } \! 
	{1\over (2m_1)!} \cdots {1\over (2m_{N-1})!} 
	\sum_{\sigma_2} \dots \sum_{\sigma_{N}} 
	\nonumber\\
&&{} \ 
\times 
	\Big[
		\prod_{\{ n | \eta_n=+1 \} } 
		\alpha_{n}
	\Big] 
	\Big[
		\prod_{ \{ n | \eta_n= 0 \} } 
		\beta_{n}
		\epsilon^{j_{2n-1} j_{\sigma_n(k_n)}}
	\Big] 
	\Big[
		\prod_{ \{ n | \eta_n= -1 \} } 
		\gamma_{n}
		\epsilon^{j_{2n} j_{\sigma_n(l_n)}}
		\epsilon^{j_{2n-1} j_{\sigma_n(k_n)}}
	\Big] 
\label{eq:f}
	\nonumber\\
&=&    
	{1\over \sqrt{2}} C_s 
	+ {2\over 3} C_2 \bar{\phi}
	+ {2 \sqrt{2} \over 3} C_3 \bar{\phi}^2 
	+ {\cal O}(C^4)
, 
\end{eqnarray}
where a series $\{ \eta_2 , \eta_3, \dots, \eta_{N} \}$ is 
the similar one defined by (\ref{eq:Series}) 
but the condition (\ref{eq:Series:Condition:Lambda(1,1)}) replaced with 
\begin{equation} 
m_n \ge 1  
	\quad 
	\mbox{for } {}^{\forall} n \le N-1 
, \qquad 
m_N = 0  
, 
\label{eq:Series:Condition:DeltaAmu}
\end{equation}
and 
\begin{eqnarray}
&&
{\rm Tr} \left[ 
	C_{(j_1 j_2)} \cdot C_{(j_3 j_4)} \cdots C_{(j_{2N-1} j_{2N})} 
	\right] 
\equiv 
	\varepsilon_{\alpha\beta} \left( 
	C_{(j_1 j_2)} \cdot C_{(j_3 j_4)} \cdots C_{(j_{2N-1} j_{2N})} 
	\right){}^{\beta\alpha} 
	, \\
&&
C_2 
\equiv 
	{\rm Tr} \left[
		C_{(ij)} \cdot C_{(kl)}
	\right] 
	\epsilon^{jk} \epsilon^{li}
	, \\
&&
C_3
\equiv 
	{\rm Tr} \left[
		C_{(ij)} \cdot C_{(kl)} \cdot C_{(mn)} 
	\right] 
	\epsilon^{jk} \epsilon^{lm} \epsilon^{ni}
	. 
\end{eqnarray}

Note that, the second term of eq.(\ref{eq:DeltaAmu:Gen}) 
may be naively written with a factor  
$$
	\left(
		C_{(j_{2N} j_{2N-1})} \cdot C_{(j_{2N-2} j_{2N-3})} 
		\cdots  C_{(j_2 j_1)}
	\right){}\!^{\alpha\beta}
	(\sigma_\mu \bar{\sigma}^\nu \varepsilon )_{\alpha\beta}
	\partial_\nu \lambda \  ( - 2 \sqrt{2} \bar{\phi} )^N
	,   
$$
in accordance with the general form of $(\lambda_\mu^{(-2)}{}_{(N)})_{(ij)}$
given in (\ref{eq:Lambda(1,1):GenStr}), 
but this factor can be simplified as written in (\ref{eq:DeltaAmu:Gen})
 and (\ref{eq:f}) 
though it is possible only in the sum, 
as explained below. 
Let us write a specific term in the sum 
corresponding to a given series 
$\{\eta_2, \dots , \eta_{N-1} , \eta_{N} \}$ as 
\begin{equation}
c \times {\cal C}_{(N)}^{\alpha\beta} 
	(\sigma_\mu \bar{\sigma}^\nu \varepsilon )_{\alpha\beta}
	\partial_\nu \lambda \ (- 2 \sqrt{2} \bar{\phi})^N
, 
\end{equation} 
where ${\cal C}_{(N)}^{\alpha\beta}$ is 
defined below eq.(\ref{eq:Lambda(1,1):GenStr}), 
and $c$ is a constant. 
We can easily see that another series given by 
$\{-\eta_{N-1}, -\eta_{N-2}, \dots , -\eta_{2} , \eta_{N} \}$ 
will produce a sign reversed, the spinor indices of ${\cal C}$ exchanged 
contribution: 
\begin{equation}
- c \times {\cal C}_{(N)}^{\beta\alpha} 
	(\sigma_\mu \bar{\sigma}^\nu \varepsilon )_{\alpha\beta}
	\partial_\nu \lambda \ (- 2 \sqrt{2} \bar{\phi})^N
. 
\end{equation} 
For example, for $N=5$, there is a contribution corresponding to 
a series $\{ 0, +1 , -1 , -1 \}$ which can be written as  
\begin{equation}
+ \frac14 \ (- \frac12) \ \frac13 \ {3\over 20} \ \frac13 
	{\cal C}_{(5)}^{\alpha\beta} 
	(\sigma_\mu \bar{\sigma}^\nu \varepsilon )_{\alpha\beta}
	\partial_\nu \lambda 
	 (- 2 \sqrt{2} \bar{\phi} )^{5}
	, 
\end{equation}
where 
\begin{equation}
{\cal C}_{(5)}^{\alpha\beta} 
\equiv 
\epsilon^{j_{10} j_{6}}
\epsilon^{j_{9} j_{5}}
\epsilon^{j_{8} j_{4}}
\epsilon^{j_{7} j_{2}}
\epsilon^{j_{3} j_{1}}
\left( 
C_{(j_{10} j_{9})} \cdot 
C_{(j_{8} j_{7})} \cdot 
C_{(j_{6} j_{5})} \cdot 
C_{(j_{4} j_{3})} \cdot 
C_{(j_{2} j_{1})} 
\right)^{\alpha\beta} 
.  
\end{equation}
Note that here we have suppressed the possible permutations 
of the indices $j_2$, $j_4$, $j_5$ and $j_6$ by representing them 
with the one contribution (whose each contraction is taken 
with the rightmost possible $SU(2)$ index of the factor 
$(C \cdots C)^{\alpha\beta}$).
On the other hand, $\{ +1 , -1 , 0 , -1 \}$ gives 
\begin{equation}
+ \frac14 \ \frac13 \ {3\over 20} \ (-\frac12) \ \frac13 
	{\cal C}'{}_{(5)}^{\alpha\beta}
	(\sigma_\mu \bar{\sigma}^\nu \varepsilon )_{\alpha\beta} 
	\partial_\nu \lambda 
	(- 2 \sqrt{2} \bar{\phi} )^{5}
, 
\end{equation}
where 
\begin{eqnarray}
{\cal C}'{}_{(5)}^{\alpha\beta}
&\equiv& 
\epsilon^{j_{10} j_{8}}
\epsilon^{j_{9} j_{4}}
\epsilon^{j_{7} j_{3}}
\epsilon^{j_{6} j_{2}}
\epsilon^{j_{5} j_{1}}
\left( 
C_{(j_{10} j_{9})} \cdot 
C_{(j_{8} j_{7})} \cdot 
C_{(j_{6} j_{5})} \cdot 
C_{(j_{4} j_{3})} \cdot 
C_{(j_{2} j_{1})} 
\right)^{\alpha\beta} 
\nonumber\\
&=& 
\epsilon^{j_{1} j_{4}}
\epsilon^{j_{2} j_{7}}
\epsilon^{j_{4} j_{8}}
\epsilon^{j_{5} j_{9}}
\epsilon^{j_{6} j_{10}}
\left(
C_{(j_{1} j_{2})} \cdot 
C_{(j_{3} j_{4})} \cdot 
C_{(j_{5} j_{6})} \cdot 
C_{(j_{7} j_{8})} \cdot 
C_{(j_{9} j_{10})} \cdot 
\right)^{\alpha\beta} 
\nonumber\\
&=& 
(-1)^5 
\epsilon^{j_{10} j_{8}}
\epsilon^{j_{9} j_{4}}
\epsilon^{j_{7} j_{3}}
\epsilon^{j_{6} j_{2}}
\epsilon^{j_{5} j_{1}}
(-1)^4
\left( 
C_{(j_{10} j_{9})} \cdot 
C_{(j_{8} j_{7})} \cdot 
C_{(j_{6} j_{5})} \cdot 
C_{(j_{4} j_{3})} \cdot 
C_{(j_{2} j_{1})} 
\right)^{\beta\alpha} 
\nonumber\\
&=& 
	- {\cal C}_{(5)}^{\beta\alpha} 
	. 
\end{eqnarray}
Here we have renamed the dummy indices in the second line 
and used a property of the quantity (\ref{eq:CCCC}), 
\begin{equation}
\left( 
C_{(j_1 j_2)} \cdot C_{(j_3 j_4)} \cdots C_{(j_{2N-1} j_{2N})} 
\right){}^{\alpha\beta} 
= 
	(-1)^{N-1} \left( 
C_{(j_{2N} j_{2N-1})} \cdot C_{(j_{2N-3} j_{2N-4})} \cdots C_{(j_{2} j_{1})} 
	\right){}^{\beta\alpha} 
. 
\label{eq:CCCC:Property}
\end{equation}
in the third line. 
Therefore, taken into account the summation over the series, 
$(\sigma_\mu \bar{\sigma}^\nu \varepsilon )_{\alpha\beta}$ above can be 
reduced to $-\eta_\mu{}^{\nu} \varepsilon_{\alpha\beta}$: 
\begin{equation}
c \times {\rm Tr} [{\cal C}_{(N)}] (- 2 \sqrt{2} \ \bar{\phi})^N  
\partial_\mu \lambda 
. 
\end{equation}

By a direct calculation with the use of eq.(\ref{eq:DeltaAmu2}),  
together with $( \delta^{*}_{\lambda_C} A_\mu )_{(0)}$ 
and $( \delta^{*}_{\lambda_C} A_\mu )_{(1)}$ given 
in eq.(\ref{eq:DeltaAmu0}) and (\ref{eq:DeltaAmu1}) respectively, 
we find that $\delta^{*}_{\lambda_C} A_\mu $ is given by 
\begin{eqnarray}
\delta^{*}_{\lambda_C} A_\mu 
&=& 
	- \partial_\mu \lambda 
	- {1\over \sqrt{2}} C_s 
		\partial_\mu \lambda \ \bar{\phi} 
	- {2\over 3} C_2  
		\partial_\mu \lambda \ \bar{\phi}^2
	- {2 \sqrt{2} \over 3} C_3
		\partial_\mu \lambda \ \bar{\phi}^3 
	+ {\cal O}(C^4)
	, 
\end{eqnarray}
and we can check that 
this is actually coincide with the result given by the formula 
(\ref{eq:DeltaAmu:Gen}). 

\subsection{The deformed gauge transformations of the other fields}

Next we consider the $(\bar{\theta}^{+})^2$-component 
of $\delta^{*}_{\lambda_C} V_{WZ}^{++}$ 
to determine $\delta^{*}_{\lambda_C} \phi$ 
and $\lambda^{(-2)}$. 
The equation to solve is read from 
(\ref{eq:DeformedTransf2}) and (\ref{eq:DeformedTransf3}): 
\begin{eqnarray}
\sqrt{2} i \delta^{*}_{\lambda_C} \phi 	
&=&
	i C_s \partial_\mu \lambda \ A^\mu 
	- 2 i C^{(+-)}{}^{\alpha\beta} 
		(\sigma_\mu \bar{\sigma}^\nu \varepsilon )_{\alpha\beta} 
		\partial_\nu \lambda \ A^\mu 
	+ C^{++}{}^{\alpha\beta} 
		(\sigma_\mu \bar{\sigma}^\nu \varepsilon )_{\alpha\beta} 
		\lambda_\nu^{(-2)} A^\mu \nonumber\\ 
&&	- \partial^{++} \lambda^{(-2)}
	, 
\label{eq:(0,2)Comp}
\end{eqnarray}
where $\delta^{*}_{\lambda_C} \phi$ is identified 
with the lowest order term in the harmonic expansion of the left hand side 
(i.e. the $U(1)$ charge 0, irreps of isospin 0 part).  
Expanding $\delta^{*}_{\lambda_C} \phi$ and $\lambda^{(-2)}$ as 
\begin{eqnarray}
\delta^{*}_{\lambda_C} \phi
&\equiv& 
	\sum_{N=0}^{\infty} (\delta^{*}_{\lambda_C} \phi )_{(N)} 
	, 
\end{eqnarray}
where $(\delta^{*}_{\lambda_C} \phi )_{(N)}$ is an ${\cal O}(C^N)$-quantity, 
eq.(\ref{eq:(0,2)Comp}) gives the following equations 
to determine $\delta^{*}_{\lambda_C} \phi$: 
\begin{eqnarray}
(\delta^{*}_{\lambda_C} \phi )_{(1)}
&=& 
	{1\over\sqrt{2}} C_s \partial_\mu\lambda \ A^\mu 
	\label{eq:DeltaPhi:1}
, \\
( \delta^{*}_{\lambda_C} \phi )_{(N)}
&=& 
	{i \over 3 \sqrt{2}} 
	\epsilon^{ik} \epsilon^{jl} 
	C_{(ij)}^{\alpha\beta} (\lambda_{\nu}^{(-2)}{}_{(N-1)} )_{(kl)} 
	(\sigma_\mu \bar{\sigma}^\nu \varepsilon )_{\alpha\beta} 
	A^\mu 
	\quad (N\ge 2)
	, 
\label{eq:DeltaPhi:N}
\end{eqnarray}
where we have used the reduction formula (\ref{eq:Reduction}). 
Comparing this equation to eq.(\ref{eq:DeltaAmu2}), 
we find that $\delta^{*}_{\lambda_C} \phi$ can be obtained by 
replacing one of $\bar{\phi}$ in $\delta^{*}_{\lambda_C} A_\mu$ 
with $- A^\mu$: 
indeed, we have 
\begin{eqnarray}
\delta^{*}_{\lambda_C} \phi 
= 
	{1\over \sqrt{2}} C_s 
		\partial_\mu \lambda \ A^\mu  
	+ {2\over 3} C_2  
		\partial_\mu \lambda \ A^\mu \bar{\phi}
	+ {2 \sqrt{2} \over 3} C_3
		\partial_\mu \lambda \ A^\mu \bar{\phi}^2 
	+ {\cal O}(C^4)
	.
\end{eqnarray}
Thus the full form of $\delta^{*}_{\lambda_C} \phi$ can be formulated as 
the one quite similar to eq.(\ref{eq:DeltaAmu:Gen}): 
\begin{equation}
\delta^{*}_{\lambda_C} \phi 
= 
	f(\bar{\phi}) \partial_\mu \lambda \ A^\mu 
, 
\end{equation}
where $f(\bar{\phi})$ is defined in eq.(\ref{eq:f}). 

Eq.(\ref{eq:(0,2)Comp}) also gives the equations to determine 
$\lambda^{(-2)}$: 
\begin{eqnarray}
\partial^{++} \lambda^{(-2)}{}_{(1)}
&=&
	- 2 i C^{(+-)}{}^{\alpha\beta} 
		(\sigma_\mu\bar{\sigma}^\nu \varepsilon )_{\alpha\beta}
		\partial_\nu \lambda \ A^\mu 
		, \\
\partial^{++} \lambda^{(-2)}{}_{(N)}
&=& 
	C^{++}{}^{\alpha\beta} 
		(\sigma_\mu \bar{\sigma}^\nu \varepsilon )_{\alpha\beta} 
		\lambda_\nu^{(-2)}{}_{(N-1)} A^\mu  
	+ \sqrt{2} i (\delta^{*}_{\lambda_C} \phi )_{(N)}
	\quad ( N\ge 2 )
. 
\label{eq:Determine(0,2)}
\end{eqnarray}
The first equation is easily solved to give $\lambda^{(-2)}{}_{(1)}$, 
%solved by 
%\begin{equation}
%\lambda^{(-2)}{}_{(1)} 
%= - i C^{--}{}^{\alpha\beta} 
%	( \sigma_\mu \bar{\sigma}^\nu \varepsilon )_{\alpha\beta} 
%	\partial_\nu \lambda \ A^\mu 
%	. 
%\label{eq:Lambda(0,2)(1)}
%\end{equation}
and the second equation can be used to determine $\lambda^{(-2)}{}_{(N)}$ 
from $\lambda_\mu^{(-2)}{}_{(N-1)}$ 
and $(\delta^{*}_{\lambda_C} \phi)_{(N)}$ given above. 
Comparing these equations to eqs.(\ref{eq:Recursion(1,1):Initial}) 
and (\ref{eq:Recursion(1,1)}), 
it is found that $\lambda^{(-2)}$ can be obtained 
by replacing one of $\phi$ in $\lambda_\mu^{(-2)}$ with $-A^\mu/\sqrt{2}$. 
Of course, with the harmonic expansion of $\lambda^{(-2)}$, we can construct 
the recursion relation in terms of the coefficient functions in the 
harmonic expansion, 
which is similar to eq.(\ref{eq:Recursion(1,1)4}). 
The expansion is 
\begin{eqnarray} 
\lambda^{(-2)} 
&\equiv& 
	\sum_{N=1}^{\infty} \lambda^{(-2)}{}_{(N)}
, 
	\\
\lambda^{(-2)}{}_{(N)}
&=& 
	\sum_{n=0}^{\infty} 
	( \lambda^{(-2)}{}_{(N)} )_{(i_1 \dots i_{n} k_1 \dots k_{n+2})}
	u^{+}{}^{i_1} \dots u^{+}{}^{i_n} u^{-}{}^{k_1} \dots u^{-}{}^{k_{n+2}}
	, 
\end{eqnarray} 
where $\lambda^{(-2)}{}_{(N)}$ is an ${\cal O}(C^N)$-quantity. 
Note again that the $C$-independent part of $\lambda^{(-2)}$ is absent, 
because of its absence in the $C \rightarrow 0$ limit. 
Then $\lambda^{(-2)}$ is determined as  
\begin{eqnarray}
\lambda^{(-2)}
&=& 
	- i C^{--}{}^{\alpha\beta} 
	(\sigma_\mu \bar{\sigma}^\nu \varepsilon )_{\alpha\beta} 
	\partial_\nu \lambda \ A^\mu 
	\nonumber\\
&&{} 
	+ 2 \sqrt{2} i \left( C_{(ij)} \cdot C_{(kl)} \right)^{\alpha\beta} 
		( \sigma_\mu \bar{\sigma}^\nu \varepsilon )_{\alpha\beta} 
		\partial_\nu \lambda \ A^\mu \bar{\phi} 
		\left( 
			\frac13 u^{+(i} u^{-j} u^{-k} u^{-l)} 
			- \frac12 \epsilon^{jl} u^{-(i} u^{-k)}
		\right) 
	\nonumber\\
&&{} 
	+ {\cal O}(C^3)
. 
\label{eq:Lambda(0,2):DirectCal}
\end{eqnarray}

For 
$\delta^{*}_{\lambda_C} \psi^i$, 
the equation to solve is read from 
(\ref{eq:DeformedTransf2}) and (\ref{eq:DeformedTransf3}): 
\begin{eqnarray}
&&
	-2 C_s \partial_\mu \lambda \ (\sigma^\mu \bar{\psi}^{-})_\alpha  
	\nonumber\\
&&{}
	- 4 C^{(+-)}{}^{\beta\gamma} 
		(\sigma_\mu \bar{\sigma}^\nu \varepsilon )_{\beta\gamma} 
		\partial_\nu \lambda \ (\sigma^\mu \bar{\psi}^{-})_\alpha 
	- 2 i C^{++}{}^{\beta\gamma} 
		(\sigma_\mu \bar{\sigma}^\nu \varepsilon )_{\beta\gamma} 
		\lambda_\nu^{(-2)} (\sigma^\mu \bar{\psi}^{-})_\alpha 
		\nonumber\\
&&{}
	- \partial^{++} \lambda_\alpha^{(-3)}
	- 2 \sqrt{2} (\varepsilon C^{++} \lambda^{(-3)} )_\alpha \bar{\phi}
	\nonumber\\
&&= 
	4 \delta^{*}_{\lambda_C} \psi^{-}_{\alpha}  
	, 
\label{eq:(1,2)Comp}
\end{eqnarray}
where $\delta^{*}_{\lambda_C} \psi^{-}$ is identified 
with the lowest order term in the harmonic expansion of the left hand side 
(i.e. the $U(1)$ charge $-1$, irreps of isospin $1/2$ part).  
Expand $\delta^{*}_{\lambda_C} \psi^{-}$ and $\lambda_\alpha^{(-3)}$ as 
\begin{eqnarray}
\delta^{*}_{\lambda_C} \psi^{-}
&\equiv& 
	\sum_{N=0}^{\infty} (\delta^{*}_{\lambda_C} \psi^{-} )_{(N)} 
	, 
	\\
\lambda_\alpha^{(-3)} 
&\equiv& 
	\sum_{N=1}^{\infty} \lambda_\alpha^{(-3)}{}_{(N)}
, 
	\\
\lambda_\alpha^{(-3)}{}_{(N)}
&=& 
	\sum_{n=0}^{\infty} 
	( \lambda_\alpha^{(-3)}{}_{(N)} )_{(i_1 \dots i_{n} k_1 \dots k_{n+3})}
	u^{+}{}^{i_1} \dots u^{+}{}^{i_n} u^{-}{}^{k_1} \dots u^{-}{}^{k_{n+3}}
. 
\end{eqnarray}
where $(\delta^{*}_{\lambda_C} \psi^{-} )_{(N)}$ and 
$\lambda_\alpha^{(-3)}{}_{(N)}$ are ${\cal O}(C^N)$-quantities. 
Then eq.(\ref{eq:(1,2)Comp}) gives the following equations: 
\begin{eqnarray}
(\delta^{*}_{\lambda_C} \psi^{-}_{\alpha} )_{(1)}
&=& 
- \frac12 C_s \partial_\mu\lambda \ (\sigma^\mu \bar{\psi}^{-})_\alpha  
	- \frac23 C_{(ik)}^{\gamma\delta} 
		(\sigma_\mu \bar{\sigma}^\nu \varepsilon )_{\gamma\delta} 
		\partial_\nu \lambda (\sigma^\mu \bar{\psi}^{k} )_\alpha u^{-i}
	\label{eq:DeltaPsi1}
, \\
\partial^{++} \lambda_\alpha^{(-3)}{}_{(1)}
&=&
	- 4 C_{(ij)}^{\beta\gamma} 
		(\sigma_\mu\bar{\sigma}^\nu \varepsilon )_{\beta\gamma}
		\partial_\nu \lambda \ (\sigma^\mu \bar{\psi}_k)_\alpha 
		u^{+ (i} u^{-j} u^{-k)}
		, \\
\partial^{++} \lambda_\alpha^{(-3)}{}_{(N)}
&=& 
	- 2 i C^{++}{}^{\gamma\delta} 
		(\sigma_\mu \bar{\sigma}^\nu \varepsilon )_{\gamma\delta} 
	\lambda_\nu^{(-2)}{}_{(N-1)} (\sigma^\mu \bar{\psi}^{-} )_\alpha  
		\nonumber\\
&&{}
- 2 \sqrt{2} ( \varepsilon C^{++} \lambda^{(-3)}{}_{(N-1)} )_\alpha \bar{\phi} 
	- 4 (\delta^{*}_{\lambda_C} \psi_\alpha^{-} )_{(N)}
	\quad ( N\ge 2 )
. 
\label{eq:Determine(2,1)}
\end{eqnarray}
%The second equation is solved by 
%\begin{equation}
%\lambda_\alpha^{(-3)}{}_{(1)} 
%= - \frac43 C^{--}{}^{\alpha\beta} 
%	( \sigma_\mu \bar{\sigma}^\nu \varepsilon )_{\alpha\beta} 
%	\partial_\nu \lambda \ (\sigma^\mu \bar{\psi}^{-})_\alpha  
%	. 
%\label{eq:Lambda(0,2)(1)}
%\end{equation}
%The third equation can be used to determine $\lambda_\alpha^{(-3)}{}_{(N)}$ 
%from $\lambda_\mu^{(-2)}{}_{(N-1)}$ and $\lambda_\alpha^{(-3)}{}_{(N-1)}$: 
%With the use of the previous results, 
The second and the third equation will lead to 
$\lambda_\alpha^{(-3)}$ as  
\begin{eqnarray}
\lambda_\alpha^{(-3)}
&=& 
	+ {8 \over 3} (\varepsilon C^{--} \sigma^\mu \bar{\psi}^{-} )_\alpha 
		\partial_\mu \lambda 
		\nonumber\\
&&{}
	+ \frac{8\sqrt{2}}{3} \, 
	{\rm Tr} \left[ 
               C_{(mn)} \cdot C_{(ij)}  
              \right] 
		\partial_{\mu} \lambda (\sigma^{\mu} \bar{\psi}_k )_\alpha 
		\bar{\phi} 
		\nonumber\\
&&{}
	\times 
	\left( 
       u^-{}^{(m} u^-{}^n u^-{}^i u^-{}^j u^-{}^{k)} 
      -\frac{2}{5} \epsilon^{mk} u^+{}^{(n} u^-{}^i u^-{}^{j)} 
     \right) 
	+ {\cal O}(C^3) 
	.  
\end{eqnarray}
Noting again the reduction formula (\ref{eq:Reduction}), 
we find from the third equation 
that $(\delta^{*}_{\lambda_C} \psi^{-} )_{(N)}$ is given by 
\begin{eqnarray} 
( \delta^{*}_{\lambda_C} \psi^{-}_{\alpha} )_{(N)}
&=& 
	- {i \over 6} 
		\epsilon^{jn} 
		C_{(ij)}^{\gamma\delta} (\lambda_{\nu}^{(-2)}{}_{(N-1)} )_{(mn)} 
		(\sigma_\mu \bar{\sigma}^\nu \varepsilon )_{\gamma\delta} 
		(\sigma^\mu \bar{\psi}_{k})_\alpha 
		(\epsilon^{k(m} u^{-i)} - \epsilon^{im} u^{-k})
		\nonumber\\
&&{} 
	+ {i \over 20} 
		\epsilon^{ip} \epsilon^{jq} 
	C_{(ij)}^{\gamma\delta} (\lambda_{\nu}^{(-2)}{}_{(N-1)} )_{(mnpq)} 
		(\sigma_\mu \bar{\sigma}^\nu \varepsilon )_{\gamma\delta} 
		(\sigma^\mu \bar{\psi}_{k})_\alpha 
		\epsilon^{kn} u^{-m} 
		\nonumber\\
&&{} 
	- {\sqrt{2} \over 4} 
		\epsilon^{il} \epsilon^{jm} 
		\left(
		\varepsilon C_{(ij)} \cdot (\lambda^{(-3)}{}_{(N-1)})_{(klm)} 
		\right)_\alpha 
		\bar{\phi} \ u^{-k} 
		. 
	\quad (N\ge 2)
\label{eq:DeltaPsi2}
\end{eqnarray}
Although we do not write down general formula 
for $\delta^{*}_{\lambda_C}\psi^-$ in the present work, 
the lower order terms are given by
\begin{eqnarray}
\delta^{*}_{\lambda_C} \psi^{-}_{\alpha} 
&=& 
	- \frac12 C_s (\sigma^\mu \bar{\psi}^{-} )_\alpha \, \partial_{\mu} \lambda 
- \frac23 (\varepsilon C_{(ij)} \sigma^\mu \bar{\psi}^j )_\alpha 
	\, \partial_{\mu} \lambda \, u^{-i} 
	\nonumber\\
&&{} 
	- {5\sqrt{2} \over 3} 
	(\varepsilon C_{(kl)} \ \varepsilon C_{(ij)} \sigma^\mu \bar{\psi}_k )_\alpha 
		\, \bar{\phi} \, \partial_{\mu} \lambda 
		\left( 
		\frac23 \epsilon^{mi} \epsilon^{n(j} u^{-k)}
		+ \frac13 \epsilon^{mk} \epsilon^{ni} u^{-j} 
		\right)
     \nonumber\\
&&{}+(\varepsilon C_{(ij)} \ \varepsilon C_{(kl)} \ \varepsilon C_{(mn)} 
      \sigma^{\mu}\bar{\psi}_p)_{\alpha} \, \bar{\phi}^2 \, \partial_{\mu} \lambda 
      \nonumber\\
&&{} \times 
     \left\{ 
         \frac{1}{5} \epsilon^{mm^{\prime}} \epsilon^{nn^{\prime}} 
                     \delta_{m^{\prime}}{}^{(i} 
                     \delta_{n^{\prime}}{}^{j} 
                     \epsilon^{pk} 
                     u^-{}^{l)} 
     \right.                                        \nonumber\\ 
&&{} \left. 
        +\frac{1}{15} \epsilon^{mm^{\prime}} \epsilon^{nn^{\prime}} 
                      \left( 
                        \delta_{m^{\prime}}{}^{(i} 
                        \delta_{n^{\prime}}{}^{j)} 
                        \epsilon^{p(k} u^-{}^{l)} 
                       +\delta_{m^{\prime}}{}^{(k} 
                        \delta_{n^{\prime}}{}^{l)} 
                        \epsilon^{p(i} u^-{}^{j)} 
                      \right) 
       \right.                                    \nonumber\\ 
&&{}   \left. 
        +2 \epsilon^{mm^{\prime}} \epsilon^{pp^{\prime}} 
         \epsilon^{ki} \delta_{m^{\prime}}{}^{(l} 
         \delta_{p^{\prime}}{}^{n} u^-{}^{j)} 
        +\frac{32}{9} \epsilon^{kk^{\prime}} 
         \epsilon^{mm^{\prime}} \epsilon^{nn^{\prime}} 
         \delta_{k^{\prime}}{}^{(p} \delta_{m^{\prime}}{}^{i} 
         \delta_{n^{\prime}}{}^{j} u^-{}^{l)} 
       \right\}
        \nonumber\\
&&{}+ {\cal O}(C^4) 
	.
\end{eqnarray}

For 
$\delta^{*}_{\lambda_C} D^{ij}$, 
from (\ref{eq:DeformedTransf2}) and (\ref{eq:DeformedTransf3}), 
the equation to solve is given by  
\begin{eqnarray}
&&
4 \sqrt{2} C^{--}{}^{\mu\nu} \partial_\mu \lambda \ \partial_\nu \bar{\phi} 
	- i \partial^\mu \lambda_\mu^{(-2)} 
	- \sqrt{2} i C^{+-}{}^{\alpha\beta} 
		(\sigma^\mu \bar{\sigma}^\nu \varepsilon )_{\alpha\beta} 
		\partial_\nu (\lambda_\mu^{(-2)} \bar{\phi} )
	- \partial^{++} \lambda^{(-4)}
	\nonumber\\
&&= 
	3 \delta^{*}_{\lambda_C} D^{--}
	. 
\label{eq:(2,2)Comp}
\end{eqnarray}
where $\delta^{*}_{\lambda_C} D^{--}$ is identified 
with the lowest order term in the harmonic expansion of the left hand side 
(i.e. the $U(1)$ charge $-2$, irreps of isospin $1$ part).  
Expanding $\delta^{*}_{\lambda_C} D^{--}$ as 
\begin{equation}
\delta^{*}_{\lambda_C} D^{--}
\equiv 
	\sum_{N=0}^{\infty} (\delta^{*}_{\lambda_C} D^{--} )_{(N)} 
	, 
\end{equation}
where $(\delta^{*}_{\lambda_C} D^{--} )_{(N)}$ is an ${\cal O}(C^N)$-quantity. 
then eq.(\ref{eq:(0,2)Comp}) gives the following equations 
to determine $\delta^{*}_{\lambda_C} D^{--}$: 
\begin{eqnarray}
( \delta^{*}_{\lambda_C} D^{--} )_{(1)}
&=& 
2 \sqrt{2} C^{--}{}^{\mu\nu} \partial_\mu \lambda \ \partial_\nu\bar{\phi} 
%	= - i \partial^\mu (\lambda_\mu^{(-2)}{}_{(1)})
	\label{eq:DeltaD:1}
	, \\
( \delta^{*}_{\lambda_C} D^{--} )_{(N)} 
&=& 
	\left[ 
	- {i \over 3} 
		\partial^\mu (\lambda_\mu^{(-2)}{}_{(N)})_{(ij)}
	- { \sqrt{2} \over 6} i 
		C_s \partial^\mu \left( 
			(\lambda_\mu^{(-2)}{}_{(N-1)})_{(ij)} 
			\bar{\phi} 
		\right)
		\right.\nonumber\\
&&{}
	- {i \over \sqrt{2}} 
		\epsilon^{kl} C_{(ik)}^{\alpha\beta} 
		(\sigma^\mu \bar{\sigma}^\nu \varepsilon )_{\alpha\beta} 
		\partial_\mu \left( 
			(\lambda_\nu^{(-2)}{}_{(N-1)})_{(jl)} 
			\bar{\phi} 
		\right) 
		\nonumber\\
&&\left. {}
	- {3\sqrt{2} \over 20} i 
		\epsilon^{km} \epsilon^{ln} C_{(kl)}^{\alpha\beta} 
		(\sigma^\mu \bar{\sigma}^\nu \varepsilon )_{\alpha\beta} 
		\partial_\mu \left( 
			(\lambda_\nu^{(-2)}{}_{(N-1)})_{(ijmn)} 
			\bar{\phi} 
		\right) 
	\right] 
	u^{-i} u^{-j}
	\quad (N\ge 2). 
	\label{eq:DeltaD:N}
	\nonumber\\
\end{eqnarray}
Here we have used another reduction formula similar to eq.
(\ref{eq:Reduction}): 
\begin{eqnarray}
&&
	C^{(+-)}{}^{\alpha\beta} 
	(\lambda_\mu^{(-2)}{}_{(N)})_{( j_{2m} j_{2m-1} \dots j_{1} )}
	\overbrace{
	u^{+}{}^{(j_{2m}} \dots u^{+}{}^{j_{m+2}} 
	}^{m-1}
	\overbrace{
	u^{-}{}^{j_{m+1}} \dots u^{-}{}^{j_{1})} 
	}^{m+1}
	\nonumber\\
&&= 
	C_{(j_{2m+2} j_{2m+1})}^{\alpha\beta} 
	(\lambda_\mu^{(-2)}{}_{(N)})_{( j_{2m} j_{2m-1} \dots j_{1} )} 
	\left(
	\overbrace{
u^{+}{}^{(j_{2m+2}} u^{+}{}^{j_{2m+1}} u^{+}{}^{j_{2m}} 
\dots u^{+}{}^{j_{m+3}} 
		}^{m}
		\overbrace{
		u^{-}{}^{j_{m+2}} \dots u^{-}{}^{j_{1})}
		}^{m+2}
		\right.\nonumber\\
&&\left. {}
	- {1\over m+1} \epsilon^{j_{2m+2} j_{1}} 
		\overbrace{ 
		u^{+}{}^{(j_{2m+1}} u^{+}{}^{j_{2m}} \dots u^{+}{}^{j_{m+3}} 
		}^{m-1}
		\overbrace{
		u^{-}{}^{j_{m+2}} \dots u^{-}{}^{j_{2})}
		}^{m+1}
	\right.\nonumber\\
&&\left. {}
	- {(m-1) (m+1) \over 2m (2m+1)}  
		\epsilon^{j_{2m+2} j_{2}} \epsilon^{j_{2m+1} j_{1}} 
		\overbrace{
		u^{+}{}^{(j_{2m}} \dots u^{+}{}^{j_{m+3}}
		}^{m-2} 
		\overbrace{
		u^{-}{}^{j_{m+2}} \dots u^{-}{}^{j_{3})}
		}^{m}
	\right) 
	. 
\label{eq:Reduction2}
\end{eqnarray} 
Note that for $m=1$ the last term vanishes. 

With the use of eq.(\ref{eq:Recursion(1,1)4}), 
the last term in eq.(\ref{eq:DeltaD:N}) can be rewritten with 
$(\lambda_\mu^{(-2)}{}_{(N)})_{(ij)}$ and $(\lambda_\mu^{(-2)}{}_{(N-1)})_{(ij)}$, 
thus the equation becomes 
\begin{eqnarray}
( \delta^{*}_{\lambda_C} D^{--} )_{(N)} 
&=& 
	- { \sqrt{2} \over 6} i 
		C_s \partial^\mu \left( 
			(\lambda_\mu^{(-2)}{}_{(N-1)})_{(ij)} 
			\bar{\phi} 
		\right)
		u^{-i} u^{-j}
		\nonumber\\
&&{}
	- {i \over \sqrt{2}} 
		\epsilon^{kl} C_{(ik)}^{\alpha\beta} 
		(\sigma^\mu \bar{\sigma}^\nu \varepsilon )_{\alpha\beta} 
		\partial_\mu \left( 
			(\lambda_\nu^{(-2)}{}_{(N-1)})_{(jl)} 
			\bar{\phi} 
		\right) 
		u^{-i} u^{-j}
. 
\label{eq:DeltaD:Gen}
\end{eqnarray}
Recall that $\lambda_\mu^{(-2)}{}_{(N)}$ has been already explicitly given 
in eq.(\ref{eq:Lambda(1,1):N:ij}). 
Thus, substituting the previous result, 
a full form of $\delta^{*}_{\lambda_C} D^{--}$ is given by 
\begin{eqnarray}
\delta^{*}_{\lambda_C} D^{--} 
&=& 
2 \sqrt{2} C^{--}{}^{\mu\nu} \partial_\mu \lambda \partial_\nu \bar{\phi} 
	+ {\sqrt{2} \over 6} i C_s h^{--}{}_\mu{}^{\nu} (\bar{\phi}) 
		\partial^\mu (\partial_\nu \lambda \ \bar{\phi} ) 
	\nonumber\\
&&{} 
	+ { i \over \sqrt{2} } \epsilon^{kl} C_{(ik)}^{\alpha\beta} 
		(\sigma^\mu \bar{\sigma}^\nu \varepsilon )_{\alpha\beta} 
		h_{(jl)}{}_\nu{}^\rho (\bar{\phi}) 
	\partial_\mu (\partial_\rho \lambda \ \bar{\phi} ) \ u^{-i} u^{-j} 
		\nonumber\\
&=&  
2 \sqrt{2} C^{--}{}^{\mu\nu} \partial_\mu \lambda \partial_\nu \bar{\phi} 
	- \frac23 C_s C^{--}{}^{\mu\nu} \bar{\phi} 
		\partial_\mu (\partial_\nu \lambda \ \bar{\phi} ) 
	\nonumber\\
&&{} 
	+ 8 \epsilon^{kl} 
		\left( C_{(ik)} \cdot C_{(jl)} \right){}\!^{\alpha\beta} 
		(\sigma^{\mu \nu} \varepsilon )_{\alpha\beta} 
		\bar{\phi} \ \partial_\mu (\partial_\nu \lambda \ \bar{\phi}) 
		\ u^{-i} u^{-j}
	\nonumber\\
&&{} 
	+ {2 \sqrt{2} \over 3} C_s \epsilon^{mn} 
		\left(C_{(im)} \cdot C_{(jn)} \right){}\!^{\alpha\beta} 
		(\sigma^{\mu\nu} \varepsilon )_{\alpha\beta} 
		\bar{\phi}^2 \partial_\mu (\partial_\nu \lambda \ \bar{\phi} ) 
		u^{-i}u^{-j} 
	\nonumber\\
&&{} 
	+ {4 \sqrt{2}} \epsilon^{mn} 
	\left(C_{(ik)} \cdot C_{(jm)} \cdot C_{(ln)} \right){}\!^{\alpha\beta} 
		(\sigma^{\mu\nu} \varepsilon )_{\alpha\beta} 
		\bar{\phi}^2 \partial_\mu (\partial_\nu \lambda \ \bar{\phi} ) 
		u^{-i} \epsilon^{k(l} u^{-j)} 
	\nonumber\\
&&{} 
	+ {\cal O}(C^4)
.  
\label{eq:DeltaD}
\end{eqnarray}

Expanding $\lambda^{(-4)}$ as 
\begin{eqnarray}
\lambda^{(-4)} 
&\equiv& 
	\sum_{N=1}^{\infty} \lambda^{(-4)}{}_{(N)}
, 
	\\
\lambda^{(-4)}{}_{(N)}
&=& 
	\sum_{n=0}^{\infty} 
	( \lambda^{(-4)}{}_{(N)} )_{(i_1 \dots i_{n} k_1 \dots k_{n+4})}
	u^{+}{}^{i_1} \dots u^{+}{}^{i_n} u^{-}{}^{k_1} \dots u^{-}{}^{k_{n+4}}
, 
\end{eqnarray}
where $\lambda^{(-4)}{}_{(N)}$ is an ${\cal O}(C^N)$-quantity, 
eq.(\ref{eq:(2,2)Comp}) also gives the equations to determine 
$\lambda^{(-4)}{}_{(N)}$: 
\begin{equation}
\partial^{++} \lambda^{(-4)}{}_{(N)} 
= 
	- i \partial^\mu \lambda_\mu^{(-2)}{}_{(N)} 
	- \sqrt{2} i C^{+-}{}^{\alpha\beta} 
		(\sigma^\mu \bar{\sigma}^\nu \varepsilon )_{\alpha\beta} 
	\partial_\nu \left(\lambda_\mu^{(-2)}{}_{(N-1)} \bar{\phi} \right) 
	- 3 ( \delta^{*}_{\lambda_C} D^{--} )_{(N)}
. 
\end{equation} 
In principle, we can calculate $\lambda^{(-4)}$ order by order in $C$ 
with the use of the previous results. 
The lower terms are given by 
\begin{equation}
\lambda^{(-4)}
= 
	\frac{1}{3} \, 
	{\rm Tr} \left[ 
               C^{--} \cdot C^{--}  
              \right] 
	\partial_{\mu} (\bar{\phi}^2 \, \partial^{\mu} \lambda) 
	+ {\cal O}(C^3)
. 
\end{equation}
We will not, however, write down the full form here. 
What is important is that it is possible to completely gauge away 
the left hand side of eq.(\ref{eq:(2,2)Comp}) with $\lambda^{(-4)}$, 
after the subtraction of $\delta^{*}_{\lambda_C} D^{--}$ given above. 

To summarize, the gauge transformations at the order ${\cal O}(C^3)$ 
are given by
\begin{eqnarray}
\delta^{*}_{\lambda_C} A_\mu 
&=& 
        - \partial_\mu \lambda 
        - {1\over \sqrt{2}} C_s 
                \partial_\mu \lambda \ \bar{\phi} 
        - {2\over 3} C_2  
                \partial_\mu \lambda \ \bar{\phi}^2
        - {2 \sqrt{2} \over 3} C_3
                \partial_\mu \lambda \ \bar{\phi}^3 
        + {\cal O}(C^4),
\nonumber\\
\delta^{*}_{\lambda_C} \phi 
&=&
        {1\over \sqrt{2}} C_s 
                \partial_\mu \lambda \ A^\mu  
        + {2\over 3} C_2  
                \partial_\mu \lambda \ A^\mu \bar{\phi}
        + {2 \sqrt{2} \over 3} C_3
                \partial_\mu \lambda \ A^\mu \bar{\phi}^2 
        + {\cal O}(C^4),
\nonumber\\
\delta^{*}_{\lambda_C} \psi^{-}_{\alpha} 
&=& 
        - \frac12 C_s (\sigma^\mu \bar{\psi}^{-} )_\alpha \, \partial_{\mu} \lambda 
- \frac23 (\varepsilon C_{(ij)} \sigma^\mu \bar{\psi}^j )_\alpha 
	\, \partial_{\mu} \lambda \, u^{-i} 
        \nonumber\\
&&{} 
        - {5\sqrt{2} \over 3} 
        (\varepsilon C_{(mn)} \ \varepsilon C_{(ij)} \sigma^\mu \bar{\psi}_k )_\alpha 
			\, \bar{\phi} \, \partial_{\mu} \lambda 
                \left( 
                \frac23 \epsilon^{mi} \epsilon^{n(j} u^{-k)}
                + \frac13 \epsilon^{mk} \epsilon^{ni} u^{-j} 
                \right)
     \nonumber\\
&&{}+(\varepsilon C_{(ij)} \ \varepsilon C_{(kl)} \ \varepsilon C_{(mn)} 
      \sigma^{\mu}\bar{\psi}_p)_{\alpha} \, \bar{\phi}^2 \, \partial_{\mu} \lambda 
      \nonumber\\
&&{} \times 
     \left\{ 
         \frac{1}{5} \epsilon^{mm^{\prime}} \epsilon^{nn^{\prime}} 
                     \delta_{m^{\prime}}{}^{(i} 
                     \delta_{n^{\prime}}{}^{j} 
                     \epsilon^{pk} 
                     u^-{}^{l)} 
     \right.                                        \nonumber\\ 
&&{} \left. 
        +\frac{1}{15} \epsilon^{mm^{\prime}} \epsilon^{nn^{\prime}} 
                      \left( 
                        \delta_{m^{\prime}}{}^{(i} 
                        \delta_{n^{\prime}}{}^{j)} 
                        \epsilon^{p(k} u^-{}^{l)} 
                       +\delta_{m^{\prime}}{}^{(k} 
                        \delta_{n^{\prime}}{}^{l)} 
                        \epsilon^{p(i} u^-{}^{j)} 
                      \right) 
       \right.                                    \nonumber\\ 
&&{}   \left. 
        +2 \epsilon^{mm^{\prime}} \epsilon^{pp^{\prime}} 
         \epsilon^{ki} \delta_{m^{\prime}}{}^{(l} 
         \delta_{p^{\prime}}{}^{n} u^-{}^{j)} 
        +\frac{32}{9} \epsilon^{kk^{\prime}} 
         \epsilon^{mm^{\prime}} \epsilon^{nn^{\prime}} 
         \delta_{k^{\prime}}{}^{(p} \delta_{m^{\prime}}{}^{i} 
         \delta_{n^{\prime}}{}^{j} u^-{}^{l)} 
       \right\}
        \nonumber\\
&&{}+ {\cal O}(C^4) ,
\nonumber\\
\delta^{*}_{\lambda_C} D^{--}
&=&
2 \sqrt{2} C^{--}{}^{\mu\nu} \partial_\mu \lambda \partial_\nu \bar{\phi} 
	- \frac23 C_s C^{--}{}^{\mu\nu} \bar{\phi} 
		\partial_\mu (\partial_\nu \lambda \ \bar{\phi} ) 
	\nonumber\\
&&{} 
	+ 8 \epsilon^{kl} 
		\left( C_{(ik)} \cdot C_{(jl)} \right){}\!^{\alpha\beta} 
		(\sigma^{\mu \nu} \varepsilon )_{\alpha\beta} 
		\bar{\phi} \ \partial_\mu (\partial_\nu \lambda \ \bar{\phi}) 
		\ u^{-i} u^{-j}
	\nonumber\\
&&{} 
	+ {2 \sqrt{2} \over 3} C_s \epsilon^{mn} 
		\left(C_{(im)} \cdot C_{(jn)} \right){}\!^{\alpha\beta} 
		(\sigma^{\mu\nu} \varepsilon )_{\alpha\beta} 
		\bar{\phi}^2 \partial_\mu (\partial_\nu \lambda \ \bar{\phi} ) 
		u^{-i}u^{-j} 
	\nonumber\\
&&{} 
	+ {4 \sqrt{2}} \epsilon^{mn} 
	\left(C_{(ik)} \cdot C_{(jm)} \cdot C_{(ln)} \right){}\!^{\alpha\beta} 
		(\sigma^{\mu\nu} \varepsilon )_{\alpha\beta} 
		\bar{\phi}^2 \partial_\mu (\partial_\nu \lambda \ \bar{\phi} ) 
		u^{-i} \epsilon^{k(l} u^{-j)} 
	\nonumber\\
&&{} 
	+ {\cal O}(C^4)
	, 
	\nonumber\\
\delta_{\lambda_C}^{*} (\mbox{others}) 
&=& 
	0
	. 
\end{eqnarray}
One can confirm that the terms at the order ${\cal O}(C^1)$ 
in the action $S_{*,2}+S_{*,3}$ are invariant under
these transformations. 

\subsection{Field redefinitions and the action} 
We can redefine (reparameterize)  
the component fields in the gauge superfield $V^{++}_{WZ}$ 
in order to make the gauge transformation become a convenient form. 
There are possibly many choices to achieve this. 
Here we will absorb all of the $C$-dependent terms 
emerging in the deformed gauge transformation 
by the reparameterization procedure 
so as to make the resulted component gauge transformation laws 
canonical 
(i.e. the ordinary Abelian gauge transformations). 
Presumably we could reparameterize the coefficients 
so as not to fully extinguish 
the $C$-dependent terms from the resulted component gauge transformation 
laws.  
Then the procedure to absorb terms emerging above into the definition of the 
coefficient functions in the gauge superfield could be a very difficult task, 
because we have to find a parameterization which gives 
the deformed gauge transformation of the gauge superfield 
under the resulted partially deformed gauge transformation laws.

In order to make the component gauge transformation canonical, 
the following field redefinitions are sufficient
: 
\begin{eqnarray}
A_\mu 
&\longrightarrow& 
	\left( 1 + f(\bar{\phi}) \bar{\phi} \right) A_\mu 
	\nonumber\\
&&= 
	A_\mu 
	+ {1\over \sqrt{2}} C_s 
		A_\mu \bar{\phi} 
	+ {2\over 3} C_2  
		A_\mu \bar{\phi}^2
	+ {\cal O}(C^3) 
%	+ {2 \sqrt{2} \over 3} C_3
%		A_\mu \bar{\phi}^3 
%	+ {\cal O}(C^4)
	,\\ 
\phi 
&\longrightarrow& 
	\phi 
	- \frac12 f(\bar{\phi}) \left( 1 + f(\bar{\phi}) \bar{\phi} \right) 
		A_\mu A^\mu 
	\nonumber\\
&&= 
	\phi 
	- {1\over 2 \sqrt{2}} C_s 
		A_\mu A^\mu  
	- \Big( 
		{1\over 4} C_s^2 
		+ {1\over 3} C_2  
		\Big)
		A_\mu A^\mu \bar{\phi}
	+ {\cal O}(C^3)
%	\nonumber\\
%&&{} \quad 
%	- \Big( 
%		{\sqrt{2} \over 3} C_s C_2 
%		+ {\sqrt{2} \over 3} C_3
%		\Big) 
%		A_\mu A^\mu \bar{\phi}^2 
%	+ {\cal O}(C^4)
	, \\ 
\psi^{-}_{\alpha} 
&\longrightarrow& 
	\psi^{-}_{\alpha} 
	- \frac12 C_s (\sigma^\mu \bar{\psi}^{-} )_\alpha A_{\mu} 
- \frac23 (\varepsilon C_{(ij)} \sigma^\mu \bar{\psi}^j )_\alpha A_{\mu} u^{-i} 
	\nonumber\\
&&{} 
	- {5\sqrt{2} \over 3} 
	(\varepsilon C_{(mn)} \cdot C_{(ij)} \sigma^\mu \bar{\psi}_k )_\alpha 
	\bar{\phi} A_{\mu} 
		\left( 
		\frac23 \epsilon^{mi} \epsilon^{n(j} u^{-k)}
		+ \frac13 \epsilon^{mk} \epsilon^{ni} u^{-j} 
		\right)
	 \nonumber\\
&&{}+ {\cal O}(C^3) 
	, \\
D^{--} 
&\longrightarrow&  
	D^{--} 
	- 2 \sqrt{2} C^{--}{}^{\mu\nu} A_\mu \partial_\nu \bar{\phi} 
	- {\sqrt{2} \over 6} i C_s h^{--}{}_\mu{}^{\nu} (\bar{\phi}) 
		\partial^\mu (A_\nu  \bar{\phi} ) 
	\nonumber\\
&&{} 
	- { i \over \sqrt{2} } \epsilon^{kl} C_{(ik)}^{\alpha\beta} 
		(\sigma^\mu \bar{\sigma}^\nu \varepsilon)_{\alpha\beta} 
		h_{(jl)}{}_\nu{}^\rho (\bar{\phi}) 
		\partial_\mu (A_\rho  \bar{\phi} ) \ u^{-i} u^{-j} 
	\nonumber\\
&&=  
	D^{--} 
	- 2 \sqrt{2} C^{--}{}^{\mu\nu} A_\mu \partial_\nu \bar{\phi} 
+ \frac23 C_s C^{--}{}^{\mu\nu} \bar{\phi} \partial_\mu (A_\nu \bar{\phi} ) 
	\nonumber\\
&&{} 
	-8 \epsilon^{kl} 
		\left( C_{(ik)} \cdot C_{(jl)} \right){}\!^{\alpha\beta} 
		(\sigma^{\mu\nu} \varepsilon)_{\alpha\beta} 
		\bar{\phi} \ \partial_\mu (A_\nu \bar{\phi}) \ u^{-i} u^{-j}
	+ {\cal O}(C^3)
	, 
\end{eqnarray}
where $f(\bar{\phi})$ and $h^{--}{}_\mu{}^\nu (\bar{\phi})$ are defined by 
eq.(\ref{eq:f}) and (\ref{eq:h--}) respectively. 
There is no need to redefine the rest of the fields. 
We will denote the reparameterized gauge superfield as 
$V_C^{++}$. 
After this reparameterization, it becomes to hold that 
\begin{equation}
\delta^{*}_{\tilde{\lambda}_C} V_C^{++}
= 
	\delta^{0}_{\lambda} V_C^{++}
, 
\label{eq:Requirement}
\end{equation}
where $\tilde{\lambda}_C$ is given by replacing $A_\mu$ in $\lambda_C$ with 
$\left( 1 + f(\bar{\phi}) \bar{\phi} \right) A_\mu$ 
%\begin{eqnarray}
%\tilde{\lambda}_C (x_A, \theta^{+}, \bar{\theta}^{+}, u)
%&=& 
%	, 
%\end{eqnarray}
and $\delta^{0}_{\lambda}$ denotes the ordinary component gauge transformations 
with gauge parameter $\lambda(x_A)$: 
\begin{equation}
\delta^{0}_{\lambda} A_\mu 
\equiv 
	- \partial_\mu \lambda 
, 
\qquad 
\delta^{0}_{\lambda}(\mbox{others}) 
\equiv  
	0 
.
\end{equation}
{}From eq.(\ref{eq:Requirement}),  we are allowed to consider 
that the gauge transformations of the component fields satisfy 
\begin{equation}
\delta^{*}_{\tilde{\lambda}_C} A_\mu 
= \delta^{0}_{\lambda} A_\mu
= - \partial_\mu \lambda 
, 
\qquad 
\delta^{*}_{\tilde{\lambda}_C}(\mbox{others}) 
= \delta^{0}_{\lambda}(\mbox{others}) 
= 0 
. 
\end{equation}
This means that the component gauge transformations can be thought of as 
the ordinary one. 

After the field redefinition, 
the action in the WZ gauge will become to have manifestly gauge invariant form 
(in the sense of the ordinary component Abelian gauge transformations), 
$S_{*C}$. 
Let us denote the contribution to this action from the $n$-th order in 
$V_C^{++}$ 
as $S_{*C,n}$. 
It is seen that 
$S_{*,2}$ given in (\ref{eq:S_2}) changes to 
\begin{eqnarray}
S_{*C,2}
&=&
	\int d^4x \left[
	- {1\over 4} F_{\mu\nu} ( F^{\mu\nu} + \tilde{F}^{\mu\nu} ) 
	- {1\over \sqrt{2}} C_s \partial_\mu ( A_\nu \bar{\phi} ) 
		( F^{\mu\nu} + \tilde{F}^{\mu\nu}) 
		\right. \nonumber\\
&&{}
	-i \psi^i \sigma^\mu \partial_\mu \bar{\psi}_i
	+{i \over 2} C_s \varepsilon^{\alpha\beta} A_\mu 
		(\sigma^\mu \bar{\psi}^k )_\alpha 
		(\sigma^\nu \partial_\nu \bar{\psi}_k )_\beta 
	-{2\over 3}i C_{(ij)}^{\alpha\beta} A_\mu 
		(\sigma^\mu \bar{\psi}^i )_\alpha 
		(\sigma^\nu \partial_\nu \bar{\psi}^j )_\beta 
	\nonumber\\
&&{}
	+ \phi \partial^2 \bar{\phi}
	- {\sqrt{2} \over 4} C_s A_\mu A^\mu \partial^2 \bar{\phi}
	\nonumber\\
&&\left.{}
	+ \frac14 D_{ij} D^{ij}
	- {\sqrt{2}} C_{(ij)}^{\mu\nu} D^{ij} A_\mu \partial_\nu \bar{\phi} 
	+ {\cal O}(C^2)
	\right] 
	. 
\end{eqnarray}
Thus we find 
\begin{eqnarray}
S_{*C,2} + S_{*C,3} 
&=& 
	\int d^4x \left[
	- {1\over 4} ( 1 + \sqrt{2} C_s \bar{\phi} ) 
		F_{\mu\nu} ( F^{\mu\nu} + \tilde{F}^{\mu\nu} ) 
		\right. \nonumber\\
&&{}
	-i \Bigl( 1- {1\over \sqrt{2}} C_s \bar{\phi} \Bigr) 
		\psi^i \sigma^\mu \partial_\mu \bar{\psi}_i
	+ { i \over \sqrt{2}} C_s \partial_\mu \bar{\phi} 
		(\psi^i \sigma^\mu \bar{\psi}_i)
	\nonumber\\ 
&&{}
	- 2\sqrt{2} i C_{(ij)}^{\alpha\beta}
		\psi^i_\alpha (\sigma^\mu \bar{\psi}^j)_\beta 
		\partial_\mu \bar{\phi}
	- { 2 \sqrt{2} \over 3} i C_{(ij)}^{\alpha\beta}
		\psi^i_\alpha ( \sigma^\mu \partial_\mu \bar{\psi}^j)_\beta 
		\bar{\phi}
	\nonumber\\
&&{}
	+ {i \over 2} C_s \bar{\psi}^i \bar{\psi}^j  D_{ij}
	- i C_{(ij)}^{\mu\nu} \bar{\psi}^i \bar{\psi}^j F_{\mu\nu}
	\nonumber\\ 
&&{}
	+ \phi \partial^2 \bar{\phi}
	\nonumber\\
&&\left.{}
	+ \frac14 ( 1 - \sqrt{2} C_s \bar{\phi} ) D_{ij} D^{ij}
	+ {1\over \sqrt{2}} C_{(ij)}^{\mu\nu} D^{ij} F_{\mu\nu} \bar{\phi} 
	+ {\cal O}(C^2)
	\right] 
	. 
\end{eqnarray}
All the terms emerging here are manifestly gauge invariant 
as expected from the field redefinition procedure. 
We note that 
the kinetic terms of the fields show similar $\bar{\phi}$ dependence as 
in \cite{FeSo}.
We expect that 
the full action can be  written in terms of a function of
$\bar{\phi}$,  which is left for future study.

\section{Conclusions and Discussion}
In this paper, we have studied ${\cal N}=2$ supersymmetric $U(1)$ gauge theory
on the noncommutative harmonic superspace.
We investigated the deformed gauge transformation
which preserves the WZ gauge in detail.
We also calculated the deformed Lagrangian explicitly up to the third
order in component fields.
The general gauge transformation includes the arbitrary power of 
anti-holomorphic scalar fields, which does not appear in noncommutative
${\cal N}=1$  superspace.
Generalization to ${\cal N}=2$ $U(N)$ gauge theory is straightforward.
Since the deformation is rather different from the ${\cal N}=1$ theory,
it would be an  interesting problem to study the 
perturbative and nonperturbative structure of 
deformed ${\cal N}=2$ theory.

In this paper we have studied the gauge symmetry but not supersymmetry.
In order to study supersymmetry of the theory, one must act the
supercharges $Q^{i}_{\alpha}$ on the gauge superfield $V^{++}_{WZ}$.
This transformation does not preserve the WZ gauge. 
Therefore we need to make further gauge transformation to recover the
WZ gauge.
Because this gauge transformation is  very complicated, we left this
problem in a subsequent paper. 

Another interesting problem is to study the deformed ${\cal N}=2$ theory from
the $D$-brane viewpoint. 
In order to study this problem , it would be natural to investigate 
the hybrid formalism compactified on the $K3$ surface or
four-dimensional torus.
The theory on the $D$-brane
provides the ${\cal N}=2$ supersymmetric Yang-Mills theory in four dimensions.
Since the six-dimensional hybrid superstring includes the harmonic superspace
structure\cite{BeVa}, it is also interesting to clarify the relation
between both harmonic superspace structures.

{\bf Acknowledgements}\\
The works of T.A and A.O were supported in part 
by a 21st Century COE Program at
TokyoTech "Nanometer-Scale Quantum Physics" by the
Ministry of Education, Culture, Sports, Science and Technology.

\section*{Appendix A: Star Product for Analytic Superfields}
\renewcommand{\theequation}{A.\arabic{equation}}
\setcounter{equation}{0}
The $*$-product $A*B$ for analytic superfields $A$ and $B$ is given by
\begin{eqnarray}
A*B&=& AB-(-1)^{|A|}
{1\over2}Q^{i}_{\alpha}A C^{\alpha\beta}_{ij} Q^{j}_{\beta}B
-{1\over8} (Q^{i_2}_{\alpha_2} Q^{i_1}_{\alpha_1} A)
C^{\alpha_1\beta_1}_{i_1j_1} C^{\alpha_2\beta_2}_{i_2j_2}
(Q^{j_2}_{\beta_2} Q^{j_1}_{\beta_1}) B\nonumber\\
&& 
+{1\over48}(-1)^{|A|}
(Q^{i_3}_{\alpha_3}  Q^{i_2}_{\alpha_2} Q^{i_1}_{\alpha_1}  A)
C^{\alpha_1\beta_1}_{i_1j_1}
C^{\alpha_2\beta_2}_{i_2 j_2}
C^{\alpha_3\beta_3}_{i_3 j_3}
(Q^{j_3}_{\beta_3} Q^{j_2}_{\beta_2} Q^{j_1}_{\beta_1}B)  \nonumber\\
&&
+{1\over384}
(Q^{i_4}_{\alpha_4} Q^{i_3}_{\alpha_3} Q^{i_2}_{\alpha_2} Q^{i_1}_{\alpha_1} A)
C^{\alpha_1\beta_1}_{i_1j_1}
C^{\alpha_2\beta_2}_{i_2 j_2}
C^{\alpha_3\beta_3}_{i_3 j_3}
C^{\alpha_4\beta_4}_{i_4 j_4}
(Q^{j_4}_{\beta_4}Q^{j_3}_{\beta_3}Q^{j_2}_{\beta_2} Q^{j_1}_{\beta_1}B) .
\end{eqnarray}
Here for an analytic superfield
$\Phi$ of the form 
\begin{eqnarray}
\Phi&=& \phi(x_A,u)+\theta^{+}\psi(x_A,u)+
\bar{\theta}^{+}\bar{\chi}(x_A,u)
+(\theta^{+})^2 M(x_A,u)
+(\bar{\theta}^{+})^2 N(x_A,u)\nonumber\\
&& +\theta^{+}\sigma^{m}\bar{\theta}^{+}A_{m}(x_A,u)
+(\bar{\theta}^{+})^2\theta^{+}\lambda(x_A,u)
+(\theta^{+})^2\bar{\theta}^{+}\bar{\kappa}(x_A,u)
+(\theta^{+})^2(\bar{\theta}^{+})^2 D(x_A,u),\nonumber\\
\label{eq:analytic1a}
\end{eqnarray}
$Q^{i}_{\alpha}\Phi$ etc. are given by
\begin{eqnarray}
 Q^{i}_{\alpha}\Phi
&=&
-u^{+i}\psi_{\alpha}
-2u^{+i}\theta^{+}_{\alpha}M
+\bar{\theta}^{+\dot{\alpha}}\left\{-u^{+i}
\sigma^{\mu}_{\alpha\dot{\alpha}}A_{\mu}
+2i u^{-i}\sigma^{\mu}_{\alpha\dot{\alpha}}\partial_{\mu}\phi\right\}
\nonumber\\
&&
\!\!\!\!
-(\bar{\theta}^{+})^2\left\{u^{+i}\lambda_{\alpha}
+iu^{-i}\sigma^{\mu}_{\alpha\dot{\alpha}}
\partial_{\mu}\bar{\chi}^{\dot{\alpha}}
\right\}
+(\theta^{+}\sigma^{\mu}\bar{\theta}^{+})
\left\{
u^{+i} (\sigma^{\mu}\bar{\kappa})_{\alpha}
-iu^{-i}
(\partial_{\nu}\psi\sigma^{\mu}\bar{\sigma}^{\nu})^{\gamma}
\varepsilon_{\alpha\gamma}
\right\}
\nonumber\\
&&
+(\bar{\theta}^{+})^2\theta^{+\beta}
\left\{-u^{+i}2\varepsilon_{\alpha\beta}D
-i u^{-i}(\sigma^{\nu}\bar{\sigma}^{\mu})_{\beta}{}^{\gamma}
\varepsilon_{\alpha\gamma}\partial_{\mu}A_{\nu}
\right\}
+2i u^{-i} (\theta^+)^2 \sigma^{\mu}_{\alpha\dot{\alpha}}
\bar{\theta}^{+\dot{\alpha}}\partial_{\mu}M
\nonumber\\
&&-u^{-i}i
(\theta^+)^2(\bar{\theta}^{+})^2 \sigma^{\mu}_{\alpha\dot{\alpha}}
\partial_{\mu}\bar{\kappa}^{\dot{\alpha}},
\label{eq:susy1a}
\end{eqnarray}
\begin{eqnarray}
Q^{i_2}_{\alpha_2}Q^{i_1}_{\alpha_1}\Phi  &=&
2 u^{+i_1}u^{+i_2} \varepsilon_{\alpha_1\alpha_2} M\nonumber\\
&& -\bar{\theta}^{+\dot{\alpha}}
\left\{
2u^{+i_1}u^{+i_2}\varepsilon_{\alpha_1\alpha_2} \bar{\kappa}_{\dot{\alpha}}
-2i u^{-i_1}u^{+i_2}\partial_{\nu}\psi_{\alpha_2}
\sigma^{\nu}_{\alpha_1\dot{\alpha}}
-2i u^{+i_1}u^{-i_2}\sigma^{\mu}_{\alpha_2\dot{\alpha}}
\partial_{\mu}\psi_{\alpha_1}
\right\}
\nonumber\\
&&
+(\bar{\theta}^{+})^2
\Bigl\{
2 u^{+i_1} u^{+i_2}\varepsilon_{\alpha_1 \alpha_2}D
+i u^{-i_1}u^{+i_2} 
(\sigma^{\nu}\bar{\sigma}^{\mu})_{\alpha_2}{}^{\gamma}
\varepsilon_{\alpha_1 \gamma}\partial_{\mu}A_{\nu}\nonumber\\
&&+i u^{+i_1}u^{-i_2} \varepsilon_{\alpha_1\gamma}
(\sigma^{\mu}\bar{\sigma}^{\nu})_{\alpha_2} {}^{\gamma}\partial_{\mu}A_{\nu}
+2 u^{-i_1}u^{-i_2}
\varepsilon_{\alpha_1\gamma}
(\sigma^{\mu}\bar{\sigma}^{\nu})_{\alpha_2} {}^{\gamma}
\partial_{\mu}\partial_{\nu}\phi
\Bigr\}\nonumber\\
&&+(\theta^{+}\sigma^{\mu}\bar{\theta}^{+})
\Bigl\{
2i u^{-i_1}u^{+i_2}
\varepsilon_{\alpha_1\alpha} (\sigma^{\mu}\bar{\sigma}^{\nu})_{\alpha_2}
{}^{\alpha}\partial_{\nu}M
-2i u^{+i_1}u^{-i_2}
\varepsilon_{\alpha_2\alpha} (\sigma^{\mu}\bar{\sigma}^{\nu})_{\alpha_1}
{}^{\alpha}\partial_{\nu}M
\Bigr\}\nonumber\\
&&\hspace{-2cm}
+(\bar{\theta}^{+})^2 \theta^{+\beta}
\Bigl\{
2i u^{-i_1}u^{+i_2} \varepsilon_{\alpha_2\beta}
(\sigma^{\mu}\partial_{\mu}\bar{\kappa})_{\alpha_1}
-2 i u^{+i_1} u^{-i_2} \varepsilon_{\alpha_1\beta}
(\sigma^{\mu}\partial_{\mu}\bar{\kappa})_{\alpha_2}
+2 u^{-i_1}u^{-i_2} \varepsilon_{\alpha_1\alpha_2}\partial^{\mu}
\partial_{\mu}\psi_{\beta}
\Bigr\}\nonumber\\
&&
+(\theta^{+})^2 (\bar{\theta}^{+})^2
\Bigl\{
-2 u^{-i_1}u^{-i_2}\varepsilon_{\alpha_1\alpha_2}\partial^{\mu}\partial_{\mu}M
\Bigr\},
\end{eqnarray}
\begin{eqnarray}
Q^{i_3}_{\alpha_3} Q^{i_2}_{\alpha_2}Q^{i_1}_{\alpha_1}\Phi
&=&
-\bar{\theta}^{+\dot{\alpha}}
\Bigl\{
4i u^{-i_1}u^{+i_2}u^{+i_3}\varepsilon_{\alpha_3\alpha_2}
\sigma^{\mu}_{\alpha_1\dot{\alpha}}\partial_{\mu}M
-4i u^{+i_1}u^{-i_2}u^{+i_3}
\varepsilon_{\alpha_3\alpha_1}\sigma^{\mu}_{\alpha_2\dot{\alpha}}
\partial_{\mu}M\nonumber\\
&&+4i u^{+i_1}u^{+i_2}u^{-i_3}
\varepsilon_{\alpha_1\alpha_2}\sigma^{\mu}_{\alpha_3\dot{\alpha}}
\partial_{\mu}M
\Bigr\}\nonumber\\
&&
+(\bar{\theta}^{+})^2
\Bigl\{
-2i u^{-i_1}u^{+i_2} u^{+i_3}\varepsilon_{\alpha_2\alpha_3}
(\sigma^{\mu}\partial_{\mu}\bar{\kappa})_{\alpha_1}
+2i u^{+i_1} u^{-i_2} u^{+i_3} \varepsilon_{\alpha_1\alpha_3}
(\sigma^{\mu}\partial_{\mu}\bar{\kappa})_{\alpha_2}\nonumber\\
&& -2i u^{+i_1} u^{+i_2} u^{-i_3} \varepsilon_{\alpha_1\alpha_2}
(\sigma^{\mu}\partial_{\mu}\bar{\kappa})_{\alpha_3}
-2 u^{-i_1}u^{-i_2}u^{+i_3}\varepsilon_{\alpha_1\alpha_2}
\partial^{\mu}\partial_{\mu}\psi_{\alpha_3}\nonumber\\
&&
+2 u^{-i_1}u^{+i_2}u^{-i_3}\varepsilon_{\alpha_1\alpha_3}
\partial^{\mu}\partial_{\mu}\psi_{\alpha_2}
+2 u^{+i_1}u^{-i_2}u^{-i_3}\varepsilon_{\alpha_2\alpha_3}
\partial^{\mu}\partial_{\mu}\psi_{\alpha_2}
\Bigr\}
\nonumber\\
&&
+(\bar{\theta}^{+})^2 \theta^{+\beta}
\Bigl\{
4 u^{-i_1}u^{-i_2}u^{+i_3}
\varepsilon_{\alpha_3\beta}\varepsilon_{\alpha_1\alpha_2}
\partial^{\mu}\partial_{\mu}M
-4 u^{-i_1}u^{+i_2}u^{-i_3}\varepsilon_{\beta\alpha_2}
\varepsilon_{\alpha_1\alpha_3}\partial^{\mu}\partial_{\mu}M\nonumber\\
&&+4 u^{+i_1}u^{-i_2}u^{-i_3}\varepsilon_{\beta\alpha_1}
\varepsilon_{\alpha_2\alpha_3}\partial^{\mu}\partial_{\mu}M
\Bigr\},
\end{eqnarray}
\begin{eqnarray}
Q^{i_4}_{\alpha_4} Q^{i_3}_{\alpha_3}
Q^{i_2}_{\alpha_2} Q^{i_1}_{\alpha_1}\Phi  
&=& (\bar{\theta}^{+})^2
\Bigl\{
-4 u^{-i_1}u^{-i_2}u^{+i_3} u^{+i_4} 
\varepsilon_{\alpha_3\alpha_4}\varepsilon_{\alpha_1\alpha_2}
+4 u^{-i_1}u^{+i_2}u^{-i_3}u^{+i_4}
\varepsilon_{\alpha_4\alpha_2}\varepsilon_{\alpha_1\alpha_3}\nonumber\\
&&-4 u^{+i_1}u^{-i_2}u^{-i_3}u^{+i_4}
\varepsilon_{\alpha_4\alpha_1} \varepsilon_{\alpha_2\alpha_3}
\nonumber\\
&& 
-4 u^{-i_1}u^{+i_2}u^{+i_3} u^{-i_4}
\varepsilon_{\alpha_3\alpha_2} \varepsilon_{\alpha_1\alpha_4}
+4 u^{+i_1}u^{-i_2}u^{+i_3} u^{-i_4}
\varepsilon_{\alpha_3\alpha_1}  \varepsilon_{\alpha_2\alpha_4}\nonumber\\
&& -4 u^{+i_1}u^{+i_2}u^{-i_3} u^{-i_4}
\varepsilon_{\alpha_1\alpha_2} \varepsilon_{\alpha_3\alpha_4}
\Bigr\}\partial^{\mu}\partial_{\mu}M.
\end{eqnarray}

\section*{Appendix B: Calculation of Third Order Lagrangian}
\renewcommand{\theequation}{B.\arabic{equation}}
\setcounter{equation}{0}
In this appendix we calculate the action which is third order of $V^{++}_{WZ}$:
\begin{equation}
 S_{*,3}
=
{i \over 6} 
\int d^4xd^8\theta du_1 du_2 du_3 {
	V^{++}_{WZ}(1) * V^{++}_{WZ}(2) * V^{++}_{WZ}(3)
	\over (u^{+}_1 u^{+}_2)(u^{+}_2 u^{+}_3)(u^{+}_3 u^{+}_1)}.
\end{equation}
We use the chiral basis to compute this action.
In the chiral basis, the gauge superfield $V^{++}_{WZ}$ 
is expressed as 
\begin{eqnarray}
V^{++}_{WZ} (x^{\mu}_L,u) 
&=& -2i\theta^+ \sigma^{\mu} \bar{\theta}^+ A_{\mu} 
    -2(\bar{\theta}^+)^2 
     \theta^+ \sigma^{\mu}\bar{\sigma}^{\nu} \theta^- 
     \partial_{\nu} A_{\mu} 
    +i\sqrt{2}(\bar{\theta}^+)^2 \phi                    \nonumber\\ 
&&{}-i\sqrt{2}(\theta^+)^2 \bar{\phi} 
    -2\sqrt{2} (\theta^+)^2 \theta^- \sigma^{\mu} \bar{\theta}^+ 
     \partial_{\mu} \bar{\phi} 
    -\sqrt{2}i(\theta^+)^2(\theta^-)^2(\bar{\theta}^+)^2 
     \partial^2 \bar{\phi}                               \nonumber\\ 
&&  +4(\bar{\theta}^+)^2\theta^+ \psi^- 
    -4(\theta^+)^2\bar{\theta}^+ \bar{\psi}^- 
    -4i(\theta^+)^2(\bar{\theta}^+)^2 
     \theta^- \sigma^{\mu} \partial_{\mu} \bar{\psi}^-   \nonumber\\ 
&&{}+3(\theta^+)^2(\bar{\theta}^+)^2 D^{--} . 
\end{eqnarray}
We investigate what kind of contributions from 
$(V^{++}_{WZ})^3_{*}$ exist in the Lagrangian. 
As seen from the arguments in sect. 3, we need to compute
the contributions from $P$ and $P^3$ in the $*$-product.

Firstly we study the contributions from single $P$.
The $*$-product formulas for first order of $P$ is summarized as
follows:
\begin{eqnarray}
\theta^+_1{}^{\alpha} \ P \ \theta^+_2{}^{\beta} 
&=& +\frac{1}{2} C^{++}_{12}{}^{\alpha\beta}, 
\label{eq:star-1}                                                 \\ 
%%%%%%%%%%% 
\theta^+_1{}^{\beta}\theta^-_1{}_{\gamma} \ P \ 
\theta^+_2{}^{\alpha} 
&=& -\frac{1}{2} C_{12}^{++}{}^{\beta\alpha} \theta^-_1{}_{\gamma} 
    +\frac{1}{2} C_{12}^{-+}{}^{\alpha^{\prime}\alpha} 
     \varepsilon_{\gamma\alpha^{\prime}} \theta^+_1{}^{\beta}, 
\label{eq:star-2}                                                 \\ 
%%%%%%%%%%% 
\theta^+_1{}^{\alpha} \ P \ (\theta^+_2)^2 
&=& +C_{12}^{++}{}^{\alpha\alpha^{\prime}} 
     \theta^+_2{}_{\alpha^{\prime}}, 
\label{eq:star-3}                                                 \\ 
%%%%%%%%%%% 
(\theta^+_1)^2 \ P \ \theta^+_2{}^{\alpha} 
&=& -C_{12}^{++}{}^{\alpha^{\prime}\alpha} 
     \theta^+_1{}_{\alpha^{\prime}}, 
\label{eq:star-4}                                                 \\ 
%%%%%%%%%%% 
\theta^+_1{}^{\alpha}\theta^-_1{}_{\beta} \ P \ 
 \theta^+_2{}^{\gamma} \theta_2^-{}_{\delta} 
&=& -\frac{1}{2} C_{12}^{++}{}^{\alpha\gamma} 
     \theta^-_1{}_{\beta} \theta^-_2{}_{\delta} 
    +\frac{1}{2} C_{12}^{+-}{}^{\alpha\alpha^{\prime}} 
     \varepsilon_{\delta\alpha^{\prime}} 
     \theta^-_1{}_{\beta} \theta^+_2{}^{\gamma}          \nonumber\\ 
&&{}+\frac{1}{2} C_{12}^{-+}{}^{\alpha^{\prime}\gamma} 
     \varepsilon_{\beta\alpha^{\prime}} 
     \theta^+_1{}^{\alpha} \theta^-_2{}_{\delta} 
    -\frac{1}{2} C_{12}^{--}{}^{\alpha^{\prime}\beta^{\prime}} 
     \varepsilon_{\beta\alpha^{\prime}} \varepsilon_{\delta\beta^{\prime}} 
     \theta^+_1{}^{\alpha} \theta^+_2{}^{\gamma}, 
\label{eq:star-5}                                                 \\ 
%%%%%%%%%%% 
\theta^+_1{}^{\alpha}\theta^-_1{}_{\beta} \ P \ (\theta^+_2)^2 
&=& -C_{12}^{++}{}^{\alpha\alpha^{\prime}} 
     \theta^-_1{}_{\beta} \theta^+_2{}_{\alpha^{\prime}} 
    +C_{12}^{-+}{}^{\alpha^{\prime}\beta^{\prime}} 
     \varepsilon_{\beta\alpha^{\prime}} 
     \theta^+_1{}^{\alpha} \theta^+_2{}_{\beta^{\prime}}, 
\label{eq:star-6}                                                 \\ 
%%%%%%%%%%%%%%%%%% 
(\theta^+_1)^2 \ P \ (\theta^+_2)^2 
&=& -2 C_{12}^{++}{}^{\alpha^{\prime}\beta^{\prime}} 
     \theta^+_1{}_{\alpha^{\prime}} \theta^+_2{}_{\beta^{\prime}}, 
\label{eq:star-7}                                                 \\ 
%%%%%%%%%%%% 
(\theta^+_1)^2 (\theta^-_1)^2 \ P \ (\theta^+_2)^2 
&=& -2 C_{12}^{++}{}^{\alpha^{\prime}\beta^{\prime}} 
     \theta^+_1{}_{\alpha^{\prime}} (\theta^-_1)^2 
     \theta^+_2{}_{\beta^{\prime}} 
    -2C_{12}^{-+}{}^{\alpha^{\prime}\beta^{\prime}} 
     (\theta^+_1)^2 \theta_1^-{}_{\alpha^{\prime}} 
     \theta^+_2{}_{\beta^{\prime}}, 
\label{eq:star-8}                                                \\ 
%%%%%%%%%%%% 
(\theta^+_1)^2 \theta^-_1{}^{\alpha} \ P \ 
 (\theta^+_2)^2 \theta^-_2{}^{\beta} 
&=& +2C_{12}^{++}{}^{\alpha^{\prime}\beta^{\prime}} 
     \theta^+_1{}_{\alpha^{\prime}} \theta^-_1{}^{\alpha} 
     \theta^+_2{}_{\beta^{\prime}} \theta^-_2{}^{\beta} 
    +C_{12}^{+-}{}^{\alpha^{\prime}\beta} 
     \theta^+_1{}_{\alpha^{\prime}} \theta^-_1{}^{\alpha} 
     (\theta^+_2)^2                                      \nonumber\\ 
&&{}+C_{12}^{-+}{}^{\alpha\alpha^{\prime}} 
     (\theta^+_1)^2 
     \theta^+_2{}_{\alpha^{\prime}} \theta^-_2{}^{\beta} 
    +\frac{1}{2} C_{12}^{--}{}^{\alpha\beta} 
     (\theta^+_1)^2 (\theta^+_2)^2. 
\label{eq:star-9} 
\end{eqnarray}
The following identities are also useful:
\begin{eqnarray}
(\theta_1^+)^2 (\theta_2^{\pm})^2 
&=& (u^+_1 u^{\pm}_2)^2 \theta^4, 
\label{eq:contruct-1}                                             \\ 
(\theta_1^+)^2 \theta_2^{\pm}{}^{\alpha} \theta_3^+{}^{\beta} 
&=& +\frac{1}{2} \varepsilon^{\alpha\beta} \theta^4 
    (u^+_1 u^{\pm}_2) (u^+_3 u^+_1), 
\label{eq:contruct-2}                                             \\ 
(\bar{\theta}_1^+)^2 \bar{\theta}_2^{\pm}{}^{\dot{\alpha}} 
\bar{\theta}_3^+{}^{\dot{\beta}} 
&=& -\frac{1}{2} \varepsilon^{\dot{\alpha}\dot{\beta}} \bar{\theta}^4 
    (u^+_1 u^{\pm}_2) (u^+_3 u^+_1). 
\label{eq:contruct-3}
\end{eqnarray}
We also use some the harmonic integral formulas, summarized as
\begin{eqnarray}
\int du \, u^+_i u^-_j &=& \frac{1}{2!} \epsilon_{ij}, 
\label{eq:u-int-1}                                               \\ 
\int du \, u^+_i u^+_j u^-_k u^-_l 
&=& \frac{1}{3!} 
     (\epsilon_{ik}\epsilon_{jl}+\epsilon_{il}\epsilon_{jk}), 
\label{eq:u-int-2}                                               \\ 
\int du_1 \, \frac{u_1^+}{u_1^+ u_2^+} &=& -u_2^-, 
\label{eq:u-int-3}                                               \\ 
\int du_3 \, 
 \frac{u_3^+{}^{(i} u_3^+{}^{j)}}{(u_3^+ u_1^+)(u_2^+ u_3^+)} 
&=& \frac{u_1^+{}^{(i} u_1^-{}^{j)} - u_2^+{}^{(i} u_2^-{}^{j)}}
         {u_1^+ u_2^+}, 
\label{eq:u-int-4}
\end{eqnarray}
%and the equation
%\begin{eqnarray}
%(u_1^+u_2^+)(u_3^+u_4^+) 
%&=& (u_1^+u_3^+)(u_2^+u_4^+) 
%   -(u_1^+u_4^+)(u_2^+u_3^+). 
%\label{eq:}
%\end{eqnarray}
{}From the single $P$,
we have thirteen types contributions to the Lagrangian, listed
as  below:
\begin{eqnarray}
L_1 
&=& \frac{i}{6} \int d^8 \theta 
     \frac{du_1 du_2 du_3}{(u_1^+u_2^+) (u_2^+u_3^+) (u_3^+u_1^+)} 
                                                        \nonumber\\ 
&&{}\qquad \times 4\sqrt{2}i 
    \left\{ 
     \theta_1^+ \sigma^{\mu} \bar{\theta}_1^+ \ P \ 
     \theta_2^+ \sigma^{\nu} \bar{\theta}_2^+ 
    \right\} 
     (\theta_3^+)^2 (\theta_3^-)^2 (\bar{\theta}_3^+)^2 
     A_{\mu} A_{\nu} \, \partial^2 \bar{\phi},           %\nonumber\\ 
%&=& -\frac{\sqrt{2}}{12} (\sigma^{\mu}\bar{\sigma}^{\nu}
%                            \varepsilon)_{\alpha\beta} 
%      C_{ij}^{\alpha\beta} \epsilon^{ij} 
%      A_{\mu} A_{\nu} \partial^2 \bar{\phi}, 
\end{eqnarray}
%where we use eqs.(\ref{eq:star-1}),(\ref{eq:contruct-3}) 
%and (\ref{eq:u-int-3}). 
\begin{eqnarray}
L_2 
&=& \frac{i}{6} \int d^8 \theta 
     \frac{du_1 du_2 du_3}{(u_1^+u_2^+) (u_2^+u_3^+) (u_3^+u_1^+)} 
                                                         \nonumber\\ 
&&{}\qquad \times 32 
    \left\{ 
     \theta_1^+ \sigma^{\mu} \bar{\sigma}^{\nu} \theta_1^- \ P \ 
     (\theta_2^+)^2 \bar{\theta}_2^+{}_{\dot{\alpha}} 
    \right\} 
    (\theta_3^+)^2 \bar{\theta}_3^+{}_{\dot{\beta}} 
    \partial_{\nu} A_{\mu} \bar{\psi}_2^-{}^{\dot{\alpha}} 
    \bar{\psi}_3^-{}^{\dot{\beta}},                       %\nonumber\\ 
%&=& -\frac{2i}{3} \, C_{(ij)}^{\mu\nu} 
%     \partial_{\mu} A_{\nu} \bar{\psi}^i \bar{\psi}^j, 
\end{eqnarray}
%where we use eqs.(\ref{eq:star-6}),(\ref{eq:contruct-2}),
%(\ref{eq:contruct-3}),(\ref{eq:u-int-1}). 
%Third we have 
\begin{eqnarray}
L_3 
&=& \frac{i}{6} \int d^8 \theta 
     \frac{du_1 du_2 du_3}{(u_1^+u_2^+) (u_2^+u_3^+) (u_3^+u_1^+)} 
                                                         \nonumber\\ 
&&{}\qquad \times (-48) 
    \left\{ 
     (\theta_1^+)^2 \bar{\theta}_1^+{}_{\dot{\alpha}} \ P \ 
     (\theta_2^+)^2 \bar{\theta}_2^+{}_{\dot{\beta}} 
    \right\} 
    (\theta_3^+)^2 (\bar{\theta}_3^+)^2 
    \bar{\psi}_1^-{}^{\dot{\alpha}} \bar{\psi}_2^-{}^{\dot{\beta}} 
    D_{33}^{--},                                       %   \nonumber\\ 
%&=& +\frac{i}{6} C_s \bar{\psi}^i \bar{\psi}^j D_{ij}, 
\end{eqnarray}
%where we use eqs.(\ref{eq:star-7}),(\ref{eq:contruct-2}),
%(\ref{eq:contruct-3}),(\ref{eq:u-int-1}). 
%Fourth we have 
\begin{eqnarray}
L_4 
&=& \frac{i}{6} \int d^8 \theta 
     \frac{du_1 du_2 du_3}{(u_1^+u_2^+) (u_2^+u_3^+) (u_3^+u_1^+)} 
                                                         \nonumber\\ 
&&{}\times (-4\sqrt{2}i) 
  \left\{ 
   (\bar{\theta}_1^+)^2 
    \theta_1^+ \sigma^{\mu} \bar{\sigma}^{\nu} \theta_1^- \ P \ 
   \theta_2^+ \sigma^{\rho} \bar{\sigma}^{\sigma} \theta_2^-
   (\bar{\theta}_2^+)^2 
  \right\} 
   (\theta_3^+)^2 
   \partial_{\nu} A_{\mu} \partial_{\sigma} A_{\rho} \, \bar{\phi}, 
\end{eqnarray}
%where we use eqs.(\ref{eq:star-5}),(\ref{eq:contruct-1}),
%(\ref{eq:contruct-2}),(\ref{eq:contruct-3}),(\ref{eq:u-int-1}) 
%and (\ref{eq:u-int-4}). 
%Fiveth we have 
\begin{eqnarray}
L_5 
&=& \frac{i}{6} \int d^8 \theta 
     \frac{du_1 du_2 du_3}{(u_1^+u_2^+) (u_2^+u_3^+) (u_3^+u_1^+)} 
                                                        \nonumber\\ 
&&{}\qquad \times (-9\sqrt{2}i) 
   \left\{ 
    (\theta_1^+)^2 (\bar{\theta}_1^+)^2 \ P \ 
    (\theta_2^+)^2 
   \right\} 
   (\theta_3^+)^2 (\bar{\theta}_3^+)^2 
   D_{11}^{--} \bar{\phi} D_{33}^{--},                %    \nonumber\\ 
%&=& -\frac{\sqrt{2}}{12} C_s \, D_{ij} \, D^{ij} \, \bar{\phi}, 
\end{eqnarray}
%where we use eqs.(\ref{eq:star-9}),(\ref{eq:contruct-1}),
%(\ref{eq:contruct-3}) and (\ref{eq:u-int-2}). 
%Sixth we have 
\begin{eqnarray}
L_6 
&=& \frac{i}{6} \int d^8 \theta 
     \frac{du_1 du_2 du_3}{(u_1^+u_2^+) (u_2^+u_3^+) (u_3^+u_1^+)} 
                                                        \nonumber\\ 
&&{}\qquad \times (-32)
    \left\{ 
     \theta_1^+ \sigma^{\mu} \bar{\theta}_1^+ \ P \ 
     (\theta_2^+)^2 \bar{\theta}_2^+{}_{\dot{\alpha}} 
    \right\} 
    (\theta_3^+)^2 (\bar{\theta}_3^+)^2 
    (\theta_3^- \sigma^{\nu})_{\dot{\beta}} 
    A_{\mu} \bar{\psi}_2^-{}^{\dot{\alpha}} 
    \partial_{\nu} \bar{\psi}_3^-{}^{\dot{\beta}},   %     \nonumber\\ 
%&=&-\frac{i}{12} C_s \varepsilon^{\alpha\beta} A_{\mu} 
%    (\sigma^{\mu} \bar{\psi}^i)_{\alpha} 
%    (\sigma^{\nu} \partial_{\nu} \bar{\psi}_i)_{\beta} 
%   +\frac{i}{9} \, C_{(ij)}^{(\alpha\beta)} A_{\mu} 
%    (\sigma^{\mu} \bar{\psi}^i)_{\alpha} 
%    (\sigma^{\nu} \partial_{\nu} \bar{\psi}^j)_{\beta}, 
\end{eqnarray}
%where we use eqs.(\ref{eq:star-3}),(\ref{eq:contruct-2}),
%(\ref{eq:u-int-1}),(\ref{eq:u-int-2}) and (\ref{eq:u-int-3}). 
%Seventh we have 
\begin{eqnarray}
L_7 
&=& \frac{i}{6} \int d^8 \theta 
     \frac{du_1 du_2 du_3}{(u_1^+u_2^+) (u_2^+u_3^+) (u_3^+u_1^+)} 
                                                         \nonumber\\ 
&&{}\qquad \times 16\sqrt{2} 
    \left\{ 
     (\bar{\theta}_1^+)^2 \theta_1^+{}^{\alpha} \ P \ 
     (\theta_2^+)^2 
    \right\} 
    (\theta_3^+)^2 (\theta_3^- \sigma^{\mu})_{\dot{\alpha}} 
    (\bar{\theta}_3^+)^2 
    \psi_1^-{}_{\alpha} \, \bar{\phi} \, 
    \partial_{\mu} \bar{\psi}_3^-{}^{\dot{\alpha}}, %      \nonumber\\ 
%&=& -\frac{\sqrt{2}}{12}i \, C_s \, 
%     \psi^i \sigma^{\mu} \partial_{\mu} \bar{\psi}_i \, \bar{\phi} 
%    +\frac{\sqrt{2}}{9}i \, C_{(ij)}^{(\alpha\beta)} \, 
%     \bar{\phi} \, \psi^i_{\alpha} 
%     (\sigma^{\mu}\partial_{\mu} \bar{\psi}^j)_{\beta}, 
\end{eqnarray}
%where we use eqs.(\ref{eq:star-3}),(\ref{eq:contruct-1}),
%(\ref{eq:contruct-2}),(\ref{eq:u-int-1}),(\ref{eq:u-int-2}) 
%and (\ref{eq:u-int-3}). 
%Eighth we have 
\begin{eqnarray}
L_8 
&=& \frac{i}{6} \int d^8 \theta 
     \frac{du_1 du_2 du_3}{(u_1^+u_2^+) (u_2^+u_3^+) (u_3^+u_1^+)} 
                                                         \nonumber\\ 
&&{}\qquad \times (-32) 
   \left\{ 
    \sqrt{2}i (\bar{\theta}_1^+)^2 \theta_1^+{}^{\alpha} \ P \ 
    (\theta_2^+)^2 \bar{\theta}^+_2{}_{\dot{\alpha}} 
   \right\} 
   (\theta_3^+)^2 \theta_3^- \sigma^{\nu} \bar{\theta}_3^+ 
   \psi_1^-{}_{\alpha} \bar{\psi}_2^-{}^{\dot{\alpha}} \, 
   \partial_{\mu} \bar{\phi}, %                            \nonumber\\ 
%&=& +\frac{\sqrt{2}}{12}i \, C_s 
%     \psi^i \sigma^{\mu} \bar{\psi}_i \, \partial_{\mu} \bar{\phi} 
%    -\frac{\sqrt{2}i}{3} \, C_{(ij)}^{\alpha\beta} 
%     \psi^i_{\alpha} (\sigma^{\mu}\bar{\psi}^j)_{\beta} \, 
%     \partial_{\mu} \bar{\phi}, 
\end{eqnarray}
%where we use eqs.(\ref{eq:star-3}),(\ref{eq:contruct-2}),
%(\ref{eq:contruct-3}),(\ref{eq:u-int-1}). 
%Nineth we have 
\begin{eqnarray}
L_9 
&=& \frac{i}{6} \int d^8 \theta 
    \frac{du_1 du_2 du_3}{(u_1^+u_2^+) (u_2^+u_3^+) (u_3^+u_1^+)} 
                                                         \nonumber\\ 
&&{} \times (-8\sqrt{2}i) 
    \left\{ 
     (\bar{\theta}_1^+)^2 
     \theta_1^+ \sigma^{\mu} \bar{\sigma}^{\nu} \theta_1^- \ P \ 
     \theta_2^+ \sigma^{\rho} \bar{\theta}_2^+ 
    \right\} 
    (\theta_3^+)^2 \theta_3^- \sigma^{\sigma} \bar{\theta}_3^+
    \partial_{\nu} A_{\mu} A_{\rho} \partial_{\sigma} \bar{\phi},
%                                                         \nonumber\\ 
%&=& -\frac{\sqrt{2}}{6} \, C_s {\rm tr} \, 
%     (\sigma^{\mu\nu}\sigma^{\rho\sigma}) 
%     \partial_{\mu} A_{\nu} A_{\sigma} \partial_{\rho} \bar{\phi}, 
\end{eqnarray}
%where we use eqs.(\ref{eq:star-2}),(\ref{eq:contruct-2}),
%(\ref{eq:contruct-3}),(\ref{eq:u-int-1}) and (\ref{eq:u-int-3}). 
%Tenth we have 
\begin{eqnarray}
L_{10} 
&=& \frac{i}{6} \int d^8 \theta 
     \frac{du_1 du_2 du_3}{(u_1^+u_2^+) (u_2^+u_3^+) (u_3^+u_1^+)} 
                                                         \nonumber\\ 
&&{}\qquad \times 12\sqrt{2}i 
   \left\{ 
    (\theta_1^+)^2 (\bar{\theta}_1^+)^2 \ P \ 
    \theta_2^+ \sigma^{\mu} \bar{\theta}_2^+ 
   \right\} 
    (\theta_3^+)^2 \theta_3^- \sigma^{\nu} \bar{\theta}_3^+ 
    D_{11}^{--} A_{\mu} \partial_{\nu} \bar{\phi},   %     \nonumber\\ 
\end{eqnarray}
\begin{eqnarray}
L_{11} 
&=& \frac{i}{6} \int d^8 \theta 
     \frac{du_1 du_2 du_3}{(u_1^+u_2^+) (u_2^+u_3^+) (u_3^+u_1^+)} 
                                                         \nonumber\\ 
&&{}\qquad \times 6\sqrt{2}i 
  \left\{ 
   (\bar{\theta}_1^+)^2 
   \theta_1^+ \sigma^{\mu} \bar{\sigma}^{\nu} \theta_1^- \ P \ 
   (\theta_2^+)^2 
  \right\} 
   (\theta_3^+)^2 (\bar{\theta}_3^+)^2 
   \partial_{\nu} A_{\mu} \bar{\phi} D_{33}^{--}, %        \nonumber\\ 
\end{eqnarray}
\begin{eqnarray}
L_{12} 
&=& \frac{i}{6} \int d^8 \theta 
     \frac{du_1 du_2 du_3}{(u_1^+u_2^+) (u_2^+u_3^+) (u_3^+u_1^+)} 
                                                        \nonumber\\ 
&&{}\qquad \times 8\sqrt{2}i 
  \left\{ 
   (\theta_1^+)^2 \theta_1^- \sigma^{\mu} \bar{\theta}_1^+ 
\ P \ (\theta_2^+)^2 \theta_2^- \sigma^{\nu} \bar{\theta}_2^+ 
  \right\} 
   (\bar{\theta}_3^+)^2 
   \partial_{\mu}\bar{\phi} \, \partial_{\nu}\bar{\phi} \, \phi ,
\end{eqnarray}
\begin{eqnarray}
L_{13} 
&=& \frac{i}{6} \int d^8 \theta 
     \frac{du_1 du_2 du_3}{(u_1^+u_2^+) (u_2^+u_3^+) (u_3^+u_1^+)} 
                                                        \nonumber\\ 
&&{}\qquad \times (-2\sqrt{2}i) 
  \left\{ 
   (\theta_1^+)^2 (\theta_1^-)^2 (\bar{\theta}_1^+)^2 
\ P \ (\theta_2^+)^2 
  \right\} 
   (\bar{\theta}_3^+)^2 
   \partial^2 \bar{\phi} \, \bar{\phi} \, \phi.      %     \nonumber\\ 
\end{eqnarray}
These terms are computed by applying the formulas (\ref{eq:star-1})-
(\ref{eq:u-int-4}). 
The results are as follows:
\begin{eqnarray}
 L_{1}&=&-\frac{\sqrt{2}}{12} (\sigma^{\mu}\bar{\sigma}^{\nu}
                            \varepsilon)_{\alpha\beta} 
      C_{ij}^{\alpha\beta} \epsilon^{ij} 
      A_{\mu} A_{\nu} \partial^2 \bar{\phi},\nonumber\\
 L_{2}&=&-\frac{2i}{3} \, C_{(ij)}^{\mu\nu} 
     \partial_{\mu} A_{\nu} \bar{\psi}^i \bar{\psi}^j, 
\nonumber\\
 L_{3}&=& \frac{i}{6} C_s \bar{\psi}^i \bar{\psi}^j D_{ij},\nonumber\\
% L_{4}&=& 0\nonumber\\
 L_{5}&=& -\frac{\sqrt{2}}{12} C_s \, D_{ij} \, D^{ij} \, \bar{\phi},
\nonumber\\
 L_{6}&=&-\frac{i}{12} C_s \varepsilon^{\alpha\beta} A_{\mu} 
    (\sigma^{\mu} \bar{\psi}^i)_{\alpha} 
    (\sigma^{\nu} \partial_{\nu} \bar{\psi}_i)_{\beta} 
   +\frac{i}{9} \, C_{(ij)}^{(\alpha\beta)} A_{\mu} 
    (\sigma^{\mu} \bar{\psi}^i)_{\alpha} 
    (\sigma^{\nu} \partial_{\nu} \bar{\psi}^j)_{\beta}, \nonumber\\
 L_{7}&=&\frac{\sqrt{2}}{12}i \, C_s \, 
     \psi^i \sigma^{\mu} \partial_{\mu} \bar{\psi}_i \, \bar{\phi} 
    -\frac{\sqrt{2}}{9}i \, C_{(ij)}^{(\alpha\beta)} \, 
     \bar{\phi} \, \psi^i_{\alpha} 
     (\sigma^{\mu}\partial_{\mu} \bar{\psi}^j)_{\beta},
\nonumber\\
 L_{8}&=&\frac{\sqrt{2}}{12}i \, C_s 
     \psi^i \sigma^{\mu} \bar{\psi}_i \, \partial_{\mu} \bar{\phi} 
    -\frac{\sqrt{2}i}{3} \, C_{(ij)}^{\alpha\beta} 
     \psi^i_{\alpha} (\sigma^{\mu}\bar{\psi}^j)_{\beta} \, 
     \partial_{\mu} \bar{\phi},\nonumber\\
 L_{9}&=&-\frac{\sqrt{2}}{6} \, C_s {\rm Tr} \, 
     (\sigma^{\mu\nu}\sigma^{\rho\sigma}) 
     \partial_{\mu} A_{\nu} A_{\sigma} \partial_{\rho} \bar{\phi},
\nonumber\\
 L_{10}&=&-\frac{\sqrt{2}}{6} \, C_{(ij)}^{\mu\nu} 
    D^{ij} A_{\nu} \partial_{\mu} \bar{\phi},
\nonumber\\
 L_{11}&=&\frac{\sqrt{2}}{6} \, C_{(ij)}^{\mu\nu} 
     \partial_{\mu} A_{\nu} \, \bar{\phi} \, D^{ij}, 
\nonumber\\
L_{4}&=&L_{12}=L_{13}=0.
\end{eqnarray}
The contributions to the Lagrangian coming  from 
$P^3$ are two types:
\begin{eqnarray}
L_{14} 
&=& \frac{i}{6} \int d^8 \theta 
     \frac{du_1 du_2 du_3}{(u_1^+u_2^+) (u_2^+u_3^+) (u_3^+u_1^+)} 
                                                        \nonumber\\ 
&&{}\qquad \times (-8\sqrt{2}i) 
   \left\{ 
    (\theta_1^+)^2\theta_1^-\sigma^{\mu}\bar{\theta}_1^+ 
\ \frac{1}{3!} P^3 \ 
    (\theta_2^+)^2 \theta_2^- \sigma^{\nu} \bar{\theta}_2^+)^2 
   \right\} 
   (\theta_3^+)^2 (\theta_3^-)^2 (\bar{\theta}_3^+)^2 \, 
   \partial_{\mu} \bar{\phi} \, \partial_{\nu} \bar{\phi} \, 
   \partial^2 \bar{\phi},                          \nonumber\\ 
\end{eqnarray} 
\begin{eqnarray}
L_{15} 
&=& \frac{i}{6} \int d^8 \theta 
     \frac{du_1 du_2 du_3}{(u_1^+u_2^+) (u_2^+u_3^+) (u_3^+u_1^+)} 
                                                         \nonumber\\ 
&&{}\times 2\sqrt{2}i 
   \left\{ 
    (\theta_1^+)^2 (\theta_1^-)^2 (\bar{\theta}_1^+)^2 
\ \frac{1}{3!} P^3 \ 
    (\theta_2^+)^2 (\theta_2^-)^2 (\bar{\theta}_2^+)^2 
   \right\} 
   (\theta_3^+)^2 
   \partial^2 \bar{\phi} \, \partial^2 \bar{\phi} \, \bar{\phi}. 
\end{eqnarray}
Using the formulas
\begin{eqnarray}
(\theta^+_1)^2 \theta^-_1{}^{\alpha} \ P^3 \ 
 (\theta^+_2)^2 \theta^-_2{}^{\beta} 
&=& -3\det{C_{12}^{++}} C_{12}^{--}{}^{\alpha\beta} 
    +3C_{12}^{++}{}^{\gamma^{\prime}\delta^{\prime}} 
     C_{12}^{+-}{}^{\beta^{\prime}\beta} 
     C_{12}^{-+}{}^{\alpha\alpha^{\prime}} 
     \varepsilon_{\gamma^{\prime}\beta^{\prime}} 
     \varepsilon_{\delta^{\prime}\alpha^{\prime}},         \nonumber\\ 
(\theta^{+}_1)^2 (\theta^{-}_{1})^2 P^3
(\theta^{+}_{2})^2 (\theta_{2}^{-})^2&=&
-6C^{\alpha\delta}_{12} C^{\beta\epsilon}_{21} C^{\gamma\zeta}_{22}
\varepsilon_{\alpha\epsilon}\varepsilon_{\delta\zeta}
\varepsilon_{\beta\gamma} (\theta^{1})^2
+6 C^{\alpha\delta}_{11} C^{\beta\epsilon}_{12} C^{\gamma\zeta}_{21}
\varepsilon_{\alpha\beta}\varepsilon_{\delta\zeta}
\varepsilon_{\epsilon\gamma} (\theta^{2})^2,
\nonumber\\
\int du_1 \, 
 \frac{u_1^+{}^i u_1^+{}^k u_1^-{}^m}{u_1^+ u_2^+} 
&=& -\frac{1}{2} 
     \left( 
       u_2^+{}^{(i} u_2^-{}^{k} u_2^-{}^{m)} 
      +\frac{4}{3} \epsilon^{m(k} u_2^-{}^{i)} 
     \right),                                           \nonumber\\ 
\int du_1 du_2 \, 
 \frac{u_1^+{}_i u_1^+{}_k u_1^-{}^m u_2^+{}^j u_2^+{}^l u_2^-{}^n}
      {u_1^+ u_2^+} 
&=& +\frac{1}{24} \epsilon^{m(l} \delta_k{}^{j} \delta_i{}^{n)} 
    +\frac{2}{9} \epsilon^{n(j} \delta_{(i}{}^{l)} \delta_{k)}{}^{m}. 
\end{eqnarray}
we find $L_{14}=L_{15}=0$.
Including the combinatorial factor, 
the third order Lagrangian ${\cal L}_3$ becomes 
\begin{eqnarray}
{\cal L}_3 
&=& 3 \times \sum_{i=1}^6 L_i 
   +6 \times \sum_{i=7}^{15} L_i                         \nonumber\\ 
&=&  +\frac{\sqrt{2}}{4} C_s A_{\mu} A^{\mu} \partial^2 \bar{\phi}
	     -\frac{i}{2} C_s \varepsilon^{\alpha\beta} A_{\mu}
	      (\sigma^{\mu} \bar{\psi}^i)_{\alpha}
	      (\sigma^{\nu} \partial_{\nu} \bar{\psi}_i)_{\beta}
	     +\frac{2}{3}i \, C_{(ij)}^{\alpha\beta} A_{\mu}
	      (\sigma^{\mu} \bar{\psi}^i)_{\alpha}
	      (\sigma^{\nu} \partial_{\nu} \bar{\psi}^j)_{\beta} \nonumber\\ 
&&{}-2i C_{(ij)}^{\mu\nu} \partial_{\mu} A_{\nu} 
                           \bar{\psi}^i \bar{\psi}^j 
    +\frac{i}{2} C_s \bar{\psi}^i \bar{\psi}^j D_{ij} 
    +\frac{\sqrt{2}}{2}i \, C_s 
     \psi^i \sigma^{\mu} \partial_{\mu} \bar{\psi}_i \bar{\phi} 
    -\frac{2\sqrt{2}}{3}i \, C_{(ij)}^{\alpha\beta} 
     \psi^i_{\alpha} (\sigma^{\mu}\partial_{\mu}\bar{\psi}^j)_{\beta} 
     \bar{\phi}                                          \nonumber\\ 
& & +\frac{\sqrt{2}}{2}i \, C_s 
     \psi^i \sigma^{\mu} \bar{\psi}_i \partial_{\mu} \bar{\phi} 
    -2\sqrt{2}i \, C_{(ij)}^{\alpha\beta} 
     \psi^i_{\alpha} (\sigma^{\mu}\bar{\psi}^j)_{\beta} 
     \partial_{\mu} \bar{\phi} 
    -\sqrt{2} \, C_s {\rm Tr} \, (\sigma^{\mu\nu}\sigma^{\rho\sigma}) 
     \partial_{\mu} A_{\nu} A_{\sigma} \partial_{\rho} \bar{\phi} 
                                                         \nonumber\\ 
&&{}-\sqrt{2} \, C_{(ij)}^{\mu\nu} 
     D^{ij} A_{\nu} \partial_{\mu} \bar{\phi} 
    +\sqrt{2} \, C_{(ij)}^{\mu\nu} \partial_{\mu} A_{\nu} \bar{\phi} 
     D^{ij} 
    -\frac{\sqrt{2}}{4} C_s D_{ij} D^{ij} \bar{\phi}
.  
\end{eqnarray}
This is shown to be equal to (\ref{eq:thirdorderLagrangian}) 
with the use of the formula 
${\rm Tr} \, (\sigma^{\mu\nu}\sigma^{\rho\sigma})
= - \frac12 (\eta^{\mu\rho} \eta^{\nu\sigma} -\eta^{\mu\sigma} \eta^{\nu\rho})
	- {i\over 2} \varepsilon^{\mu\nu\rho\sigma}
$.

\end{document}